\documentclass[11pt,a4paper]{article}
\usepackage{pdflscape}
\usepackage{caption}
\usepackage{tabularx}
\usepackage{adjustbox}

\usepackage[T1]{fontenc}
\usepackage[utf8]{inputenc}
\usepackage{lmodern}
\usepackage{authblk}

\usepackage[a4paper,margin=2.2cm]{geometry}
\usepackage{microtype}

\usepackage{graphicx}
\usepackage{booktabs}
\usepackage{longtable}
\usepackage{listings}
\usepackage{array}

\usepackage[hidelinks]{hyperref}

\title{AI-Driven Multiscenario Interest Rate Forecasting in Banks: A Proof-of-Concept Prototype}
\author[1]{Dr.\ Ekkehardt\ Bauer}
\author[1,2]{Dr.\ Dirk\ Holl\"ander}
\author[1]{David\ Scholz}
\author[1]{Linus\ Wolff}
\author[1]{Dr.\ Christoph\ Ostermair}
\author[1]{Kyrillus\ Aiad}
\author[2]{Prof.\ Dr.\ Joachim\ Hasebrook}

\affil[1]{zeb.rolfes.schierenbeck.associates Ltd.}
\affil[2]{zeb Institute for Financial Services, University Witten/Herdecke}
\date{}

\newcolumntype{L}[1]{>{\raggedright\arraybackslash}p{#1}}
\begin{document}
\setcounter{secnumdepth}{3}
\maketitle
\begin{abstract}
This study focuses on the development of an AI-supported prototype for multiperspective interest rate forecasting, which combines classical econometric models with modern artificial intelligence methods. The system, which was tested in a major European bank, enables a more precise and flexible prediction of interest rate developments and supports strategic decision-making in Asset-Liability Management (ALM). The prototype integrates topic modelling, sentiment analysis, econometric forecasting, and market-based analyses within an interactive platform. By leveraging AI to analyze large volumes of financial documents and market data, monetary policy trends and sentiment signals can be identified at an early stage. The core econometric model, a Bayesian vector autoregression (BVAR), enables simulation-based scenario analyses that evaluate economic developments from multiple perspectives. The innovation of the system lies in the integration of several forecasting approaches that consolidate previously separate information sources and present them in a transparent and interpretable manner. Financial analysts and risk managers thus gain an improved decision-making basis, allowing for a more accurate assessment of interest rate risks and more forward-looking management of market movements. While the prototype already demonstrates how AI can transform interest rate management in banking, further development is required to optimize real-time data integration and regulatory compliance. Even at this stage, the study shows that multi-perspective AI-driven forecasting provides substantial added value for banks, by increasing transparency, strengthening evidence-based decision-making, and improving risk steering.
\end{abstract}
\section{Introduction and Background}
\subsection{Relevance of Multiscenario Interest Rate Forecasting in Banking}
Accurate interest rate forecasts are essential for banks—particularly for asset and liability management (ALM), internal risk assessment, and the effective response to monetary policy measures (Bohn \& Schneider, 2024). Traditional econometric models, such as vector autoregression (VAR) and autoregressive integrated moving average (ARIMA) models, can account for nonlinear economic relationships, structural breaks, and exogenous shocks only to a limited extent (Wiriyawit \& Wong, 2016; Herwartz, Lange \& Maxand, 2022). The complexity of financial markets and the increasing volatility of macroeconomic conditions require more adaptive forecasting methods, also from a regulatory perspective (European Central Bank, 2025).

From a management perspective, financial institutions require forecasting models that integrate multiple data sources, enable scenario-based risk assessments, and improve the efficiency of decision-making (Cao et al., 2024). AI-driven approaches, particularly those incorporating sentiment analysis, topic modelling (topic maps), and Bayesian vector autoregression (BVAR), enable banks to process both structured and unstructured data in order to refine interest rate forecasts (Pillai, 2023). This allows for more responsive portfolio adjustments, improved liquidity management, and compliance with regulatory requirements (Polak et al., 2020; Sutiene et al., 2024).

From a scientific perspective, the integration of AI into traditional macroeconomic models improves forecasting accuracy by mitigating overparameterization and uncertainty (Chen, 2024). BVAR, when complemented with machine learning, enables probabilistic forecasts instead of single-point estimates, while AI-driven sentiment analysis improves real-time responsiveness to economic signals (McKinsey, 2024). Hybrid approaches that combine econometric modelling with AI methods contribute to the development of more robust and interpretable forecasting frameworks (Wiriyawit \& Wong, 2016; Herwartz, Lange \& Maxand, 2022).

This study examines how financial institutions can use AI-driven multiscenario frameworks to improve the accuracy of interest rate forecasts. Such frameworks support structured risk management, increase adaptability to volatile market conditions, and promote evidence-based, data-driven decision-making. The objective is to improve banking operations and contribute to financial stability.

\subsection{Overview of Interest Rate Forecasting Methods in Banking}
Interest rate forecasting is a fundamental component of financial decision-making and influences ALM, risk assessment, and portfolio management. Traditional forecasting models remain widely used in banking, but their limitations in handling nonlinear dynamics, structural breaks, and evolving market conditions have encouraged research into machine learning and AI-driven approaches (Cerniglia \& Fabozzi, 2020).

Banks primarily rely on econometric models, including term structure models such as the Nelson–Siegel and Svensson models, which capture level, slope, and curvature effects, but often struggle to adapt to sudden changes in financial markets (Delucchi \& Giribone, 2023). VAR and its Bayesian counterpart (BVAR) remain widely used for modelling interdependencies among macroeconomic variables, incorporating prior distribution assumptions to improve stability when working with limited datasets (Liu \& Suzuki, 2024). Time-series models such as ARIMA and GARCH are also frequently applied to interest rate forecasting; however, their dependence on historical data and stationarity assumptions reduces their ability to capture dynamic market shifts (Li, Wang \& Chen, 2024; Salem \& Albourawi, 2024). Principal component analysis (PCA) is commonly used to extract dominant factors from interest rate fluctuations, although its effectiveness is constrained by the assumption of linear relationships (Liu \& Suzuki, 2024).

More recent research focuses on AI-driven methods that provide greater flexibility in forecasting complex interest rate dynamics. Gaussian process regression (GPR) has emerged as a non-parametric alternative to traditional term structure models, enabling flexible yield curve modelling without requiring a predefined functional form and addressing market anomalies such as negative interest rates (Delucchi \& Giribone, 2023). Deep learning techniques, particularly long short-term memory (LSTM) networks and transformer-based models, improve forecasting accuracy by capturing nonlinear dependencies and integrating unstructured data sources, including sentiment analysis of central bank communications (Li, Wang \& Chen, 2024).

The dynamic mode decomposition (DMD) method has been investigated as a tool for analyzing high-frequency financial data, decomposing time-series data into dominant modes that enhance forecast robustness (Karimkhani et al., 2023). Hybrid models combining Bayesian methods with machine learning have also been studied, integrating AI-generated sentiment indices into BVAR frameworks to refine economic scenario analysis (Cerniglia \& Fabozzi, 2020). In addition, the Malliavin–Mancino method has been introduced as an approach for estimating interest rate volatility structures, providing a more flexible alternative to previous volatility models (Liu \& Suzuki, 2024).

While AI-driven methods improve adaptability and accuracy, challenges remain regarding interpretability, computational efficiency, and regulatory compliance (Bohn \& Schneider, 2024). The opacity of deep learning models limits their direct application in financial institutions, where transparency and explainability are critical for risk assessment. Concerns about overfitting and robustness, particularly under volatile economic conditions, highlight the need for rigorous validation techniques (Delucchi \& Giribone, 2023).

Future advances in interest rate forecasting are likely to arise from the integration of explainable AI (XAI) methods, the optimization of hybrid econometric–machine–learning frameworks, and the enhancement of real-time forecasting capabilities, enabling better alignment with regulatory requirements and financial stability objectives.

Forecasting methodologies in banking continue to evolve as AI and machine learning techniques gain importance alongside established econometric approaches. While traditional models such as VAR, BVAR, and Nelson–Siegel remain widely used, the integration of GPR, LSTM, DMD, and Fourier-based methods represents a shift toward more data-driven and adaptive forecasting frameworks. The adoption of hybrid models can balance predictive accuracy and interpretability, guide future research, and ensure that financial institutions can manage interest rate risk more effectively while maintaining methodological transparency.

\subsection{Existing Information Sources for Interest Rate Forecasting in Banking}
The accuracy of interest rate forecasts in banking depends on the integration of multiple data sources, including market-based indicators, survey-based expectations, central bank communications, and econometric models. These sources provide complementary perspectives, enhancing forecast robustness by capturing both market expectations and macroeconomic fundamentals (Cerniglia \& Fabozzi, 2020; Delucchi \& Giribone, 2023).

Market-based indicators play a central role in interest rate forecasting. Inflation swap rates reflect inflation expectations, which are particularly reliable over short horizons (Ang et al., 2008). Forward rates, derived from yield curves, serve as market-based expectations of future interest rates, although they include term premia that may introduce bias (Fama \& Bliss, 1987). Swap curves, such as EU swap rates (vs. 6-month Euribor) and U.S. swap rates (vs. SOFR), reflect liquidity and credit risks embedded in interest rate markets and are frequently used in comparative forecasting models (Diebold \& Li, 2006).

Survey-based forecasts provide qualitative insights that complement market-based measures. The Survey of Professional Forecasters (SPF) is widely regarded as a benchmark for inflation and interest rate expectations, capturing expert assessments of forward-looking macroeconomic conditions (Ang et al., 2007). Consensus Economics surveys and Blue Chip Financial Forecasts aggregate the views of leading economists and analysts, offering structured long-term expectations that extend econometric models (Faust \& Wright, 2013).

Central bank communication, including Federal Open Market Committee (FOMC) statements, minutes, ECB decisions, and ECB reports, is critical for understanding the monetary policy stance and its impact on interest rate expectations. Empirical evidence shows that central bank guidance significantly influences market expectations, and that sentiment extracted from policy statements improves the predictive power of interest rate models (Ehrmann \& Fratzscher, 2005).

Econometric models continue to play a central role in interest rate forecasting by using structured statistical frameworks to estimate yield curve movements. Linear interpolation techniques are frequently applied to generate smooth interest rate projections, while structured change models forecast interest rate movements based on historical patterns (Diebold \& Li, 2006). Implied forward models extend this approach by deriving expected interest rate paths from current term structures, incorporating policy-driven adjustments (Ang et al., 2008).

Alternative and composite forecasting methods provide additional perspectives. Random-walk and mean-reversion models assume that interest rates either remain at current levels or revert to a long-term mean, offering baseline scenarios for comparative analysis (Rudebusch, 2002). Combining market-based, survey-based, and econometric models can yield more robust forecasts by leveraging the strengths of each approach (Faust \& Wright, 2013). Risk-adjusted models, which incorporate inflation risk premium, provide deeper insights into term structure dynamics (Chernov \& Mueller, 2012).

Nonlinear and macroeconomic trend-based models further refine interest rate forecasting methodologies. Black-linear and Black-ordered models account for the zero lower bound and discrete interest rate increments, improving forecast accuracy during periods of near-zero interest rates (Kim \& Orphanides, 2012). Macroeconomic forecasting models integrate GDP growth, inflation, and unemployment data, bridging fundamental economic indicators and interest rate projections (Rudebusch, 2002).

Finally, consensus government bond forecasts, compiled from Bloomberg analyst expectations, enable scenario-based modelling of yield curves. By clustering analyst forecasts according to expected yield curve developments, banks can assign probability weights to different interest rate scenarios, thereby improving risk-adjusted decision-making (Diebold et al., 2006).

By integrating market-based indicators, survey data, econometric models, and policy analysis, banks can enhance the accuracy and reliability of interest rate forecasts. A multi-source approach provides deeper insights into future interest rate trends and reduces uncertainty in ALM and strategic investment decisions.

\section{Methodological and Technical Background}
This section presents the methodological and technical foundations of the developed AI-supported forecasting model and explains the key procedures that contribute to improving interest rate forecasting. First, it addresses the use of topic maps, which enable the structured analysis of large volumes of unstructured financial text data and contributes to identifying macroeconomic trends, market sentiment, and regulatory changes. It then explains the role of sentiment analysis, which allows a more precise capture of monetary policy expectations and market reactions by extracting sentiments from central bank reports, analyst commentaries, and financial news. In addition, Bayesian vector autoregression (BVAR) is introduced as a probabilistic econometric method that quantifies uncertainty in interest rate forecasting and models multiple scenarios for strategic decision-making in ALM. Finally, the multiscenario approach is discussed, combining classical econometric procedures with AI-supported analytical methods to enable a more robust and adaptive interest rate forecast through the integration of diverse data sources. These methodological components form the basis of the developed prototype and make a decisive contribution to data-driven decision-making in banking.

\subsection{Application of Topic Maps in Interest Rate Forecasting for Banking}
In financial forecasting, topic maps have become an indispensable tool for structuring and analyzing large volumes of unstructured financial text data. Topic maps facilitate the identification of macroeconomic trends, policy changes, and changes in market sentiment that influence interest rate movements (Salem \& Albourawi, 2024). By dividing financial documents, such as central bank reports, analyst commentaries, and regulatory publications, into coherent topics, these models enable data-driven decision-making in ALM and risk assessment.

Early implementations of topic modelling relied on bag-of-words (BoW) and naïve Bayes classifiers, which categorize documents based on word-frequency distributions but do not capture contextual meaning (Albalawi, Yeap \& Benyoucef, 2020). More advanced techniques (e.g., latent Dirichlet allocation (LDA) for predefined topics and hierarchical Dirichlet processes (HDP) for dynamic topic generation) improved topic extraction by modelling latent structures within texts (Delucchi \& Giribone, 2023). However, these probabilistic models remain constrained by their reliance on predefined word distributions, which limits adaptability to the evolving financial discourse.

Contemporary AI-driven approaches use deep learning techniques such as Sentence-BERT (SBERT) in combination with clustering algorithms such as UMAP and HDBScan to improve thematic categorization. These models generate semantically embedded representations of textual content, enabling finer topic differentiation and improved document-classification accuracy. Interactive visualization platforms have been integrated into banking workflows, providing analysts with dynamic representations of financial discourse and enabling real-time exploration of thematic linkages and sentiment trends.

A major advantage of topic maps in interest rate forecasting is their ability to integrate sentiment analysis, allowing a quantitative assessment of market sentiment fluctuations in relation to policy decisions (Li, Wang \& Chen, 2024). This is particularly important for monitoring central bank behavior and publications, where even subtle shifts in language can indicate turning points in monetary policy and affect multiscenario forecasting models. In addition, topic maps fulfil an essential function in regulatory compliance and macroprudential risk monitoring, providing systematic insights into emerging financial risks.

Challenges remain in terms of data preprocessing, interpretability, and computational efficiency. The complexity of financial texts requires advanced techniques for dimensionality reduction, while the regulatory demand for explainability underscores the need for transparent AI methods. Future research should focus on improving explainable AI (XAI) frameworks to ensure that topic-modelling techniques remain interpretable, reliable, and adaptable to financial decision-making.

By integrating NLP, sentiment analysis, and AI-driven clustering, topic maps provide financial institutions with a powerful tool for improving interest rate forecast accuracy. Their impact in macro-financial analysis, compliance monitoring, and strategic planning continues to grow, reinforcing the importance of topic maps in AI-driven banking analytics.

\subsection{Sentiment Analysis in Interest Rate Forecasting for Banking}
Sentiment analysis complements the thematic structuring of financial texts by providing a quantitative assessment of market sentiment and enables an extended analysis of macroeconomic expectations. While topic maps identify relevant themes, sentiment analysis provides a methodological basis for assessing the tone of economic policy statements, particularly in monetary policy communication. The importance of this methodology is especially evident in the interpretation of central bank reports, analyst commentaries, and financial news, which contain decisive signals for interest rate developments (Audrino \& Offner, 2024).

Earlier sentiment-analysis methods were based on lexicon-based procedures such as the Loughran--McDonald\linebreak lexicon, which was developed specifically for economics- and finance-related texts. These approaches enabled a rudimentary assessment of sentiment in financial documents but were limited in their ability to capture contextual nuances and idiomatic expressions (Loughran \& McDonald, 2011). Advances in machine learning led to the development of deep-learning models such as FinBERT, which are trained specifically on financial texts and enable more precise sentiment assessment (Huang, Wang \& Yang, 2023). Through the use of few-shot learning techniques, FinBERT can also be adapted to specialized financial terminology, further increasing classification accuracy (Delucchi \& Giribone, 2023).

Empirical research shows that sentiment indicators have significant predictive power for short-term interest rate movements. Negative sentiment trends in financial news and monetary policy statements are often associated with rising risk premia and more restrictive interest rate expectations, while positive market sentiment correlates with more expansionary monetary policy (Tetlock, 2007; Antweiler \& Frank, 2004). Incorporating this information into econometric models improves the quality of interest rate forecasts by complementing classical macroeconomic variables and providing enhanced insights into the expectation formation of market participants (Audrino \& Offner, 2024).

A key application area of sentiment analysis lies in interpreting central bank communication, because monetary policy signals are often formulated subtly and can only be fully captured through deeper semantic analysis. Documents such as FOMC statements, ECB minutes, and monetary policy reports contain key terms and linguistic nuances that indicate future interest rate policies. Empirical studies show that sentiment data from detailed central bank reports yield higher forecasting accuracy than immediate policy announcements, because they reflect the argumentation structure and macroeconomic assessments of decision-makers (Edison, 1997; Goldberg \& Klein, 2005; Audrino \& Offner, 2024).

Integrating sentiment data into econometric interest rate models leads to significant improvements in model accuracy. For example, analyses show that under negative market sentiment, short-term yields tend to decline, while increasingly positive market sentiment is associated with a steeper yield curve (Diebold \& Li, 2006). These correlations confirm that sentiment indicators can serve as early warning signals for changes in the interest rate landscape. In addition, impulse response analyses indicate that sentiment shocks can persist across multiple periods, underscoring their relevance for macroeconomic forecasting (Chudik \& Georgiadis, 2022).

Despite advances in sentiment analysis, methodological challenges remain, particularly with regard to interpretability and validation of classification results. Because sentiment analyses use probabilistic models, uncertainty in the classifications is an important factor that must be incorporated into modelling. Future developments should place greater emphasis on explainable AI (XAI) methods to ensure traceable decision-making and to increase trust in the application of sentiment analysis in the banking sector. Moreover, integrating high-frequency market data could enable improved real-time assessment of market sentiment.

By systematically incorporating sentiment analysis into interest rate forecasting, banks can not only improve forecast accuracy but also capture more precisely the dynamic interactions between market sentiment, monetary policy, and macroeconomic developments. Combining sentiment indicators with traditional yield curve models creates an expanded data basis for strategic decisions and risk management in the financial sector.

\subsection{Bayesian Vector Autoregression (BVAR) in Interest Rate Forecasting for Banking}
The Bayesian vector autoregression (BVAR) framework is an econometric AI model that is optimally suited for interest rate forecasting in banking and addresses the overparameterization and uncertainty problems of traditional vector autoregressive (VAR) models (Boeck, Feldkircher, \& Huber, 2022). By incorporating Bayesian priors, BVAR constrains parameter estimates and thereby improves forecast accuracy and stability even with limited historical data (Kuschnig \& Vashold, 2021).

BVAR models improve traditional VAR approaches by refining interest rate forecasts through the integration of economic priors and hierarchical Bayesian structures. This reduces the variance of parameter estimates by shrinking model parameters, among other things, toward a random walk (prior distribution). This increases forecast quality, particularly for non-stationary time series as typically observed in interest rate modelling.

In banking applications, BVAR is used to model swap rates across different maturities (2, 5, 10, and 30 years), using key macroeconomic predictors such as euro area inflation, industrial production, the ECB refinancing rate, U.S. inflation, and the manufacturing PMI (AI-Driven Framework for Multiscenario Interest Rate Forecasting in Banks Built on BVAR, 2024). The model applies Granger causality tests to filter statistically insignificant variables, thereby ensuring the inclusion of economically relevant criteria (Cepni et al., 2022).

A major strength of BVAR in banking is conditional forecasting, enabling scenario-based analysis in which simulated macroeconomic shocks—such as monetary policy adjustments or inflation pressure—propagate through the system. This allows banks to quantify interest rate uncertainty, assess potential yield curve developments, and optimize ALM (Diebold \& Li, 2006). The probabilistic structure of BVAR makes it particularly useful for stress testing, in which banks assess portfolio risks under different economic conditions (Boeck et al., 2022).

Despite its strengths, BVAR has limitations. Reliance on prior distributions introduces subjectivity, because forecast results depend on the specification of these priors (Koop \& Korobilis, 2010). An incorrect choice of Bayesian shrinkage priors could shrink model parameters in the wrong direction. The selection of distributions should therefore be the result of a solid macroeconomic theory (Cepni et al., 2022). In addition, the computational complexity of BVAR increases quadratically with the number of included variables, which requires careful model selection to maintain efficiency (Kuschnig \& Vashold, 2021).

For interest rate risk management in banks, BVAR provides a structured framework to forecast interest rate movements, optimize portfolio allocations, and refine hedging strategies. By integrating simulated economic scenarios, banks gain insights into the potential impact of macroeconomic fluctuations on performance, enabling data-driven planning and, where necessary, adjustments. The adoption of BVAR in AI-driven forecasting frameworks strengthens the analytical foundation for interest rate forecasting and improves decision-making in risk management and financial strategy.

\subsection{Multiscenario Approaches to Interest Rate Forecasting in Banking}
Traditional interest rate forecasting models—including econometric models (VAR, BVAR, ARIMA), market-based indicators, and survey-based expectations—are widely used in banking but are often constrained by structural weaknesses in their explanatory power (Diebold \& Li, 2006; Cerniglia \& Fabozzi, 2020). However, they serve as benchmarks for assessing the forecasting quality of the BVAR model in backtesting. The increasing complexity of financial markets requires a multi-perspective approach that integrates statistical, econometric, and AI-driven techniques to improve forecast accuracy and robustness (Delucchi \& Giribone, 2023). Despite its potential, multi-perspective forecasting remains underexplored both in academic research and in practical banking applications (Jain \& Kulkarni, 2023).

Most current models rely either on structured macroeconomic variables such as inflation rates, employment data, and central bank policy (BVAR), or on AI-driven techniques such as sentiment analysis and topic modelling that process unstructured financial data (Salem \& Albourawi, 2024). A balanced multiscenario approach integrates both structured and unstructured data and leverages machine-learning-based sentiment extraction, topic clustering, and econometric modelling (Chen, 2024). While single-method models capture either macroeconomic fundamentals (BVAR, term-structure models) or textual sentiment shifts (FinBERT, LDA topic modelling), multi-perspective frameworks enable a dynamic interplay between market sentiment, regulatory changes, and economic indicators (Ahmed \& Meenaskshi, 2024).

A key advantage of multiscenario forecasts lies in their ability to account for structural breaks and nonlinearities in financial markets. Hybrid models that combine deep learning (LSTMs) with Bayesian inference (BVAR) improve predictive accuracy by incorporating exogenous shocks and sentiment-based signals from financial narratives (Diebold et al., 2006; Delucchi \& Giribone, 2023). As already noted, time-series models assume stationarity, whereas AI-driven sentiment tracking enables real-time detection of changes in market sentiment and thereby allows more flexible scenario-based forecasts (Li, Wang \& Chen, 2024).

More recently, hybrid models have been developed that combine traditional econometric techniques with AI-driven methods. For example, models have been examined that integrate wavelet analysis, mixed-spectrum analysis, and nonlinear ARMA with Fourier coefficients in order to analyze non-stationarity and nonlinearity in financial time series (Clavel \& Nachane, 2008; Enke \& Mehdiyev, 2012).

Despite their advantages, multi-perspective forecasting faces challenges related to model complexity, data integration, and interpretability (Zhao et al., 2024). Deep learning techniques are effective in sentiment detection but are often not transparent in their decision-making, which reduces their usefulness in banks’ risk management (Cao et al., 2024). In addition, the computational effort required to integrate BVAR, sentiment analysis, and market-based forecasting tools demands high-frequency data processing capabilities, posing implementation challenges for financial institutions (Bata et al., 2024).

As regulatory frameworks adapt to AI-based forecasts, multi-perspective approaches are likely to play an increasing role in ALM, stress testing, and the assessment of investment policy. Future research should focus on optimizing hybrid AI–econometric models, incorporating explainable AI (XAI), and improving scenario-based interest rate forecasting methods to ensure reliability and transparency in banking operations (Cerniglia \& Fabozzi, 2020).

The still limited use of multiscenario approaches in interest rate forecasting leaves ample room for future research and application. Integrating AI techniques, such as machine learning algorithms, into traditional econometric models may uncover complex, nonlinear relationships within financial data that may be missed when only a single method is used. For example, incorporating sentiment analyses provides real-time insights into market dynamics, thereby improving the responsiveness and accuracy of forecasts. Multi-perspective approaches to interest rate forecasting in banking are still under development, but their potential to transform forecasting accuracy and asset management practices appears considerable.

\section{Development and Scope of the Prototype}
\subsection{Motivation and Objective of the Proof of Concept}
The increasing complexity of financial markets and the limitations of traditional interest rate forecasting models require the integration of artificial intelligence (AI) and econometric methods to improve forecasting accuracy and scenario-based analysis. In response, a major European bank developed a prototype as a proof of concept (PoC) to assess the feasibility of a multiscenario, AI-driven approach to interest rate forecasting. The primary objectives of the PoC are to improve transparency in forecasts, support strategic asset and liability management, and improve risk assessment through scenario-based methods. By integrating AI-driven topic modelling (topic analysis), sentiment analysis, Bayesian vector autoregression (BVAR), and scenario-based forecasting, the PoC provides a structured framework for analyzing macroeconomic trends, market sentiment, and the dynamics of financial risks. The study aims to create both a foundation for data-driven decision-making in interest rate forecasting and improved adaptability to changing market conditions, while ensuring methodological rigor and economic plausibility.

\subsection{Overall Approach of the Prototype}
The prototype includes a module that extracts key financial topics from central bank publications, analyst reports, and financial press articles using SBERT, UMAP, and HDBScan clustering techniques (cf. Moura et al., 2023). A complementary sentiment analysis model quantifies market sentiment with respect to inflation, economic activity, and employment, and generates predictive indicators for interest rate developments. The PoC also incorporates existing market forecasts and aggregates internal bank estimates, Bloomberg forward curves, and consensus forecasts for government bonds. A BVAR-based econometric AI model increases forecasting accuracy by taking into account macroeconomic variables such as inflation, industrial production, and monetary policy interest rates.

The PoC is implemented via an interactive dashboard that enables users to examine market trends, compare forecasting methods, and simulate macroeconomic shocks in a structured, scenario-based environment. The prototype shows substantial potential to improve forecasting accuracy and risk assessment but remains in a testing phase in which further development is required to ensure scalability, regulatory compliance, and integration into the bank’s infrastructure.

\subsection{Core Elements of the System}
\subsubsection{AI-Driven Topic Modelling in the Proof-of-Concept Prototype}
The AI-driven topic modelling module in the proof-of-concept prototype was developed to extract and analyze key financial topics from large volumes of unstructured text data. It enables users to systematically explore macroeconomic narratives by using machine-learning techniques for clustering and visualizing documents. The module uses Sentence-BERT (SBERT) for semantic representation, Uniform Manifold Approximation and Projection (UMAP) for dimensionality reduction, and Hierarchical Density-Based Spatial Clustering of Applications with Noise (HDBScan) for topic clustering. These technologies enable highly precise thematic grouping of financial documents and improve the interpretability of central bank communications, analyst reports, and financial press articles (Li, Wang \& Chen, 2024; Salem \& Albourawi, 2024).

The topic-modelling workflow of the PoC starts with text preprocessing, in which documents are broken down into their word components (tokenization, lemmatization) and converted into numerical representations. As described, SBERT is used for this purpose, a transformer-based deep-learning model that captures contextual meaning. Unlike traditional methods based on word frequencies (bag-of-words, term frequency–inverse document frequency, TF-IDF), SBERT produces text representations that preserve semantic relationships and improve the accuracy of topic assignment (Delucchi \& Giribone, 2023). The resulting representations are then condensed using UMAP. UMAP is a nonlinear dimensionality-reduction algorithm that ensures that similar documents remain positioned closely in the visualization and can be grouped into a small number of thematic fields. Clustering is then performed with HDBScan so that coherent topic groups are formed based on density estimation, without requiring a predefined number of topics (Boeck, Feldkircher, \& Huber, 2022).

The topic-modelling results are displayed via an interactive dashboard in which documents are shown on a two-dimensional map (cf. Fig. 1). Each document appears as a dot, with different colours used to distinguish clusters. The relative positioning of documents reflects their semantic similarity and allows users to identify relationships between financial topics. Users can zoom into areas of interest, filter specific topics via an interactive legend, and retrieve AI-generated topic labels and document summaries. A legend is displayed in the user interface that ranks topics by importance based on the number of documents per topic. Users can isolate or hide selected topics and thereby refine their analysis dynamically. These functions enable financial analysts to capture and interpret large volumes of macroeconomic texts efficiently, improving the transparency of interest rate forecasting.

The prototype includes a time-series component that allows users to analyze how the topic distribution evolves over time. Buttons for selecting the time horizon allow filtering by different periods and ensure that shifts in macroeconomic discourse can be tracked. The system processes three primary data sources: analyst reports from financial institutions, central bank publications of the European Central Bank (ECB) and the U.S. Federal Reserve, and articles from the financial press. These sources are converted into a standardized format for analysis to ensure consistency across document types.

The technical implementation is based on Python and the Dash library, a Python library for interactive visualization. Most components of the PoC rely on open-source technologies. However, a large language model (LLM) is used, integrated via an OpenAI-API-compatible provider, to ensure that topic labels and document summaries are readable, understandable, and substantively relevant. The modular architecture allows continuous updates as AI technologies develop and can be readily adapted to future developments.

\begin{figure}[htbp]
\centering
\includegraphics[width=0.95\linewidth]{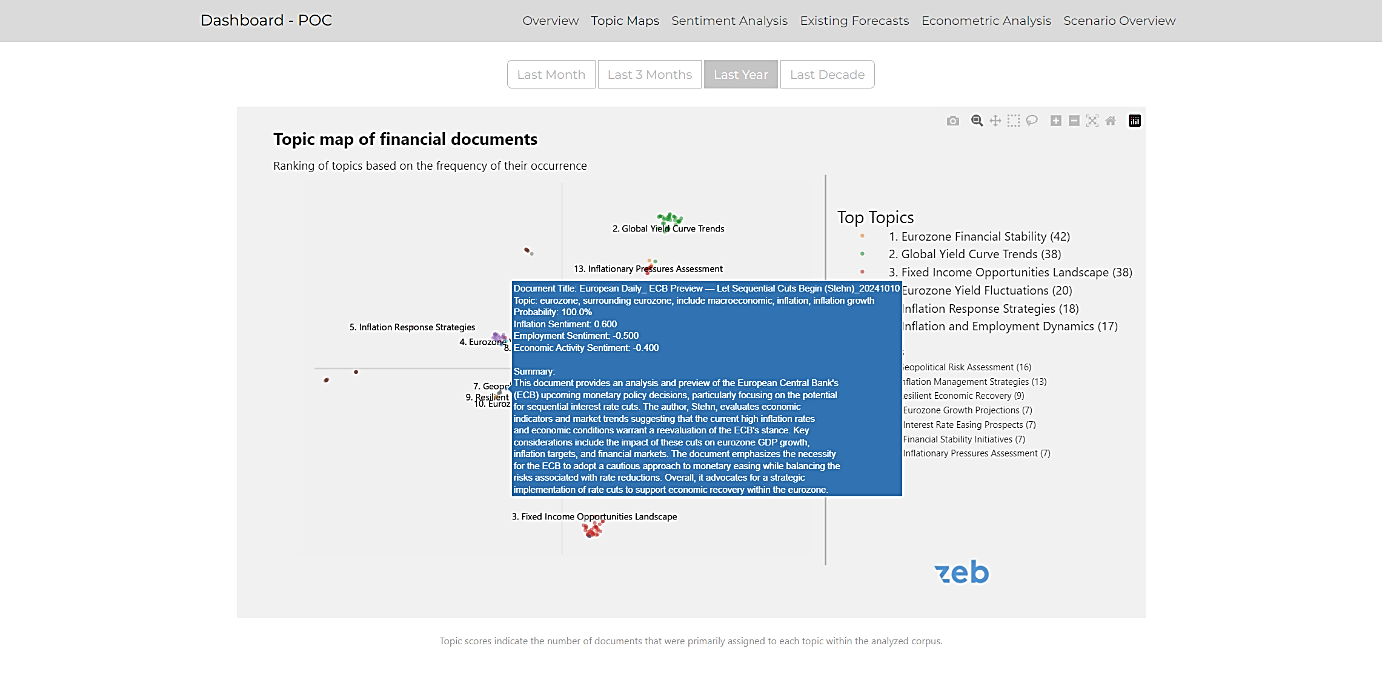}
\caption{Dashboard interface with topic clusters and time-series visualization as a two-dimensional map with topic ranking, legend, and AI-generated document insights.}
\label{fig:1}
\end{figure}
\subsubsection{Sentiment Analysis in the PoC Prototype}
The sentiment analysis module in the prototype complements topic modelling via topic maps by capturing and quantifying the sentiment expressed in financial documents. The focus is on the key macroeconomic variables that influence interest rates. The module was designed to systematically assess sentiment in central bank publications, analyst reports, and financial news, and enables relating the sentiment expressed in texts to historical and forecasted interest rate movements. The system primarily measures market sentiment with respect to inflation, economic activity, and employment, as these factors play a central role in monetary policy decisions (cf. Kashyap \& Stein, 2023). Given the complexity of financial language, especially in central bank statements, the sentiment analysis approach is tailored to address domain-specific challenges that conventional sentiment analysis tools do not capture (Audrino \& Offner, 2024).

Implementation begins with natural language processing (NLP) techniques for preprocessing text data. Documents are, as above, broken down into components (tokenization, lemmatization) and converted into numerical representations of their embedding in text (vectorization). Sentiment is then extracted from the transformed text. The system uses a few-shot learning approach that leverages a customized transformer-based language model trained on financial texts (FinBERT). This enables the model to capture context-dependent sentiment nuances in monetary policy statements and analyst forecasts with high accuracy (Huang, Wang \& Yang, 2023). Sentiment classification is performed using a three-factor approach in which the model generates independent sentiment values for inflation, economic activity, and employment. Unlike conventional polarity-based sentiment models, this system is designed to differentiate nuances in policy statements. This is particularly relevant in interpreting central bank communications on inflation, because both excessively high and excessively low inflation can lead to negative sentiment. The results of the sentiment analysis are integrated into a two-axis time-series visualization that allows users to compare sentiment fluctuations with historical and forecasted interest rate movements (cf. Fig. 2).

The sentiment scores are shown alongside key financial indicators such as the ECB refinancing rate and 10-year swap rates to facilitate identification of potential sentiment-driven interest rate fluctuations. A customizable filter system allows users to refine the dataset by document type, with options for analyzing immediate policy statements (e.g., statements of the Federal Open Market Committee, FOMC, and decisions of the European Central Bank, ECB) or more comprehensive reports (e.g., FOMC minutes and ECB accounts). The selection of documents used for the results shown here is presented in Appendix 1; an example analysis of 40 documents (ECB accounts and decisions as well as FOMC minutes and statements) for deriving sentiment values is shown in Appendix 2. Empirical validation within the PoC showed that longer central bank reports provide stronger sentiment signals for predicting interest rate decisions, highlighting the importance of detailed policy documents in predictive modelling (Audrino \& Offner, 2024).

The sentiment analysis module is a core component of the PoC analysis framework and, like the other modules, uses the Dash web application framework for visualization and integration of the different displays. The sentiment analysis module works with structured text processed via a central document input. All analysis modules are connected to this pipeline and use the same documents. Summary tables accompany the visualization and display current market sentiment values, year-over-year comparisons, and average sentiment ratings over three years to facilitate interpretation. In addition, a separate module containing Bloomberg forecast probabilities for interest rate changes provides insights into upcoming central bank meetings, enabling analysts to assess whether sentiment aligns with market expectations.

The sentiment analysis module provides an integrated, AI-driven approach to interpreting financial market sentiment within the broader context of interest rate forecasting. By structuring data for financial assessment and correlating it with market movements, the module improves the predictive quality of macroeconomic analysis and strengthens the role of scenario-based forecasting in strategic ALM.

\begin{figure}[htbp]
\centering
\includegraphics[width=0.95\linewidth]{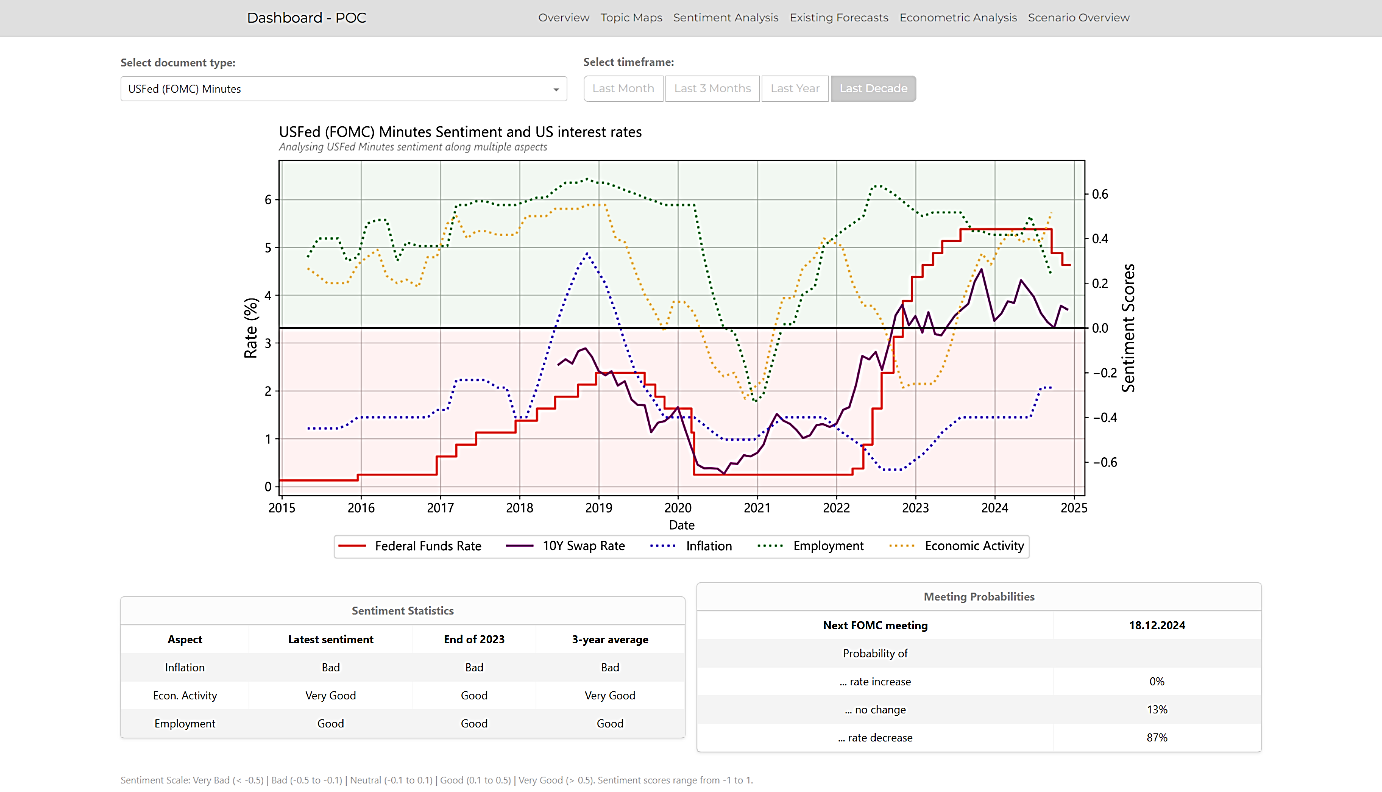}
\caption{Presentation of sentiment trend visualizations, sentiment probability distributions, and interactive filter options.}
\label{fig:2}
\end{figure}
\subsubsection{Econometric, BVAR-Based Forecasting in the Prototype}
The econometric AI model in the PoC prototype is based on Bayesian vector autoregression (BVAR) and is designed to capture macroeconomic interdependencies and forecast swap rates across different maturities (2, 5, 10, and 30 years). The model integrates so-called “shrinkage priors” (i.e., a prior distribution used in machine learning to shrink small effects and amplify larger ones). This helps overcome the overparameterization problem that affects conventional standard vector autoregression (VAR) models. As a result, more stable and robust forecasts can be produced even with limited historical data (Boeck, Feldkircher, \& Huber, 2022). The BVAR approach enables both unconditional forecasts (trend-based projections) and conditional forecasts (scenario-based simulations incorporating macroeconomic shocks; Kuschnig \& Vashold, 2021).

The module is implemented using the BVAR package in R, an open-source statistical development environment, which facilitates estimation via a hierarchical Bayesian framework. The dataset consists of seven relevant macroeconomic predictors identified in advance from more than 40 potential indicators using Granger causality tests to ensure economic relevance. These predictors include euro area inflation, industrial production, the ECB refinancing rate, import-weighted commodity prices, the EuroStoxx 50 index, U.S. inflation, and the U.S. manufacturing PMI, each selected on the basis of Granger causality tests (cf. Troster, 2018). The estimation process applies Minnesota priors, a specific form of Bayesian shrinkage prior distribution that balances model flexibility and parameter stability and thereby prevents overfitting in high-dimensional macroeconomic systems (Diebold \& Li, 2006). The R programs used for this purpose are documented in Appendices 3–8; numerical examples are provided in Tables 1 as well as 2a and 2b.

Forecast generation within the PoC follows a two-stage process. First, unconditional forecasts are generated based on past trends and assume no external shocks. Second, conditional forecasts incorporate user-defined macroeconomic shocks, enabling scenario-based simulations. Users can adjust variables such as inflation, monetary policy rates, or financial market conditions, and the model dynamically recalibrates swap forecasts to represent the impact of these changes (cf. Geweke, 1999). The interactive dashboard visualizes the yield curve forecasts generated by BVAR and compares them with market-based forecasts and sentiment-driven expectations (cf. Fig. 3; list of the data sources used for this purpose in Appendix 1). This functionality enables users to assess how different economic scenarios affect swap rate trajectories and thereby improve stress-testing capabilities for the bank’s ALM.

\begin{table}[htbp]
\centering
\caption{Forecast models for the combined euro swap rate across different tenors (2 to 30 years).}
\label{tab:1}
\begin{tabular}{lcccccccc}
\toprule
 & Date & 2Y & 3Y & 5Y & 10Y & 15Y & 20Y & 30Y \\
\midrule
Current & 29.11.2024 & 2.11 & 2.07 & 2.07 & 2.16 & 2.22 & 2.16 & 1.96 \\
ISP forecast & 02.12.2024 & 1.99 & 2.07 & 2.23 & 2.81 & 2.77 & 2.74 & 2.59 \\
Bloomberg forward curve & 02.12.2024 & 1.86 & 1.90 & 1.97 & 2.09 & 2.14 & 2.08 & 1.87 \\
Econometric model (mid rate) & 02.12.2024 & 1.94 & - & 2.00 & 2.13 & - & - & 1.92 \\
\bottomrule
\end{tabular}
\end{table}
The BVAR module interacts closely with the topic and sentiment analysis modules to improve forecast validation. The PoC prototype processes macroeconomic data via R and visualizes forecasts using the programming language and the Dash framework. In addition to forecasts generated by the PoC prototype, the system also integrates already available forecasts, as the next section shows.

\begin{figure}[htbp]
\centering
\includegraphics[width=0.95\linewidth]{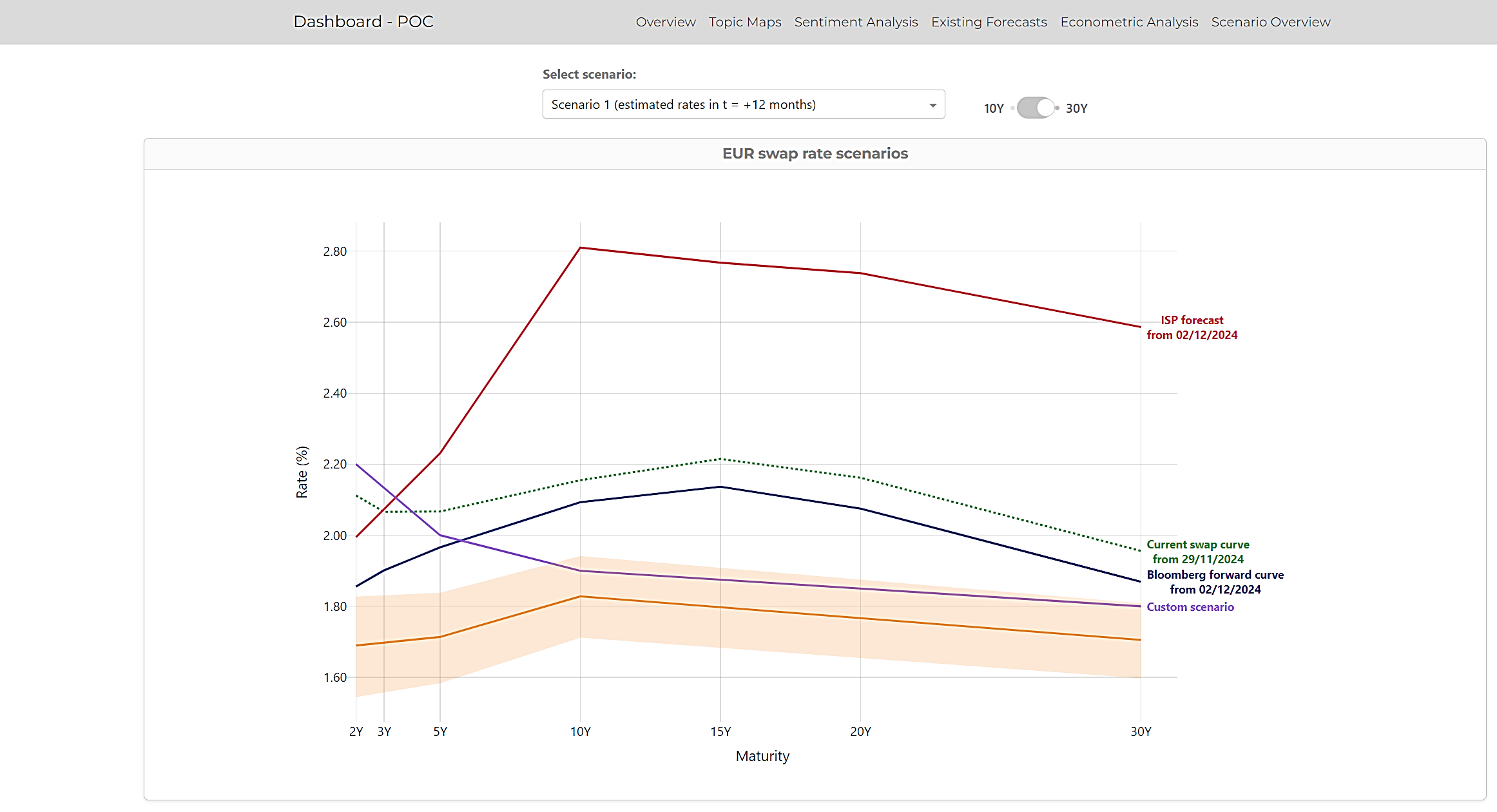}
\caption{Combined view of existing forecasts and forecasts computed in the prototype.}
\label{fig:3}
\end{figure}
\subsubsection{Compilation of Existing Forecasts in the PoC Prototype}
The compilation of existing interest rate forecasts is an essential function of the PoC prototype and serves to contextualise and evaluate already available market expectations. Whereas the preceding modules are based on generating new forecasts, this module provides a structured overview of existing interest rate forecasts from different sources to create a well-founded comparison basis for the multi-scenario framework. The goal is to present existing forecast heterogeneity transparently in order to enable banks to form a differentiated assessment of market expectations.

The module integrates four core forecast sources: (1) swap rates for the euro area (6M Euribor) and the United States (SOFR), (2) consensus forecasts for EU government bonds (Q4 2025), (3) consensus forecasts for U.S. government bonds (Q4 2025), and (4) Bloomberg forward curves (AI-Driven Framework for Multiscenario Interest Rate Forecasting in Banks, 2024). These data provide an aggregated view of market expectations and enable a differentiated analysis of different interest rate development scenarios.

A central feature of the module is the direct comparability of forecasts across different time horizons. The display is provided in the form of interpolated yield curves with maturities between 2 and 30 years, with forecast values available both for the observation date and for the expected interest rate development in 12 months. This allows a systematic evaluation of medium-term market expectations and provides decision bases for strategic adjustments in ALM and banks’ interest rate management.

For the methodological processing of consensus forecasts, analyst estimates from Bloomberg are aggregated to derive realistic scenario profiles. This comprises three main steps: (1) clustering analyst forecasts by expected yield curve change (e.g., rising or flat yield curves), (2) calculating scenario probabilities based on the frequency of forecast clusters, and (3) deriving an average yield curve per scenario. For the euro area, forecasts based on German government bonds were used, as these historically have the highest correlation with aggregated euro area government bonds (Diebold \& Li, 2006).

Particular attention is given to interpreting deviations between different forecast sources. Bloomberg’s forward curve reflects market-implied interest rate expectations already priced in, whereas analyst consensus forecasts are often based on fundamental macroeconomic assumptions. Combining these different perspectives allows the identification of systematic biases and uncertainty in forecast accuracy (Audrino \& Offner, 2024). This comparative analysis is implemented in the PoC prototype via an interactive dashboard visualization that enables users to select specific scenarios and dynamically compare their yield curve trajectories.

Table 2a (“Scenario Builder”) contains predefined assumptions about key macroeconomic drivers, including the inflation rate, industrial production, commodity prices, the ECB policy rate, the development of the Euro Stoxx 50, as well as U.S. inflation and business cycle indicators. These variables are varied across four alternatively defined scenarios, each assigned a probability of occurrence (50\%, 15\%, 20\%, 15\%). This enables probabilistic weighting, which reflects uncertainty about future macroeconomic developments. The listed standard deviations indicate dispersion within scenarios and allow conclusions about the relative sensitivity of individual drivers.

\begin{table}[htbp]
\centering
\caption{Scenario assumptions for macroeconomic drivers and probabilistic weighting of alternative future scenarios}
\label{tab:2a}
\begin{tabular}{lccccc}
\toprule
 &  & 50\% & 15\% & 20\% & 15\% \\
\midrule
 &  & Scenario 1 & Scenario 2 & Scenario 3 & Scenario 4 \\
Inflation rate &  & 0.00\% & -0.25\% & 0.00\% & 0.25\% \\
Industrial production &  & 0.00\% & -2.75\% & 2.75\% & -2.75\% \\
Commodity prices &  & -3.34\% & -3.34\% & 0.00\% & 3.34\% \\
ECB refinancing rate &  & 0.00\% & -0.13\% & -0.25\% & 0.13\% \\
Euro Stoxx 50 &  & 0.00\% & -5.05\% & 5.05\% & -5.05\% \\
US inflation rate &  & 0.00\% & 0.00\% & 0.00\% & 0.40\% \\
US manufacturing PMI &  & 0.00 & -1.62 & 1.62 & 0.00 \\
\bottomrule
\end{tabular}
\end{table}
Table 2b (“Scenario Results”) shows the effects of these scenarios on swap rates across different maturities (2, 5, 10, and 30 years). The resulting interval forecasts (lower, mid, and upper level per scenario) are derived from simulation-based model calculations using historical financial data in combination with a Bayesian vector autoregression (BVAR) approach. In addition, directional indicators for yield curve changes (rising, falling) and shape changes (steepening, flattening, inversion) are evaluated. Aggregation across all scenarios is performed using an expectation-based approach that accounts for the respective probabilities, creating a stochastic overall projection of the yield curve (cf. Gneiting \& Katzfuss, 2014).

\begin{table}[htbp]
\centering
\footnotesize
\setlength{\tabcolsep}{2.5pt}
\renewcommand{\arraystretch}{0.9}

\caption{Model-based scenario results for swap rates, yield curve trajectory, and probabilistically aggregated interest rate projections}
\label{tab:2b}
\begin{adjustbox}{max width=\textwidth}
\begin{tabular}{lccccccccccccc}
\toprule
Current & Current & Scenario 1 & Scenario 1 & Scenario 1 & Scenario 2 & Scenario 2 & Scenario 2 & Scenario 3 & Scenario 3 & Scenario 3 & Scenario 4 & Scenario 4 & Scenario 4 \\
\midrule
 &  & Lower level & Mid
level & Upper level & Lower level & Mid
level & Upper level & Lower level & Mid level & Upper level & Lower level & Mid level & Upper level \\
2Y swap & 2.11 & 2.04 & 2.06 & 2.08 & 1.52 & 1.71 & 1.89 & 2.15 & 2.20 & 2.25 & 2.20 & 2.28 & 2.36 \\
5Y swap & 2.07 & 2.01 & 2.03 & 2.04 & 1.59 & 1.77 & 1.95 & 2.09 & 2.17 & 2.24 & 2.12 & 2.20 & 2.29 \\
10Y swap & 2.16 & 2.11 & 2.13 & 2.14 & 1.80 & 1.94 & 2.08 & 2.17 & 2.25 & 2.32 & 2.19 & 2.27 & 2.34 \\
30Y swap & 1.96 & 1.92 & 1.94 & 1.95 & 1.69 & 1.80 & 1.91 & 1.97 & 2.03 & 2.09 & 1.96 & 1.99 & 2.01 \\
Slope
10Y-2Y & 0.04 & 0.07 & 0.07 & 0.07 & 0.28 & 0.23 & 0.19 & 0.02 & 0.05 & 0.07 & -0.01 & -0.01 & -0.02 \\
Direction &  & down & down & down & down & down & down & up & up & up & up & up & up \\
Slope &  & steep & steep & steep & steep & steep & steep & flat & steep & steep & invers & invers & invers \\
Overall &  & down steep & down steep & down steep & down steep & down steep & down steep & up
flat & up steep & up steep & up
invers & up
invers & up
invers \\
Probabilities &  &  & 0.50 &  &  & 0.15 &  &  & 0.20 &  &  & 0.15 &  \\
ECB
refinan-cing rate & 3.15 & 3.10 & 3.12 & 3.13 & 2.52 & 2.70 & 2.88 & 2.86 & 2.94 & 3.01 & 3.47 & 3.63 & 3.81 \\
\bottomrule
\end{tabular}
\end{adjustbox}
\end{table}
Overall, the tables illustrate how the developed prototype models scenario-based macroeconomic shocks and translates them into differentiated interest rate projections. Uncertainty is explicitly taken into account, enabling forward-looking, quantitatively robust statements about the level, trajectory, and direction of yield curves—an essential instrument for asset-liability management in banks (cf. Boeck et al., 2022; Cerniglia \& Fabozzi, 2020).

The technical implementation uses Excel-based data sources that are regularly synchronized with Bloomberg updates. Scenario calculation and clustering analysis of analyst forecasts are performed in a separate analysis layer, the results of which are integrated directly into the PoC prototype’s forecasting framework. Probabilistic methods are used for scenario formation to better reflect uncertainty in aggregated forecasts.

The forecast comparison module is integrated into an interactive dashboard in which users can select specific forecast perspectives and time horizons (cf. Fig. 4). The visualization component shows the current swap curve together with alternative forecasts so that users can assess discrepancies between market-implied expectations and analyst-based forecasts. In a separate panel, scenarios for government bond yields are linked to a probability-weighted estimate of the yield curve for the selected forecast horizon. In this way, users can analyze how expert forecasts align with macroeconomic sentiment trends and econometric model results.

\begin{figure}[htbp]
\centering
\includegraphics[width=0.95\linewidth]{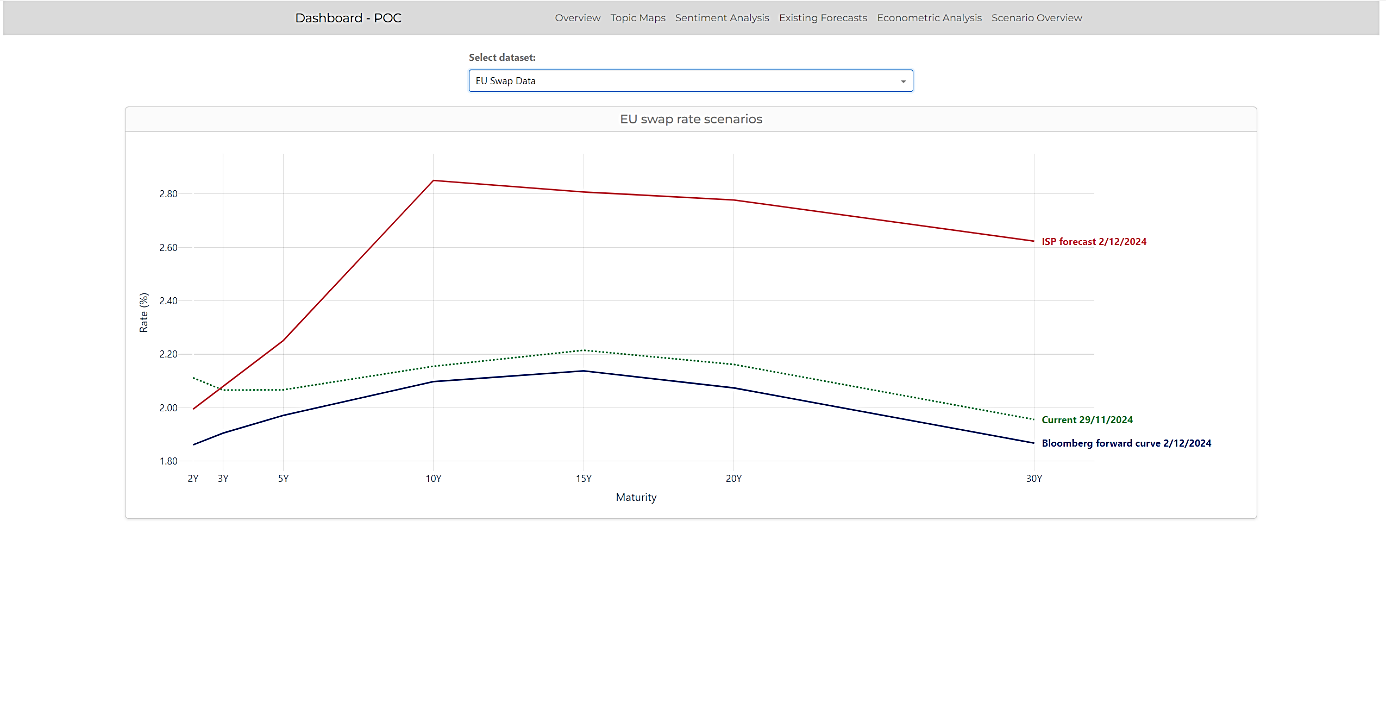}
\caption{Forecast comparison charts, clustering-based probability distributions of the yield curve, and scenario-based sensitivity analyses.}
\label{fig:4}
\end{figure}
\subsubsection{Scenario-Based Forecasting and Dashboard Visualization Module in the Prototype}
The scenario-based forecasting and dashboard visualization module is the central analytical interface of the prototype and integrates AI-driven topic analysis, sentiment analysis, BVAR-based econometric forecasts, and market-based swap rate forecasts into a structured environment for interest rate analysis (Kuschnig \& Vashold, 2021; Diebold \& Li, 2006). This module enables users to simulate macroeconomic scenarios, compare forecasting methods, and assess yield curve developments under different economic conditions, thereby improving the bank’s risk-adjusted ALM and facilitating regulatory compliance. By aligning market sentiment indicators with macroeconomic expectations, the system highlights potential discrepancies between forecasted yield curves and investor sentiment, increasing the robustness of scenario analysis (Huang, Wang \& Yang, 2023; Delucchi \& Giribone, 2023). In addition to integrating perspectives from the various modules, the system estimates the probability of steepening, flattening, or parallel shifts of the yield curve by applying clustering techniques, thereby providing a risk-adjusted perspective on future interest rate developments (Boeck, Feldkircher, \& Huber, 2022).

A key feature of the module is the graphical visualization of forecasts, which enables intuitive interaction with scenario results (cf. Fig. 5). Interest rate developments are displayed as interactive yield curves that allow users to analyze the effects of alternative economic developments on future interest rate levels. These visualizations enable not only direct comparison between scenarios but also a decomposition of influencing factors to better understand the underlying mechanisms of interest rate movements (Audrino \& Offner, 2024).

The dashboard integrates three core visualization functions: (1) dynamic yield curves showing aggregated forecasts for different time periods, (2) scenario comparison contrasting alternative economic developments, and (3) sentiment indices correlating expectation values generated by sentiment analysis with macroeconomic variables. The implementation is based on the Dash framework, enabling interactive parameter adjustments by the user. Analysts can modify individual variables to perform ad hoc sensitivity analyses and assess the effects of market changes on interest rate developments.

\begin{figure}[htbp]
\centering
\includegraphics[width=0.95\linewidth]{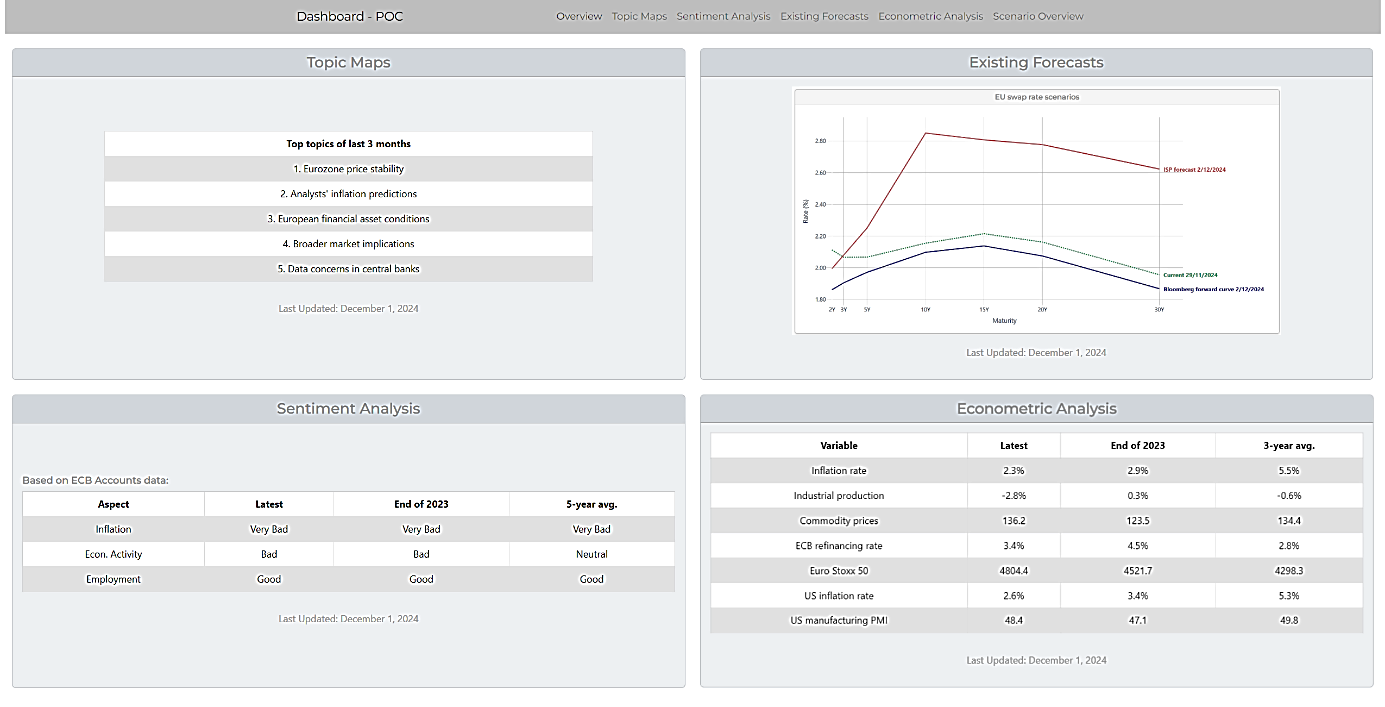}
\caption{Scenario-based forecasting and dashboard visualization module.}
\label{fig:5}
\end{figure}
The scenario-based forecasting and dashboard visualization module constitutes a comprehensive AI–econometric framework that advances quantitative interest rate forecasting by integrating structured macroeconomic modelling with AI-driven sentiment analysis. By providing a robust, interpretable, and data-driven forecasting approach, this module improves strategic financial decision-making and macroeconomic risk assessment in banking.

\section{Discussion}
\subsection{Summary of Findings}
The proof-of-concept prototype developed for AI-driven multi-scenario interest rate forecasting in banking integrates multiple analytical techniques to improve interest rate predictions. By combining AI-supported topic analysis for topic modelling, sentiment analysis to assess market sentiment, structured market forecasts, and scenario-based visualization, the system enables a comprehensive, data-driven approach to financial forecasting. The PoC aims to improve transparency in interest rate forecasting, support strategic Asset-Liability Management (ALM), and enable structured risk assessment through macroeconomic scenario simulations.

A central insight from the PoC is the ability to consolidate structured macroeconomic data and unstructured financial text sources within a single analytical framework. The topic modelling component extracts thematic structures from financial documents, sentiment analysis quantifies the sentiment expressed by central banks and analysts, and the BVAR model ensures statistical accuracy by integrating key macroeconomic indicators. In addition, the system consolidates existing market forecasts and provides benchmarking tools that compare AI-driven insights with conventional economic forecasts. These modules interact seamlessly in a dashboard-based environment in which users can adjust macroeconomic variables and observe their effects on swap rate expectations.

The BVAR model, serving as the econometric core component, introduces a scenario-based forecasting mechanism that enables both unconditional projections based on historical data and conditional forecasts incorporating user-defined macroeconomic shocks. This approach improves financial institutions’ ability to stress-test different economic conditions, refine risk assessments, and support forward-looking ALM strategies (Kuschnig \& Vashold, 2021; Boeck, Feldkircher, \& Huber, 2022). Moreover, integrating sentiment analysis into forecast validation represents a novel feature, enabling the identification of misalignments between market sentiment and macroeconomic fundamentals (Audrino \& Offner, 2024; Huang, Wang \& Yang, 2023).

\subsection{Limitations and Further Development}
Despite its methodological progress, the PoC is still a prototype and not yet a production-ready system. To reach that stage, model calibration and robustness testing must first be further expanded, especially the validation of signals derived from sentiment analysis against historical interest rate movements. This is necessary because AI-driven sentiment analysis offers interpretive possibilities whose effectiveness depends on the stability of financial sentiment trends and their actual, quantifiable relationship to market outcomes such as interest rate changes (Delucchi \& Giribone, 2023).

The BVAR-based scenario engine improves interpretability but relies on prior distributions that introduce subjectivity into forecast results (Diebold \& Li, 2006): shrinkage priors mitigate overfitting risk but can restrict model flexibility when confronted with structural changes in the macroeconomic environment (Cepni et al., 2022). In addition, the computational complexity of the model increases monotonically with the number of macroeconomic variables included, requiring careful feature selection to ensure fast and efficient data processing.

From a technical and operational perspective, the dashboard visualizes scenario results effectively and clearly. Further improvements in user interactivity, real-time data integration, and automated reporting would be helpful to enhance usability for financial analysts. Data input is currently structured around static datasets, meaning that continuous real-time updates from external sources such as Bloomberg and ECB data feeds would require additional automation to ensure forecasts remain up to date.

Finally, regulatory and compliance challenges remain an important consideration: AI-driven forecasting methods must align with banking risk and governance frameworks and ensure the explainability and interpretability of AI-based decision-making (Bohn \& Schneider, 2024). The opacity of deep-learning models and the probabilistic nature of Bayesian forecasts require transparent validation frameworks in order to gain acceptance within financial institutions.

\subsection{Potentials and Practical Relevance}
The PoC demonstrates the potential for AI-driven, multi-perspective interest rate forecasting to improve financial decision-making and ALM in banking. By integrating macroeconomic models with AI-supported sentiment tracking and structured market forecasts, the system introduces a holistic analytical framework that supports risk-adjusted investment strategies.

The ability to quantify sentiment deviations from macroeconomic expectations provides banks with an early warning mechanism for market shifts, enabling proactive monetary policy analysis and investment realignment. Scenario-based simulations allow finely granulated stress testing, improve the assessment of liquidity risk, and support regulatory-compliant capital planning (Cerniglia \& Fabozzi, 2020).

In addition, the scalability of the PoC architecture indicates its potential for future enhancements, including integrating high-frequency trading models, adaptive AI learning mechanisms, and real-time macroeconomic monitoring tools. While the prototype currently relies on static datasets and structured economic assumptions, advances in real-time data streaming, automated model training, and improved explainability techniques could significantly expand its practical applicability in financial risk management and portfolio optimization.

Overall, the PoC shows that AI-supported, multi-perspective forecasting methods can provide substantial added value for financial analysis and risk management in banks. The combination of topic modelling, sentiment analysis, and econometric simulation improves the capture of market dynamics and enables a differentiated assessment of macroeconomic developments. Nevertheless, it is essential to emphasize that the PoC is not a deterministic forecasting model but is based on probabilistic estimation methods. The forecasts generated should not be interpreted as exact predictions of future interest rate developments, but as methodologically grounded estimates based on available data, known relationships, and quantified market expectations. Ultimately, it remains the responsibility of financial management to critically scrutinize this information, place it in the organizational context, and form a well-founded assessment for strategic decisions.

\appendix
\section*{Appendices}

\appendix

\section{Selection of documents for the sample calculations in this article}

\begingroup
\footnotesize
\setlength{\tabcolsep}{2pt}
\renewcommand{\arraystretch}{0.9}
\setlength{\LTleft}{0pt}
\setlength{\LTright}{0pt}
\begin{longtable}{L{3.4cm}cccccL{5.4cm}}
\captionsetup{labelformat=empty}
\caption{Appendix 1: Selection of documents used for sentiment analysis}\label{app:docs}\\
\toprule
KPI & Daily & Weekly & Monthly & Quarterly & Annually & Source \\
\midrule
\endfirsthead
\toprule
KPI & Daily & Weekly & Monthly & Quarterly & Annually & Source \\
\midrule
\endhead
\midrule
\multicolumn{7}{r}{\small\itshape Continued on next page} \\
\endfoot
\bottomrule
\endlastfoot
GDP &  &  &  & X & X & Refinitiv \\
Inflation &  &  & X & X & X & Refinitiv \\
Core Inflation ECB data &  &  & X &  &  & ECB \\
Inflation ECB data &  &  & X &  &  & ECB \\
Price level ECB &  &  & X &  &  & ECB \\
Exchange rate EUR / USD & X & X & X & X & X & Refinitiv \\
Exchange rate USD / GBP & X & X & X & X & X & Refinitiv \\
S\&P 500 COMPOSITE & X & X & X & X & X & Refinitiv \\
EURO STOXX 50 & X & X & X & X & X & Refinitiv \\
Unemployment rate &  &  & X & X & X & Refinitiv \\
Employment Ratio &  &  &  & X &  & EuroStat \\
EURO main refi ECB & X & X & X & X & X & Refinitiv \\
Federal funds rate & X & X & X & X & X & Refinitiv \\
Money supply M1 EU &  &  & X & X &  & Refinitiv \\
Money supply M2 EU &  &  & X & X &  & Refinitiv \\
Money supply M3 EU &  &  & X & X &  & Refinitiv \\
Oil price nominal &  &  & X & X & X & c \\
Oil price real &  &  & X & X & X & US Energy Information Administration \\
Commodity Index &  &  & X &  &  & ECB \\
Personal consumption EU &  &  &  & X &  & Refinitiv \\
Industrial output &  &  & X &  &  & ECB \\
Industrial production &  &  & X & X & X & BBG \\
Exports EA &  &  &  & X &  & Refinitiv \\
Imports EA &  &  &  & X &  & Refinitiv \\
Output gap &  &  &  & X & X & Refinitiv \\
ZEW Expectation Index &  &  & X &  &  & BBG \\
ZEW expectation econ growth &  &  & X & X & X & BBG \\
Economic Sentiment Indicator &  &  & X &  &  & ECB \\
Eurostat Labor Costs Nominal Values Eurozone YoY WDA &  &  &  & X & X & BBG \\
Markit Services PMI &  &  & X & X & X & BBG \\
Markit Manufacturing PMI &  &  & X &  &  & BBG \\
Public Debt to GDP EA &  &  &  &  & X & Refinitiv \\
Euro Zone Debt \% GDP 1 &  &  &  &  & X & BBG \\
Euro Zone Debt \% GDP 2 &  &  &  & X &  & EuroStat \\
Saving rate Euro zone &  &  &  & X &  & Refinitiv \\
Output gap Euro Zone &  &  &  & X & X & Refinitiv \\
Business Climate\_Euro Zone &  &  & X &  &  & Refinitiv \\
Industrial confidence indicator &  &  & X &  &  & Refinitiv \\
New housing Euro-Zone &  &  & X &  &  & Refinitiv \\
Residential property prices Euro\_Zone &  &  &  & X &  & Refinitiv \\
Global Real Production
Index &  &  & X &  &  & Dallas Fed \\
GfK Consumer Confidence &  &  & X &  &  & BBG \\
Nonfarm Payrolls &  &  & X & X & X & BBG \\
Initial jobless claims &  & X & X & X & X & BBG \\
ISM Non-Manufacturing Index &  &  & X &  &  & BBG \\
ISM Manufacturing PMI &  &  & X & X & X & BBG \\
US Average Hourly Earnings All Employees Total Private Yearly Percent Change SA &  &  & X & X & X & BBG \\
Inflation &  &  & X & X & X & Refinitiv \\
US Personal Income MoM SA &  &  & X &  &  & BBG \\
Unemployment rate &  &  & X & X & X & Refinitiv \\
Retail prices &  &  & X & X & X & BBG \\
US Labor Force Participation &  &  & X & X & X & BBG \\
ISM Business Index\_USA &  &  & X & X &  & Refinitiv \\
\end{longtable}
\endgroup

Sources: Refinitiv = renamed to LSEG (London Stock Exchange Group PLC), provider of financial data; ECB = European Central Bank, provider of financial data; US Energy Information Administration, provider of financial data; EuroStat = Statistics Bureau of the European Commission; BBG = Bloomberg Terminal, computer software system provided by the financial data vendor Bloomberg L.P.; Dallas Fed = Federal Reserve Bank of Dallas, provider of financial data.

\section{Sentiment analysis for selected document types for the time series 2014-2024}

\begin{landscape}
\begingroup
\tiny
\setlength{\tabcolsep}{1pt}
\renewcommand{\arraystretch}{0.75}
\setlength{\LTleft}{0pt}
\setlength{\LTright}{0pt}
\begin{longtable}{L{1.6cm}*{12}{L{1.55cm}}}
\captionsetup{labelformat=empty}
\caption{Appendix 2: Example sentiment analysis (40 documents)}\label{app:sent40}\\
\toprule
Time Period & Economic Activity: ECB Accounts & Economic Activity: ECB Decisions & Economic Activity: FOMC Minutes & Economic Activity: FOMC Statements & Employment Rate: ECB Accounts & Employment Rate: ECB Decisions & Employment Rate: FOMC Minutes & Employment Rate: FOMC Statements & Inflation Rate: ECB Accounts & Inflation Rate: ECB Decisions & Inflation Rate: FOMC Minutes & Inflation Rate: FOMC Statements \\
\midrule
\endfirsthead
\toprule
Time Period & Economic Activity: ECB Accounts & Economic Activity: ECB Decisions & Economic Activity: FOMC Minutes & Economic Activity: FOMC Statements & Employment Rate: ECB Accounts & Employment Rate: ECB Decisions & Employment Rate: FOMC Minutes & Employment Rate: FOMC Statements & Inflation Rate: ECB Accounts & Inflation Rate: ECB Decisions & Inflation Rate: FOMC Minutes & Inflation Rate: FOMC Statements \\
\midrule
\endhead
\midrule
\multicolumn{13}{r}{\small\itshape Continued on next page} \\
\endfoot
\bottomrule
\endlastfoot
01.09.2014 &  & 0 & 5 & 3 &  & 0 & 0 & 5 &  & -4 & -4 & -4 \\
01.10.2014 &  & 0 & 6 & 5 &  & 0 & 5 & 6 &  & 0 & -4 & -4 \\
01.11.2014 &  & 0 &  &  &  & 0 &  &  &  & 0 &  &  \\
01.12.2014 &  & 0 & 5 & 5 &  & 0 & 6 & 6 &  & 0 & -4 & -4 \\
01.01.2015 &  & 0 & 5 & 5 &  & 0 & 6 & 6 &  & 0 & -5 & -4 \\
01.02.2015 & 5 &  &  &  & 0 &  &  &  & -7 &  &  &  \\
01.03.2015 &  & 0 & 0 & 0 &  & 0 & 5 & 6 &  & 0 & -6 & -4 \\
01.04.2015 & 4 & 0 & -5 & -3 & 5 & 0 & -3 & 0 & -6 & 0 & -4 & -4 \\
01.05.2015 & 6 &  &  &  & 5 &  &  &  & 4 &  &  &  \\
01.06.2015 &  & 0 & 3 & 3 &  & 0 & 5 & 5 &  & 0 & -4 & -4 \\
01.07.2015 & 6 & 0 & 4 & 5 & 4 & 0 & 5 & 6 & 5 & 0 & -4 & -4 \\
01.08.2015 & 3 &  &  &  & 5 &  &  &  & -4 &  &  &  \\
01.09.2015 &  & 0 & 5 & 5 &  & 0 & 6 & 6 &  & 0 & -4 & -4 \\
01.10.2015 & -4 & 0 & 5 & 5 & 5 & 0 & 0 & 0 & -6 & 0 & -4 & -4 \\
01.11.2015 & 4 &  &  &  & 5 &  &  &  & -6 &  &  &  \\
01.12.2015 &  & 0 & 5 & 5 &  & 0 & 6 & 7 &  & 0 & -4 & -4 \\
01.01.2016 & 4 & 0 & -3 & 0 & 5 & 0 & 6 & 5 & -6 & 0 & -4 & -4 \\
01.02.2016 & 4 &  &  &  & 5 &  &  &  & -6 &  &  &  \\
01.03.2016 &  & 6 & 5 & 3 &  & 5 & 6 & 5 &  & -4 & -4 & -4 \\
01.04.2016 & -4 & 0 & -3 & -3 & 5 & 0 & 5 & 6 & -7 & 0 & -4 & -4 \\
01.05.2016 & 4 &  &  &  & 5 &  &  &  & -6 &  &  &  \\
01.06.2016 &  & 4 & 3 & 5 &  & 0 & -5 & -3 &  & 0 & -4 & -4 \\
01.07.2016 & 6 & 0 & 6 & 4 & 5 & 0 & 5 & 5 & -4 & 0 & -4 & -4 \\
01.08.2016 & 1 &  &  &  & 6 &  &  &  & 0 &  &  &  \\
01.09.2016 &  & 0 & 3 & 4 &  & 0 & 5 & 5 &  & 0 & -4 & -4 \\
01.10.2016 & 4 & 0 &  &  & 5 & 0 &  &  & 0 & 0 &  &  \\
01.11.2016 & 4 &  & 5 & 4 & 6 &  & 6 & 5 & 5 &  & -4 & -4 \\
01.12.2016 &  & 0 & 5 & 5 &  & 0 & 6 & 6 &  & 0 & -2 & -4 \\
01.01.2017 & 6 & 0 &  &  & 5 & 0 &  &  & -4 & 0 &  &  \\
01.02.2017 & 6 &  & 6 & 5 & 5 &  & 5 & 6 & -4 &  & -4 & -4 \\
01.03.2017 &  & 0 & 5 & 6 &  & 0 & 6 & 5 &  & 0 & 4 & -4 \\
01.04.2017 & 6 & 0 &  &  & 5 & 0 &  &  & 4 & 0 &  &  \\
01.05.2017 & 6 &  & 0 & 0 & 5 &  & 5 & 5 & -4 &  & -4 & 4 \\
01.06.2017 &  & 0 & 5 & 6 &  & 0 & 6 & 5 &  & 0 & -4 & 4 \\
01.07.2017 & 6 & 0 & 5 & 5 & 5 & 0 & 6 & 6 & -4 & 0 & -4 & 4 \\
01.08.2017 & 5 &  &  &  & 6 &  &  &  & -4 &  &  &  \\
01.09.2017 &  & 0 & 4 & 5 &  & 0 & 5 & 6 &  & 0 & -4 & 4 \\
01.10.2017 & 5 & 0 &  &  & 6 & 0 &  &  & -4 & 0 &  &  \\
01.11.2017 & 6 &  & 6 & 6 & 5 &  & 5 & 5 & -4 &  & -4 & -4 \\
01.12.2017 &  & 0 & 5 & 5 &  & 0 & 6 & 7 &  & 0 & -4 & 4 \\
01.01.2018 & 7 & 0 & 5 & 5 & 6 & 0 & 6 & 7 & -4 & 0 & -4 & 4 \\
01.02.2018 & 5 &  &  &  & 6 &  &  &  & -4 &  &  &  \\
01.03.2018 &  & 0 & 5 & 5 &  & 0 & 7 & 7 &  & 0 & 4 & 4 \\
01.04.2018 & 7 & 0 &  &  & 5 & 0 &  &  & -4 & 0 &  &  \\
01.05.2018 & 0 &  & 5 & 5 & 5 &  & 6 & 7 & 0 &  & 4 & 4 \\
01.06.2018 &  & 0 & 6 & 6 &  & 0 & 7 & 7 &  & 4 & 4 & 4 \\
01.07.2018 & 5 & 0 &  &  & 6 & 0 &  &  & 4 & 0 &  &  \\
01.08.2018 & 6 &  & 6 & 6 & 5 &  & 7 & 7 & 4 &  & 4 & 4 \\
01.09.2018 &  & 0 & 5 & 6 &  & 0 & 6 & 7 &  & 0 & 4 & 4 \\
01.10.2018 & 6 & 0 &  &  & 5 & 0 &  &  & 4 & 0 &  &  \\
01.11.2018 & 3 &  & 6 & 5 & 5 &  & 7 & 6 & 4 &  & 0 & 0 \\
01.12.2018 &  & 0 & 5 & 5 &  & 0 & 6 & 6 &  & 0 & 0 & 0 \\
01.01.2019 & -3 & 4 & 5 & 5 & 5 & 0 & 6 & 6 & -4 & 0 & 0 & 0 \\
01.02.2019 & -3 &  &  &  & 5 &  &  &  & -4 &  &  &  \\
01.03.2019 &  & 3 & -3 & -3 &  & 5 & 6 & 5 &  & 4 & -4 & 4 \\
01.04.2019 & -3 & 0 &  &  & 5 & 0 &  &  & -4 & 4 &  &  \\
01.05.2019 & -3 &  & 5 & 5 & 5 &  & 6 & 6 & -4 &  & -4 & 4 \\
01.06.2019 &  & 0 & -3 & 5 &  & 0 & 5 & 6 &  & 0 & -4 & 0 \\
01.07.2019 & -3 & 0 & 0 & 4 & 5 & 0 & 6 & 6 & -4 & -6 & -4 & 2 \\
01.08.2019 & -3 &  &  &  & 5 &  &  &  & -4 &  &  &  \\
01.09.2019 &  & 5 & 0 & 4 &  & 0 & 5 & 6 &  & -4 & -4 & 2 \\
01.10.2019 & -5 & 5 & 0 & 5 & -4 & 0 & 5 & 6 & -6 & -4 & -4 & -4 \\
01.11.2019 & -3 &  &  &  & 0 &  &  &  & -4 &  &  &  \\
01.12.2019 &  & 4 & 5 & 5 &  & 0 & 6 & 6 &  & 0 & -4 & -4 \\
01.01.2020 & -3 & 4 & 5 & 5 & -4 & 0 & 6 & 6 & 0 & 0 & -4 & -4 \\
01.02.2020 & 0 &  &  &  & 5 &  &  &  & -4 &  &  &  \\
01.03.2020 &  & 4 & -6 & 10 &  & 5 & 5 & 5 &  & 0 & -4 & 0 \\
01.04.2020 & -8 & 5 & -7 & -5 & -7 & 0 & -8 & -6 & -6 & -4 & -6 & -4 \\
01.05.2020 & -8 &  &  &  & -7 &  &  &  & -6 &  &  &  \\
01.06.2020 & -8 & 5 & -7 & -5 & -6 & 0 & -5 & -6 & -7 & -4 & -6 & -4 \\
01.07.2020 &  & 4 & -3 & -5 &  & 5 & -4 & 3 &  & -4 & -6 & -4 \\
01.08.2020 & 3 &  &  &  & -5 &  &  &  & -4 &  &  &  \\
01.09.2020 &  & 4 & 6 & 0 &  & 5 & 5 & 5 &  & -4 & -4 & -4 \\
01.10.2020 & 5 & 4 &  &  & -4 & 5 &  &  & -6 & -4 &  &  \\
01.11.2020 & -4 &  & -2 & -5 & -5 &  & -3 & -3 & -6 &  & -4 & -4 \\
01.12.2020 &  & 6 & -3 & -5 &  & 5 & -5 & -3 &  & 4 & -4 & -4 \\
01.01.2021 & -5 & 4 & -5 & -3 & -4 & 5 & -6 & -5 & -6 & -4 & -4 & -4 \\
01.02.2021 & -3 &  &  &  & -4 &  &  &  & -6 &  &  &  \\
01.03.2021 &  & 4 & 6 & 0 &  & 5 & 5 & 5 &  & -4 & -4 & -4 \\
01.04.2021 & 6 & 3 & 6 & 3 & 5 & 5 & 5 & 5 & -4 & 4 & -4 & -4 \\
01.05.2021 & 5 &  &  &  & -3 &  &  &  & -4 &  &  &  \\
01.06.2021 &  &  & 6 & 6 &  &  & 5 & 5 &  &  & -4 & -4 \\
01.07.2021 & 5 &  & 6 & 6 & -3 &  & 5 & 5 & 4 &  & -4 & -4 \\
01.08.2021 & 6 &  &  &  & 5 &  &  &  & -4 &  &  &  \\
01.09.2021 &  &  & 0 & 6 &  &  & -3 & 5 &  &  & -4 & -4 \\
01.10.2021 & 6 &  &  &  & 5 &  &  &  & -4 &  &  &  \\
01.11.2021 & 5 &  & 0 & 6 & -4 &  & 5 & 5 & -6 &  & -4 & -4 \\
01.12.2021 &  &  & 5 & 5 &  &  & 7 & 6 &  &  & -6 & -4 \\
01.01.2022 & 3 &  & 5 & 6 & 5 &  & 7 & 5 & -4 &  & -6 & -4 \\
01.02.2022 &  & -3 &  &  &  & 0 &  &  &  & 4 &  &  \\
01.03.2022 & 0 & -3 & -4 & -3 & 5 & 0 & 7 & 6 & -6 & -4 & -6 & -4 \\
01.04.2022 & -4 & -4 &  &  & 5 & 0 &  &  & -7 & -7 &  &  \\
01.05.2022 & -4 &  & 0 & -4 & 5 &  & 7 & 5 & -7 &  & -6 & -6 \\
01.06.2022 &  & 4 & 0 & 4 &  & 5 & 5 & 5 &  & -7 & -7 & -6 \\
01.07.2022 & -4 & 0 & -4 & -4 & 5 & 0 & 5 & 5 & -7 & 6 & -7 & -6 \\
01.08.2022 & -4 &  &  &  & 5 &  &  &  & -7 &  &  &  \\
01.09.2022 &  & -6 & -4 & -4 &  & -4 & 5 & 5 &  & -7 & -6 & -6 \\
01.10.2022 & -4 & 0 &  &  & 5 & 0 &  &  & -7 & -7 &  &  \\
01.11.2022 & -6 &  & -4 & -4 & 5 &  & 5 & 5 & -7 &  & -6 & -6 \\
01.12.2022 &  & -4 & -3 & -4 &  & 0 & 5 & 5 &  & -7 & -4 & -6 \\
01.01.2023 & -4 &  &  &  & 5 &  &  &  & -6 &  &  &  \\
01.02.2023 &  & 0 & 0 & 3 &  & 0 & 5 & 6 &  & -6 & -4 & -4 \\
01.03.2023 & 6 & 4 & 0 & -3 & 5 & 5 & 6 & 6 & -4 & -6 & -4 & -4 \\
01.04.2023 & 0 &  &  &  & 5 &  &  &  & -6 &  &  &  \\
01.05.2023 &  & -4 & 0 & 0 &  & 0 & 5 & 5 &  & -6 & -4 & -4 \\
01.06.2023 & 4 & -3 & 3 & 3 & 5 & 4 & 5 & 5 & -6 & -6 & -4 & -4 \\
01.07.2023 & -4 & -3 & 6 & 3 & 6 & 0 & 5 & 5 & -5 & -4 & -4 & -4 \\
01.08.2023 & -3 &  &  &  & 5 &  &  &  & -4 &  &  &  \\
01.09.2023 &  & -3 & 5 & 5 &  & 0 & 0 & 0 &  & -4 & -4 & -4 \\
01.10.2023 & -5 & 0 &  &  & -4 & 0 &  &  & -6 & -4 &  &  \\
01.11.2023 & -5 &  & 6 & 6 & -3 &  & 5 & 5 & 4 &  & -4 & -4 \\
01.12.2023 &  & 5 & -3 & -3 &  & 0 & 5 & 5 &  & -4 & -4 & -4 \\
01.01.2024 & -5 &  & 6 & 6 & -3 &  & 5 & 5 & 4 &  & -4 & -4 \\
01.02.2024 & -3 &  &  &  & 5 &  &  &  & 4 &  &  &  \\
01.03.2024 &  &  & 6 & 5 &  &  & 5 & 6 &  &  & -4 & -4 \\
01.04.2024 & -3 &  &  &  & 5 &  &  &  & -4 &  &  &  \\
01.05.2024 & -3 &  & 3 & 5 & -4 &  & 5 & 6 & 0 &  & -4 & -4 \\
01.06.2024 &  & 5 & 6 & 5 &  & 0 & 5 & 6 &  & -4 & -4 & -4 \\
01.07.2024 & 6 & 0 & 5 & 5 & 5 & 0 & -3 & 0 & -4 & -4 & 4 & 4 \\
01.08.2024 & 0 &  &  &  & 5 &  &  &  & -4 &  &  &  \\
01.09.2024 &  & -3 &  & 5 &  & 0 &  & -3 &  & -4 &  & 4 \\
\end{longtable}
\endgroup
\end{landscape}

\section{R Script Granger test to discard predictors}
\begin{lstlisting}
################################################################################
\end{lstlisting}
\begin{lstlisting}
################ Multiscenario IR forecasting utilizing AI #####################
\end{lstlisting}
\begin{lstlisting}
########## Supporting a structured view of interest rate forecasts #############
\end{lstlisting}
\begin{lstlisting}
############################## zeb business school #############################
\end{lstlisting}
\begin{lstlisting}
################################################################################
\end{lstlisting}
\begin{lstlisting}
########### Econometric AI model: Granger test to discard predictors ###########
\end{lstlisting}
\begin{lstlisting}
# In case the user needs to manually change the code to run smoothly, the following
\end{lstlisting}
\begin{lstlisting}
# message is displayed:
\end{lstlisting}
\begin{lstlisting}
#~~~~~~~~~~~~~~~~~~~~~~~~~~~~~~~~~~~~#
\end{lstlisting}
\begin{lstlisting}
#~~~ User input possibly required ~~~#
\end{lstlisting}
\begin{lstlisting}
#~~~~~~~~~~~~~~~~~~~~~~~~~~~~~~~~~~~~#
\end{lstlisting}
\begin{lstlisting}
# Step 1. Load/Install Dependencies ############################################
\end{lstlisting}
\begin{lstlisting}
#~~~~~~~~~~~~~~~~~~~~~~~~~~~~~~~~~~~~#
\end{lstlisting}
\begin{lstlisting}
#~~~ User input possibly required ~~~#
\end{lstlisting}
\begin{lstlisting}
#~~~~~~~~~~~~~~~~~~~~~~~~~~~~~~~~~~~~#
\end{lstlisting}
\begin{lstlisting}
# if one of the following packages is not yet installed, please "uncomment" and run the respective line of code
\end{lstlisting}
\begin{lstlisting}
#install.packages("dplyr")
\end{lstlisting}
\begin{lstlisting}
#install.packages("tidyr")
\end{lstlisting}
\begin{lstlisting}
#install.packages("data.table")
\end{lstlisting}
\begin{lstlisting}
#install.packages("lmtest")
\end{lstlisting}
\begin{lstlisting}
#install.packages("knitr", dependencies=T)
\end{lstlisting}
\begin{lstlisting}
#install.packages("corrplot")
\end{lstlisting}
\begin{lstlisting}
#install.packages("Hmisc", dependencies=T)
\end{lstlisting}
\begin{lstlisting}
library(dplyr)
\end{lstlisting}
\begin{lstlisting}
library(tidyr)
\end{lstlisting}
\begin{lstlisting}
library(data.table)
\end{lstlisting}
\begin{lstlisting}
library(lmtest)
\end{lstlisting}
\begin{lstlisting}
library(knitr)
\end{lstlisting}
\begin{lstlisting}
library(corrplot)
\end{lstlisting}
\begin{lstlisting}
library(Hmisc)
\end{lstlisting}
\begin{lstlisting}
# Step 2. Load Data-Set ########################################################
\end{lstlisting}
\begin{lstlisting}
# Please copy the path of where the current data-set is located and insert it here
\end{lstlisting}
Monthly\_EA <- read\_excel("EA\_data\_monthly.xlsx")

\begin{lstlisting}
# Step 3. Prepare the Data #####################################################
\end{lstlisting}
\begin{lstlisting}
# Exclude all the Data, that carry NA in Jan 1999 (Start of Swap-Data)
\end{lstlisting}
Monthly\_EA\_numeric <- Monthly\_EA[, -1]  \# Exclude the first column (dates)

Monthly\_EA\_ts <- ts(Monthly\_EA\_numeric, \#Convert into a Time-Series

start = c(1986, 12), \# Start of Data-Set in December 1986

frequency = 12)      \# Monthly data

time\_index <- time(Monthly\_EA\_ts)  \# Get the time index

years <- floor(time\_index)

months <- round((time\_index - years) * 12 + 1)

index\_start <- which(years == 1999 \& months == 01) \#Swap Data Starts 1999

index\_start \# Nr. of Rows in Data-Set where Swaps start.

columns\_with\_na <- is.na(Monthly\_EA\_ts[index\_start, ])

Monthly\_EA\_cleaned\_ts <- Monthly\_EA\_ts[, !columns\_with\_na]

Monthly\_EA\_cleaned\_ts <- Monthly\_EA\_cleaned\_ts[time(Monthly\_EA\_cleaned\_ts)

>= time(Monthly\_EA\_ts)[index\_start], ]

data <- ts(Monthly\_EA\_cleaned\_ts, start = c(1999, 1), frequency = 12)

data

\begin{lstlisting}
# Step 4. Granger-Test #########################################################
\end{lstlisting}
\begin{lstlisting}
# We conduct the Granger-Test to test for predictive power
\end{lstlisting}
\begin{lstlisting}
# Define Dependent / Desired Forecast Target Variable (e.g. Swap of different Maturity)
\end{lstlisting}
target <- "ICEIB2Y" \#You can check variable name in helper Excel and via colnames(data)

\begin{lstlisting}
# Define the maximum lag for Granger causality testing
\end{lstlisting}
max\_lag <- 1  \# Adjust this to test robustness of findings. 1-12 is a value that's sensible.

\begin{lstlisting}
# Initialize a data frame to store the results for the overall time frame
\end{lstlisting}
granger\_results <- data.frame(Variable = character(),

Lag = integer(),

Statistic = numeric(),

P\_Value = numeric(),

stringsAsFactors = FALSE)

\begin{lstlisting}
# Initialize a data frame to store the results for the high yield environment
\end{lstlisting}
granger\_results\_hy <- data.frame(Variable = character(),

Lag = integer(),

Statistic = numeric(),

P\_Value = numeric(),

stringsAsFactors = FALSE)

\begin{lstlisting}
# Initialize a data frame to store the results for the decreasing yield environment
\end{lstlisting}
granger\_results\_dy <- data.frame(Variable = character(),

Lag = integer(),

Statistic = numeric(),

P\_Value = numeric(),

stringsAsFactors = FALSE)

\begin{lstlisting}
# Initialize a data frame to store the results for the low yield environment
\end{lstlisting}
granger\_results\_ly <- data.frame(Variable = character(),

Lag = integer(),

Statistic = numeric(),

P\_Value = numeric(),

stringsAsFactors = FALSE)

\begin{lstlisting}
# Initialize a data frame to store the results for the increasing yield environment
\end{lstlisting}
granger\_results\_iy <- data.frame(Variable = character(),

Lag = integer(),

Statistic = numeric(),

P\_Value = numeric(),

stringsAsFactors = FALSE)

\begin{lstlisting}
# create dataframe for high yield period
\end{lstlisting}
data\_hy <- window(data, start = c(1999,01), end = c(2008,12))

\begin{lstlisting}
# delete columns with std = 0
\end{lstlisting}
data\_hy <- data\_hy[, !sapply(data\_hy, function(x) \{ sd(x) == 0\} )]

\begin{lstlisting}
# create dataframe for decreasing yield period
\end{lstlisting}
data\_dy <- window(data, start = c(2008,07), end = c(2016,03))

\begin{lstlisting}
# delete columns with std = 0
\end{lstlisting}
data\_dy <- data\_dy[, !sapply(data\_dy, function(x) \{ sd(x) == 0\} )]

\begin{lstlisting}
# create dataframe for low yield period
\end{lstlisting}
data\_ly <- window(data, start = c(2016,03), end = c(2022,03))

\begin{lstlisting}
# delete columns with std = 0
\end{lstlisting}
\begin{lstlisting}
#a <- ncol(data_ly)
\end{lstlisting}
\begin{lstlisting}
#a
\end{lstlisting}
data\_ly <- data\_ly[, !sapply(data\_ly, function(x) \{ sd(x) == 0\} )]

\begin{lstlisting}
#a <- ncol(data_ly)
\end{lstlisting}
\begin{lstlisting}
#a
\end{lstlisting}
\begin{lstlisting}
# create dataframe for recent past with increasing yields
\end{lstlisting}
data\_iy <- window(data, start = c(2021,01), end = c(2024,06))

data\_iy <- data\_iy[, !sapply(data\_iy, function(x) \{ sd(x) == 0\} )]

\begin{lstlisting}
#correlation plots -> eventually done by David
\end{lstlisting}
\begin{lstlisting}
#data_test <- data_hy[,c(1:5)]
\end{lstlisting}
\begin{lstlisting}
# Correlation matrix and p-values
\end{lstlisting}
\begin{lstlisting}
#res2 <- rcorr(as.matrix(data_test))
\end{lstlisting}
\begin{lstlisting}
# Insignificant correlations are crossed
\end{lstlisting}
\begin{lstlisting}
#corrplot(res2$r, type = "lower", order = "hclust",
\end{lstlisting}
\begin{lstlisting}
#         p.mat = res2$P, sig.level = 0.05, insig = "blank",
\end{lstlisting}
\begin{lstlisting}
#         tl.cex = 0.8, tl.col = "black", # Adjust text size and color
\end{lstlisting}
\begin{lstlisting}
#         tl.labels = colnames(resids)) # Use math expression labels
\end{lstlisting}
\begin{lstlisting}
# Loop the test through all variables in the data frame except the target
\end{lstlisting}
\begin{lstlisting}
# entire dataframe (1999-2024)
\end{lstlisting}
for (var in colnames(data)) \{

if (var != target) \{

\begin{lstlisting}
# Perform Granger causality test
\end{lstlisting}
granger\_test <- grangertest(data[,var], data[, colnames(data) == target], order = max\_lag) \# Check if column 39 is truly target column

\begin{lstlisting}
# Extract test statistics and p-values
\end{lstlisting}
test\_statistic <- granger\_test\$`F`[2]

p\_value <- granger\_test\$`Pr(>F)`[2]

\begin{lstlisting}
# Add the results to the table
\end{lstlisting}
granger\_results <- rbind(granger\_results,

data.frame(Variable = var,

Lag = max\_lag,

Statistic = test\_statistic,

P\_Value = p\_value))

\}

\}

granger\_results

\begin{lstlisting}
# Create a Table.
\end{lstlisting}
\begin{lstlisting}
# Reformat as "Booktabs" recommended.
\end{lstlisting}
kable(granger\_results,

format = "latex",

caption = "Granger-Test Results",

digits = 4,

align = "c",

col.names = c("Variable", "Lag","Test Statistic", "P-Value"))

\begin{lstlisting}
# high yield period (1999-2008)
\end{lstlisting}
for (var in colnames(data\_hy)) \{

if (var != target) \{

\begin{lstlisting}
# Perform Granger causality test
\end{lstlisting}
granger\_test\_hy <- grangertest(data\_hy[,var], data\_hy[, colnames(data\_hy) == target], order = max\_lag) \# Check if column 39 is truly target column

\begin{lstlisting}
# Extract test statistics and p-values
\end{lstlisting}
test\_statistic <- granger\_test\_hy\$`F`[2]

p\_value <- granger\_test\_hy\$`Pr(>F)`[2]

\begin{lstlisting}
# Add the results to the table
\end{lstlisting}
granger\_results\_hy <- rbind(granger\_results\_hy,

data.frame(Variable = var,

Lag = max\_lag,

Statistic = test\_statistic,

P\_Value = p\_value))

\}

\}

granger\_results\_hy

\begin{lstlisting}
# Create a Table.
\end{lstlisting}
\begin{lstlisting}
# Reformat as "Booktabs" recommended.
\end{lstlisting}
kable(granger\_results\_hy,

format = "latex",

caption = "Granger-Test Results",

digits = 4,

align = "c",

col.names = c("Variable", "Lag","Test Statistic", "P-Value"))

\begin{lstlisting}
# decreasing yield period (2008-2016)
\end{lstlisting}
for (var in colnames(data\_dy)) \{

if (var != target) \{

\begin{lstlisting}
# Perform Granger causality test
\end{lstlisting}
granger\_test\_dy <- grangertest(data\_dy[,var], data\_dy[, colnames(data\_dy) == target], order = max\_lag) \# Check if column 39 is trudy target column

\begin{lstlisting}
# Extract test statistics and p-values
\end{lstlisting}
test\_statistic <- granger\_test\_dy\$`F`[2]

p\_value <- granger\_test\_dy\$`Pr(>F)`[2]

\begin{lstlisting}
# Add the results to the table
\end{lstlisting}
granger\_results\_dy <- rbind(granger\_results\_dy,

data.frame(Variable = var,

Lag = max\_lag,

Statistic = test\_statistic,

P\_Value = p\_value))

\}

\}

granger\_results\_dy

\begin{lstlisting}
# Create a Table.
\end{lstlisting}
\begin{lstlisting}
# Reformat as "Booktabs" recommended.
\end{lstlisting}
kable(granger\_results\_dy,

format = "latex",

caption = "Granger-Test Results",

digits = 4,

align = "c",

col.names = c("Variable", "Lag","Test Statistic", "P-Value"))

\begin{lstlisting}
# low yield period (2016-2022)
\end{lstlisting}
for (var in colnames(data\_ly)) \{

if (var != target) \{

\begin{lstlisting}
# Perform Granger causality test
\end{lstlisting}
granger\_test\_ly <- grangertest(data\_ly[,var], data\_ly[, colnames(data\_ly) == target], order = max\_lag) \# Check if column 39 is truly target column

\begin{lstlisting}
# Extract test statistics and p-values
\end{lstlisting}
test\_statistic <- granger\_test\_ly\$`F`[2]

p\_value <- granger\_test\_ly\$`Pr(>F)`[2]

\begin{lstlisting}
# Add the results to the table
\end{lstlisting}
granger\_results\_ly <- rbind(granger\_results\_ly,

data.frame(Variable = var,

Lag = max\_lag,

Statistic = test\_statistic,

P\_Value = p\_value))

\}

\}

granger\_results\_ly

\begin{lstlisting}
# Create a Table.
\end{lstlisting}
\begin{lstlisting}
# Reformat as "Booktabs" recommended.
\end{lstlisting}
kable(granger\_results\_ly,

format = "latex",

caption = "Granger-Test Results",

digits = 4,

align = "c",

col.names = c("Variable", "Lag","Test Statistic", "P-Value"))

\begin{lstlisting}
# increasing yield period (2021-2024)
\end{lstlisting}
for (var in colnames(data\_iy)) \{

if (var != target) \{

\begin{lstlisting}
# Perform Granger causality test
\end{lstlisting}
granger\_test\_iy <- grangertest(data\_iy[,var], data\_iy[, colnames(data\_iy) == target], order = max\_lag) \# Check if column 39 is truly target column

\begin{lstlisting}
# Extract test statistics and p-values
\end{lstlisting}
test\_statistic <- granger\_test\_iy\$`F`[2]

p\_value <- granger\_test\_iy\$`Pr(>F)`[2]

\begin{lstlisting}
# Add the results to the table
\end{lstlisting}
granger\_results\_iy <- rbind(granger\_results\_iy,

data.frame(Variable = var,

Lag = max\_lag,

Statistic = test\_statistic,

P\_Value = p\_value))

\}

\}

granger\_results\_iy

\begin{lstlisting}
# Create a Table.
\end{lstlisting}
\begin{lstlisting}
# Reformat as "Booktabs" recommended.
\end{lstlisting}
kable(granger\_results\_iy,

format = "latex",

caption = "Granger-Test Results",

digits = 4,

align = "c",

col.names = c("Variable", "Lag","Test Statistic", "P-Value"))

\section{R Script Backtest of VAR and ARIMA models}
\begin{lstlisting}
################################################################################
\end{lstlisting}
\begin{lstlisting}
################ Multiscenario IR forecasting utilizing AI #####################
\end{lstlisting}
\begin{lstlisting}
########## Supporting a structured view of interest rate forecasts #############
\end{lstlisting}
\begin{lstlisting}
############################## zeb business school #############################
\end{lstlisting}
\begin{lstlisting}
################################################################################
\end{lstlisting}
\begin{lstlisting}
####################### Backtest of VAR and ARIMA models #######################
\end{lstlisting}
\begin{lstlisting}
# In case the user needs to manually change the code to run smoothly, the following
\end{lstlisting}
\begin{lstlisting}
# message is displayed:
\end{lstlisting}
\begin{lstlisting}
#~~~~~~~~~~~~~~~~~~~~~~~~~~~~~~~~~~~~#
\end{lstlisting}
\begin{lstlisting}
#~~~ User input possibly required ~~~#
\end{lstlisting}
\begin{lstlisting}
#~~~~~~~~~~~~~~~~~~~~~~~~~~~~~~~~~~~~#
\end{lstlisting}
\begin{lstlisting}
# Step 1. Load/Install Dependencies ############################################
\end{lstlisting}
\begin{lstlisting}
#~~~~~~~~~~~~~~~~~~~~~~~~~~~~~~~~~~~~#
\end{lstlisting}
\begin{lstlisting}
#~~~ User input possibly required ~~~#
\end{lstlisting}
\begin{lstlisting}
#~~~~~~~~~~~~~~~~~~~~~~~~~~~~~~~~~~~~#
\end{lstlisting}
\begin{lstlisting}
# if one of the following packages is not yet installed, please "uncomment" and run the respective line of code
\end{lstlisting}
\begin{lstlisting}
#install.packages("vars")
\end{lstlisting}
\begin{lstlisting}
#install.packages("readxl")
\end{lstlisting}
\begin{lstlisting}
#install.packages("ggplot2")
\end{lstlisting}
\begin{lstlisting}
#install.packages("reshape2")
\end{lstlisting}
\begin{lstlisting}
#install.packages("gridExtra")
\end{lstlisting}
\begin{lstlisting}
#install.packages("tidyr")
\end{lstlisting}
\begin{lstlisting}
#install.packages("dplyr")
\end{lstlisting}
\begin{lstlisting}
#install.packages("lmtest")
\end{lstlisting}
\begin{lstlisting}
#install.packages("corrplot")
\end{lstlisting}
\begin{lstlisting}
#install.packages("Hmisc")
\end{lstlisting}
\begin{lstlisting}
#install.packages("vars")
\end{lstlisting}
\begin{lstlisting}
#install.packages("forecast")
\end{lstlisting}
\begin{lstlisting}
library(vars)
\end{lstlisting}
\begin{lstlisting}
library(readxl)
\end{lstlisting}
\begin{lstlisting}
library(ggplot2)
\end{lstlisting}
\begin{lstlisting}
library(reshape2)
\end{lstlisting}
\begin{lstlisting}
library(gridExtra)
\end{lstlisting}
\begin{lstlisting}
library(tidyr)
\end{lstlisting}
\begin{lstlisting}
library(dplyr)
\end{lstlisting}
\begin{lstlisting}
library(lmtest)
\end{lstlisting}
\begin{lstlisting}
library(corrplot)
\end{lstlisting}
\begin{lstlisting}
library(Hmisc)
\end{lstlisting}
\begin{lstlisting}
library(vars)
\end{lstlisting}
\begin{lstlisting}
library(forecast)
\end{lstlisting}
\begin{lstlisting}
# Step 2. Load Data-Set ########################################################
\end{lstlisting}
\begin{lstlisting}
# Paste the path of where the current data-set is located.
\end{lstlisting}
Monthly\_EA <- read\_excel("EA\_data\_monthly.xlsx")

\begin{lstlisting}
# Step 3. Prepare the Data #####################################################
\end{lstlisting}
\begin{lstlisting}
# We will exclude all the Data, that carries NA in Jan 1999 (Start of Swap-Data)
\end{lstlisting}
Monthly\_EA\_numeric <- Monthly\_EA[, -1]  \# Exclude the first column (dates)

Monthly\_EA\_ts <- ts(Monthly\_EA\_numeric, \#Convert into a Time-Series

start = c(1986, 12), \# Start of Data-Set in December 1986

frequency = 12)      \# Monthly data

time\_index <- time(Monthly\_EA\_ts)  \# Get the time index

years <- floor(time\_index)

months <- round((time\_index - years) * 12 + 1)

index\_start <- which(years == 1999 \& months == 01) \#Swap Data Starts 1999

index\_start \# Nr. of Rows in Data-Set where Swaps start.

columns\_with\_na <- is.na(Monthly\_EA\_ts[index\_start, ])

Monthly\_EA\_cleaned\_ts <- Monthly\_EA\_ts[, !columns\_with\_na]

Monthly\_EA\_cleaned\_ts <- Monthly\_EA\_cleaned\_ts[time(Monthly\_EA\_cleaned\_ts)

>= time(Monthly\_EA\_ts)[index\_start], ]

data <- ts(Monthly\_EA\_cleaned\_ts, start = c(1999, 1), frequency = 12)

data

\begin{lstlisting}
# Save Data into individual Series and Transform them to Log-Levels (except Rates)
\end{lstlisting}
\begin{lstlisting}
# Taking Logs allows us to interpret Model output as Elasticities / % Changes
\end{lstlisting}
infl\_us <- na.omit(data[,colnames(data) == "infl\_US"]) \# US-Inflation Rate

Y <- na.omit(data[,colnames(data) == "ind\_prod"]) \# Growth-Rate Industrial Prod. (EA)

Price <- na.omit(data[,colnames(data) == "infl"])  \# Inflation (EA)

r\_s <-  na.omit(data[,colnames(data) == "EUR\_m\_refi"]) \# EZB Refinancing Rate

cdty <- log(na.omit(data[,colnames(data) == "Com\_Index"])) \# Commodity Price Index: - Taking Logs

eur\_stoxx <- log(na.omit(data[,colnames(data) == "euro\_stoxx"])) \# Euro-Stoxx50 Index: - Taking Logs

pmi <- na.omit(data[,colnames(data) == "m\_PMI\_US"]) \#PMI

r\_b\_2 <-  na.omit(data[,colnames(data) == "ICEIB2Y"]) \# 2Yr Swap

r\_b\_5 <-  na.omit(data[,colnames(data) == "ICEIB5Y"]) \# 5Yr Swap

r\_b\_10 <-  na.omit(data[,colnames(data) == "ICEIB10"]) \# 10Yr Swap

r\_b\_30 <-  na.omit(data[,colnames(data) == "ICEIB30"]) \# 30Yr Swap

dat <- cbind(infl\_us , Y, Price , r\_s , pmi ,cdty, eur\_stoxx , r\_b\_30 , r\_b\_10 , r\_b\_5 ,r\_b\_2)

colnames(dat) <- c("Inflation\_US", "Industrial\_Prod", "Inflation", "ECB\_Refi","Manufacturing\_PMI(US)", "Commodity\_Indx", "EUR\_Stoxx50", "SWAP\_Rate\_30Yr","SWAP\_Rate\_10Yr", "SWAP\_Rate\_5Yr", "SWAP\_Rate\_2Yr")

dat <- ts(dat, start = c(1999,01), frequency = 12)

\begin{lstlisting}
# Initialize a list to store forecasts
\end{lstlisting}
forecasts <- list()

forecast\_2Yr <- data.frame()

forecast\_5Yr <-  data.frame()

forecast\_10Yr <-  data.frame()

forecast\_30Yr <-  data.frame()

\begin{lstlisting}
# Parameters
\end{lstlisting}
horizon <- 12

backtest\_obs <- 240

\begin{lstlisting}
# Get the total number of observations
\end{lstlisting}
n <- nrow(dat)

\begin{lstlisting}
# Start backtesting from (n - backtest_obs + 1) to the end
\end{lstlisting}
start\_idx <- n - backtest\_obs + 1

\begin{lstlisting}
# Perform rolling forecasts
\end{lstlisting}
for (i in start\_idx:(n - horizon)) \{

\begin{lstlisting}
# Define the training data up to the current point
\end{lstlisting}
train\_data <- dat[1:i, ]

mod  <- VAR(dat, p = 1, type = "const")

\begin{lstlisting}
# To avoid excessive overfitting only one lag sensible
\end{lstlisting}
\begin{lstlisting}
# Generate forecast for the next 'horizon' periods
\end{lstlisting}
forecast <- predict(mod, n.ahead = horizon)

forecast\_2Yr <- rbind(forecast\$fcst\$SWAP\_Rate\_2Yr[12], forecast\_2Yr)

forecast\_5Yr <- rbind(forecast\$fcst\$SWAP\_Rate\_5Yr[12], forecast\_5Yr)

forecast\_10Yr <- rbind(forecast\$fcst\$SWAP\_Rate\_10Yr[12], forecast\_10Yr)

forecast\_30Yr <- rbind(forecast\$fcst\$SWAP\_Rate\_30Yr[12], forecast\_30Yr)

\}

actual\_2Yr <- dat[c(start\_idx+horizon):length(dat[,11]),11]

actual\_5Yr <- dat[c(start\_idx+horizon):length(dat[,11]),10]

actual\_10Yr <- dat[c(start\_idx+horizon):length(dat[,11]),9]

actual\_30Yr <- dat[c(start\_idx+horizon):length(dat[,11]),8]

\begin{lstlisting}
# Extract rmse for different tenors
\end{lstlisting}
rmse\_2Yr <- sqrt(1/length(actual\_2Yr)*sum((actual\_2Yr - rev(forecast\_2Yr))\textasciicircum{}2))

rmse\_2Yr

rmse\_5Yr <- sqrt(1/length(actual\_5Yr)*sum((actual\_5Yr - rev(forecast\_5Yr))\textasciicircum{}2))

rmse\_5Yr

rmse\_10Yr <- sqrt(1/length(actual\_10Yr)*sum((actual\_10Yr - rev(forecast\_10Yr))\textasciicircum{}2))

rmse\_10Yr

rmse\_30Yr <- sqrt(1/length(actual\_30Yr)*sum((actual\_30Yr - rev(forecast\_30Yr))\textasciicircum{}2))

rmse\_30Yr

\begin{lstlisting}
##### Benchmark / Backtest ARIMA Model ########################################
\end{lstlisting}
\begin{lstlisting}
# Initialize a list to store forecasts
\end{lstlisting}
forecast\_2Yr <- numeric()

forecast\_5Yr <-  numeric()

forecast\_10Yr <-  numeric()

forecast\_30Yr <-  numeric()

\begin{lstlisting}
# Perform rolling forecasts
\end{lstlisting}
for (i in start\_idx:(n - horizon)) \{

\begin{lstlisting}
# Define the training data up to the current point
\end{lstlisting}
train\_data <- dat[1:i, ]

\begin{lstlisting}
# Generate forecast for the next 'horizon' periods
\end{lstlisting}
model <- auto.arima(train\_data[,11]) \# For Maturities 2 and 5 (col 11 and 12)

model5 <- auto.arima(train\_data[,10]) \# For Maturities 2 and 5 (col 11 and 12)

model10 <- arima(order = c(1, 1, 1),train\_data[,9]) \# For Maturities 2 and 5 (col 11 and 12)

model30 <- arima(order = c(1, 1, 1),train\_data[,8]) \# For Maturities 2 and 5 (col 11 and 12)

forecast <- predict(model, horizon = horizon)

forecast5 <- predict(model5, horizon = horizon)

forecast10 <- predict(model10, horizon = horizon)

forecast30 <- predict(model30, horizon = horizon)

forecast\_2Yr <- c(forecast\$pred, forecast\_2Yr)

forecast\_5Yr <- c(forecast5\$pred, forecast\_5Yr)

forecast\_10Yr <- c(forecast10\$pred, forecast\_10Yr)

forecast\_30Yr <- c(forecast30\$pred, forecast\_30Yr)

\}

\begin{lstlisting}
# If forecast_2Yr is a list or matrix, flatten it
\end{lstlisting}
forecast\_2Yr <- unlist(forecast\_2Yr)

forecast\_5Yr <- unlist(forecast\_5Yr)

arima\_error <-  actual\_2Yr - forecast\_2Yr

\begin{lstlisting}
# Ensure it is numeric
\end{lstlisting}
forecast\_2Yr <- as.numeric(forecast\_2Yr)

forecast\_5Yr <- as.numeric(forecast\_5Yr)

actual\_2Yr <- dat[c(start\_idx+horizon):length(dat[,11]),11]

actual\_5Yr <- dat[c(start\_idx+horizon):length(dat[,11]),10]

actual\_10Yr <- dat[c(start\_idx+horizon):length(dat[,11]),9]

actual\_30Yr <- dat[c(start\_idx+horizon):length(dat[,11]),8]

\begin{lstlisting}
# Extract rmse for different tenors
\end{lstlisting}
rmse\_2Yr <- sqrt(1/length(actual\_2Yr)*sum((actual\_2Yr - rev(forecast\_2Yr))\textasciicircum{}2))

rmse\_2Yr

rmse\_5Yr <- sqrt(1/length(actual\_5Yr)*sum((actual\_5Yr - rev(forecast\_5Yr))\textasciicircum{}2))

rmse\_5Yr

rmse\_10Yr <- sqrt(1/length(actual\_10Yr)*sum((actual\_10Yr - rev(forecast\_10Yr))\textasciicircum{}2))

rmse\_10Yr

rmse\_30Yr <- sqrt(1/length(actual\_30Yr)*sum((actual\_30Yr - rev(forecast\_30Yr))\textasciicircum{}2))

rmse\_30Yr

\section{R Script Econometric AI model: High yields}
\begin{lstlisting}
################################################################################
\end{lstlisting}
\begin{lstlisting}
################ Multiscenario IR forecasting utilizing AI #####################
\end{lstlisting}
\begin{lstlisting}
########## Supporting a structured view of interest rate forecasts #############
\end{lstlisting}
\begin{lstlisting}
############################## zeb business school #############################
\end{lstlisting}
\begin{lstlisting}
################################################################################
\end{lstlisting}
\begin{lstlisting}
############## Econometric AI model: high yields (01/1999-12/2008) #############
\end{lstlisting}
\begin{lstlisting}
# In case the user needs to manually change the code to run smoothly, the following
\end{lstlisting}
\begin{lstlisting}
# message is displayed:
\end{lstlisting}
\begin{lstlisting}
#~~~~~~~~~~~~~~~~~~~~~~~~~~~~~~~~~~~~#
\end{lstlisting}
\begin{lstlisting}
#~~~ User input possibly required ~~~#
\end{lstlisting}
\begin{lstlisting}
#~~~~~~~~~~~~~~~~~~~~~~~~~~~~~~~~~~~~#
\end{lstlisting}
\begin{lstlisting}
# Step 1. Load/Install Dependencies ############################################
\end{lstlisting}
\begin{lstlisting}
#~~~~~~~~~~~~~~~~~~~~~~~~~~~~~~~~~~~~#
\end{lstlisting}
\begin{lstlisting}
#~~~ User input possibly required ~~~#
\end{lstlisting}
\begin{lstlisting}
#~~~~~~~~~~~~~~~~~~~~~~~~~~~~~~~~~~~~#
\end{lstlisting}
\begin{lstlisting}
# if one of the following packages is not yet installed, please "uncomment" and run the respective line of code
\end{lstlisting}
\begin{lstlisting}
#install.packages("dplyr")
\end{lstlisting}
\begin{lstlisting}
#install.packages("ggplot2")
\end{lstlisting}
\begin{lstlisting}
#install.packages("readxl")
\end{lstlisting}
\begin{lstlisting}
#install.packages("forecast")
\end{lstlisting}
\begin{lstlisting}
#install.packages("Metrics")
\end{lstlisting}
\begin{lstlisting}
#install.packages("tseries")
\end{lstlisting}
\begin{lstlisting}
#install.packages("BVAR")
\end{lstlisting}
\begin{lstlisting}
#install.packages("plotly")
\end{lstlisting}
\begin{lstlisting}
#install.packages("coda")
\end{lstlisting}
\begin{lstlisting}
#install.packages("tidyr")
\end{lstlisting}
\begin{lstlisting}
#install.packages("data.table")
\end{lstlisting}
\begin{lstlisting}
#install.packages("knitr")
\end{lstlisting}
\begin{lstlisting}
#install.packages("corrplot")
\end{lstlisting}
\begin{lstlisting}
#install.packages("Hmisc")
\end{lstlisting}
\begin{lstlisting}
library(dplyr)
\end{lstlisting}
\begin{lstlisting}
library(ggplot2)
\end{lstlisting}
\begin{lstlisting}
library(readxl)
\end{lstlisting}
\begin{lstlisting}
library(forecast)
\end{lstlisting}
\begin{lstlisting}
library(Metrics)
\end{lstlisting}
\begin{lstlisting}
library(tseries)
\end{lstlisting}
\begin{lstlisting}
library(BVAR)
\end{lstlisting}
\begin{lstlisting}
library(plotly)
\end{lstlisting}
\begin{lstlisting}
library(coda)
\end{lstlisting}
\begin{lstlisting}
library(tidyr)
\end{lstlisting}
\begin{lstlisting}
library(data.table)
\end{lstlisting}
\begin{lstlisting}
library(knitr)
\end{lstlisting}
\begin{lstlisting}
library(corrplot)
\end{lstlisting}
\begin{lstlisting}
library(Hmisc)
\end{lstlisting}
\begin{lstlisting}
# Step 2. Load Data-Set ########################################################
\end{lstlisting}
\begin{lstlisting}
#~~~~~~~~~~~~~~~~~~~~~~~~~~~~~~~~~~~~#
\end{lstlisting}
\begin{lstlisting}
#~~~ User input possibly required ~~~#
\end{lstlisting}
\begin{lstlisting}
#~~~~~~~~~~~~~~~~~~~~~~~~~~~~~~~~~~~~#
\end{lstlisting}
\begin{lstlisting}
# Please copy the path of where the current data-set is located and insert it here
\end{lstlisting}
Monthly\_EA <- read\_excel("EA\_data\_monthly.xlsx")

\begin{lstlisting}
# Step 3. Prepare the Data #####################################################
\end{lstlisting}
\begin{lstlisting}
# exclude all the data, that carry NA in Dec 1998 (Start of 2 Yr Swap)
\end{lstlisting}
Monthly\_EA\_numeric <- Monthly\_EA[, -1]  \# Exclude the first column

Monthly\_EA\_ts <- ts(Monthly\_EA\_numeric, \#Convert into a Time-Series

start = c(1986, 12), \# Start in December 1986

frequency = 12)      \# Monthly data

time\_index <- time(Monthly\_EA\_ts)  \# Get the time index

years <- floor(time\_index)

months <- round((time\_index - years) * 12 + 1)

index\_start <- which(years == 1999 \& months == 1) \#+ 2 \# Change Starting Date of Data-Set

index\_start

columns\_with\_na <- is.na(Monthly\_EA\_ts[index\_start, ])

Monthly\_EA\_cleaned\_ts <- Monthly\_EA\_ts[, !columns\_with\_na]

Monthly\_EA\_cleaned\_ts <- Monthly\_EA\_cleaned\_ts[time(Monthly\_EA\_cleaned\_ts)

>= time(Monthly\_EA\_ts)[index\_start], ]

data <- ts(Monthly\_EA\_cleaned\_ts, start = c(1999, 1), frequency = 12)

data

data <- window(data, end = c(2008,12))

\begin{lstlisting}
# Save data into individual series and transform them to Log-Levels (except interest-rates)
\end{lstlisting}
\begin{lstlisting}
# Taking logs allows us to interpret model output as elasticities / % changes
\end{lstlisting}
Y <- na.omit(data[,colnames(data) == "ind\_prod"]) \# Industrial Prod.

Price <- na.omit(data[,colnames(data) == "infl"])  \# Inflation\_ECB

fed <- na.omit(data[,colnames(data) == "fed\_funds\_r"]) \#Fed Funds

r\_s <-  na.omit(data[,colnames(data) == "EUR\_m\_refi"]) \# Refi Rate

eur\_stoxx <- log(na.omit(data[,colnames(data) == "euro\_stoxx"])) \# Euro-Stoxx Index: - Taking Logs

pmi <- na.omit(data[,colnames(data) == "m\_PMI\_US"]) \#PMI

r\_b\_2 <-  na.omit(data[,colnames(data) == "ICEIB2Y"]) \# 2Yr Swap

r\_b\_5 <-  na.omit(data[,colnames(data) == "ICEIB5Y"]) \# 5Yr Swap

r\_b\_10 <-  na.omit(data[,colnames(data) == "ICEIB10"]) \# 10Yr Swap

r\_b\_30 <-  na.omit(data[,colnames(data) == "ICEIB30"]) \# 30Yr Swap

eco\_sent <- na.omit(data[,colnames(data)=="eco\_sent"])

\begin{lstlisting}
# Define a Subset of Variables that are transformed, making sure the variables length matches.
\end{lstlisting}
\begin{lstlisting}
# ! The Ordering of Variables in the Data-Frame matters for Identification later on !
\end{lstlisting}
\begin{lstlisting}
# We will use exact Identification based a Cholesky Decomposition.
\end{lstlisting}
\begin{lstlisting}
# A1: Data order before the ECB Refi Rate is observed by the ECB when setting the Rate
\end{lstlisting}
\begin{lstlisting}
# A2: Data order before the ECB Refi Rate reacts with a time-lag to Monetary Policy
\end{lstlisting}
\begin{lstlisting}
# A3: Data order after the ECB Refi Rate reacts immediately to monetary policy
\end{lstlisting}
\begin{lstlisting}
# Assumptions are based on https://www.nber.org/system/files/working_papers/w6400/w6400.pdf
\end{lstlisting}
\begin{lstlisting}
# A discussion on these assumptions and alternative identification shemes are given by
\end{lstlisting}
\begin{lstlisting}
# https://www.sciencedirect.com/science/article/abs/pii/S1574004816000045
\end{lstlisting}
\begin{lstlisting}
# The Code may allow for identification via sign-restrictions, this is infeasible however
\end{lstlisting}
\begin{lstlisting}
# for models with large sets of variables.
\end{lstlisting}
dat <- cbind(pmi, Y, Price, fed, r\_s, eur\_stoxx,eco\_sent, r\_b\_10, r\_b\_2)

colnames(dat) <- c("Manufacturing\_PMI", "Industrial\_Prod", "Inflation","Fed\_Funds","ECB\_Refi", "EUR\_Stoxx","Economic\_Sentiment","SWAP\_Rate\_10Yr", "SWAP\_Rate\_2Yr")

dat <- ts(dat, start = c(1999,01), frequency = 12)

\begin{lstlisting}
# Correlation matrix and p-values
\end{lstlisting}
res2 <- rcorr(as.matrix(dat[,-c(8:9)]))

\begin{lstlisting}
# Insignificant correlations are crossed
\end{lstlisting}
\begin{lstlisting}
# In case R displays an error message, please ignore it
\end{lstlisting}
corrplot(res2\$r, type = "lower", order = "hclust",

p.mat = res2\$P, sig.level = 0.05, insig = "blank",

tl.cex = 0.8, tl.col = "black", \# Adjust text size and color

tl.labels = colnames(dat)) \# Use math expression labels

\begin{lstlisting}
####### Step 4: Building the Model  ############################################
\end{lstlisting}
\begin{lstlisting}
# Type of Prior, setting Hyperparameter Tuning up.
\end{lstlisting}
mn <- bv\_minnesota(lambda = bv\_lambda(mode = 0.2, sd = 0.4, min = 0.0001, max = 5),

alpha = bv\_alpha(mode = 2), var = 1e07)

soc <- bv\_soc(mode = 1, sd = 1, min = 1e-04, max = 50)

sur <- bv\_sur(mode = 1, sd = 1, min = 1e-04, max = 50)

priors <- bv\_priors(hyper = "auto", mn = mn, soc = soc, sur = sur)

\begin{lstlisting}
# Settings for the Methropolis-Hastings Algorithm
\end{lstlisting}
mh <- bv\_metropolis(scale\_hess = c(0.05, 0.0001, 0.0001),

adjust\_acc = TRUE, acc\_lower = 0.25, acc\_upper = 0.45)

\begin{lstlisting}
# Fitting the Model
\end{lstlisting}
run <- bvar(dat, \#Call in Dataset (no NA, inclduing only Series that should be modeled)

lags = 5, \# If you ran the optimization above, otherwise set 3 or 12.

n\_draw = 50000, n\_burn = 25000, n\_thin = 3, \# Standard Settings for the MCMC

priors = priors, mh = mh, verbose = TRUE)

\begin{lstlisting}
####### Step 5: Diagnostics   ##################################################
\end{lstlisting}
\begin{lstlisting}
# Summary of the Model
\end{lstlisting}
summary(run)

\begin{lstlisting}
# Assess convergence of the MCMC algorithm
\end{lstlisting}
\begin{lstlisting}
# The Trace Plots should concentrate around specific values and not "wander off"
\end{lstlisting}
plot(run)

run\_mcmc <- coda::as.mcmc(run) \#Convert the Output of Prior-Optimization into coda

resids <- ts(residuals(run), start = c(1998,12), frequency =12)

resids\_df <- as.data.frame(resids)

\begin{lstlisting}
# Plot the Resiudals
\end{lstlisting}
plot(resids)

\begin{lstlisting}
# MCMC Convergence Metrics
\end{lstlisting}
run\_mcmc <- as.mcmc(run)

autocorr.plot(run\_mcmc) \#Autocorrelation in Prior Hyperparameters

\begin{lstlisting}
# Strong Autocorrelation = Possibly weak convergence of MCMC chain
\end{lstlisting}
geweke.plot(run\_mcmc) \#Geweke 1999

\begin{lstlisting}
#Most z-scores should lie close to zero-line at least inside confidence bounds
\end{lstlisting}
crosscorr.plot(run\_mcmc)

\begin{lstlisting}
# High Correlation = Possible Issue with Collinearity
\end{lstlisting}
\begin{lstlisting}
# Effecitve Sample Size (ESS), accounting for Autocorrelation
\end{lstlisting}
effectiveSize(run\_mcmc)

\begin{lstlisting}
#ESS > 100: Good, sufficient for most inference.
\end{lstlisting}
\begin{lstlisting}
#ESS < 100: Concerning, suggests poor mixing or high autocorrelation.
\end{lstlisting}
\begin{lstlisting}
#ESS < 30: Serious issues; inference is unreliable.
\end{lstlisting}
\begin{lstlisting}
# In-Sample Predictions (one period foreward):
\end{lstlisting}
\begin{lstlisting}
# Give In-Sample Fitted Values and plot them to actually realized values.
\end{lstlisting}
fitted\_ts <- ts(fitted(run, type = "mean"),

start = c(2000,1), \#Actual Start + Number of Lags

frequency = 12)

actual\_ts <- ts(dat[-c(1:optim\_lag,309,310),], start = c(2000,1), frequency = 12)

\begin{lstlisting}
# Calculate RMSE:
\end{lstlisting}
as.data.frame(rmse(run))

\begin{lstlisting}
# Visual analysis of time series and fitted values
\end{lstlisting}
\begin{lstlisting}
# Convert fitted_ts and actual_ts into tidy data frames
\end{lstlisting}
fitted\_df <- as.data.frame(fitted\_ts)

actual\_df <- as.data.frame(actual\_ts)

\begin{lstlisting}
# Add a Time column (assumes identical time indices for all series)
\end{lstlisting}
fitted\_df\$Time <- time(fitted\_ts)

actual\_df\$Time <- time(fitted\_ts)

\begin{lstlisting}
# Reshape both data frames to long format and add a Type column
\end{lstlisting}
fitted\_long <- fitted\_df \%>\%

pivot\_longer(cols = -Time, names\_to = "Series", values\_to = "Value") \%>\%

mutate(Type = "Fitted")

actual\_long <- actual\_df \%>\%

pivot\_longer(cols = -Time, names\_to = "Series", values\_to = "Value") \%>\%

mutate(Type = "Actual")

\begin{lstlisting}
# Combine the two long data frames
\end{lstlisting}
combined\_long <- bind\_rows(fitted\_long, actual\_long)

\begin{lstlisting}
# Create the ggplot
\end{lstlisting}
p <- ggplot(combined\_long, aes(x = Time, y = Value, color = Type, group = Type)) +

geom\_line() +

facet\_wrap(\textasciitilde{} Series, ncol = 2, scales = "free\_y") +  \# Adjust y-axis scale for each facet

labs(title = "Fitted vs Actual Time Series",

x = "Time", y = "Value", color = "Type") +

theme\_minimal()

\begin{lstlisting}
# Convert the ggplot to an interactive plotly plot
\end{lstlisting}
interactive\_plot <- ggplotly(p)

\begin{lstlisting}
######## Step 6: Simulation / Impulse Response Analysis #######################
\end{lstlisting}
\begin{lstlisting}
# With standard Cholesky-Decomposition
\end{lstlisting}
opt\_irf <- bv\_irf(horizon = 12, identification = TRUE, fevd = TRUE)

print(opt\_irf)

irf(run) <- irf(run, opt\_irf, conf\_bands = c(0.05, 0.16))

plot(irf(run), area = TRUE)

\section{R Script Econometric AI model: Low yields}
\begin{lstlisting}
################################################################################
\end{lstlisting}
\begin{lstlisting}
################ Multiscenario IR forecasting utilizing AI #####################
\end{lstlisting}
\begin{lstlisting}
########## Supporting a structured view of interest rate forecasts #############
\end{lstlisting}
\begin{lstlisting}
############################## zeb business school #############################
\end{lstlisting}
\begin{lstlisting}
################################################################################
\end{lstlisting}
\begin{lstlisting}
############### Econometric AI model: low yields (03/2016-03/2022) ##############
\end{lstlisting}
\begin{lstlisting}
# In case the user needs to manually change the code to run smoothly, the following
\end{lstlisting}
\begin{lstlisting}
# message is displayed:
\end{lstlisting}
\begin{lstlisting}
#~~~~~~~~~~~~~~~~~~~~~~~~~~~~~~~~~~~~#
\end{lstlisting}
\begin{lstlisting}
#~~~ User input possibly required ~~~#
\end{lstlisting}
\begin{lstlisting}
#~~~~~~~~~~~~~~~~~~~~~~~~~~~~~~~~~~~~#
\end{lstlisting}
\begin{lstlisting}
# Step 1. Load/Install Dependencies ############################################
\end{lstlisting}
\begin{lstlisting}
#~~~~~~~~~~~~~~~~~~~~~~~~~~~~~~~~~~~~#
\end{lstlisting}
\begin{lstlisting}
#~~~ User input possibly required ~~~#
\end{lstlisting}
\begin{lstlisting}
#~~~~~~~~~~~~~~~~~~~~~~~~~~~~~~~~~~~~#
\end{lstlisting}
\begin{lstlisting}
# if one of the following packages is not yet installed, please "uncomment" and run the respective line of code
\end{lstlisting}
\begin{lstlisting}
#install.packages("dplyr")
\end{lstlisting}
\begin{lstlisting}
#install.packages("ggplot2")
\end{lstlisting}
\begin{lstlisting}
#install.packages("readxl")
\end{lstlisting}
\begin{lstlisting}
#install.packages("forecast")
\end{lstlisting}
\begin{lstlisting}
#install.packages("Metrics")
\end{lstlisting}
\begin{lstlisting}
#install.packages("tseries")
\end{lstlisting}
\begin{lstlisting}
#install.packages("BVAR")
\end{lstlisting}
\begin{lstlisting}
#install.packages("plotly")
\end{lstlisting}
\begin{lstlisting}
#install.packages("coda")
\end{lstlisting}
\begin{lstlisting}
#install.packages("tidyr")
\end{lstlisting}
\begin{lstlisting}
#install.packages("data.table")
\end{lstlisting}
\begin{lstlisting}
#install.packages("knitr")
\end{lstlisting}
\begin{lstlisting}
#install.packages("corrplot")
\end{lstlisting}
\begin{lstlisting}
#install.packages("Hmisc")
\end{lstlisting}
\begin{lstlisting}
library(dplyr)
\end{lstlisting}
\begin{lstlisting}
library(ggplot2)
\end{lstlisting}
\begin{lstlisting}
library(readxl)
\end{lstlisting}
\begin{lstlisting}
library(forecast)
\end{lstlisting}
\begin{lstlisting}
library(Metrics)
\end{lstlisting}
\begin{lstlisting}
library(tseries)
\end{lstlisting}
\begin{lstlisting}
library(BVAR)
\end{lstlisting}
\begin{lstlisting}
library(plotly)
\end{lstlisting}
\begin{lstlisting}
library(coda)
\end{lstlisting}
\begin{lstlisting}
library(tidyr)
\end{lstlisting}
\begin{lstlisting}
library(data.table)
\end{lstlisting}
\begin{lstlisting}
library(knitr)
\end{lstlisting}
\begin{lstlisting}
library(corrplot)
\end{lstlisting}
\begin{lstlisting}
library(Hmisc)
\end{lstlisting}
\begin{lstlisting}
# Step 2. Load Data-Set ########################################################
\end{lstlisting}
\begin{lstlisting}
#~~~~~~~~~~~~~~~~~~~~~~~~~~~~~~~~~~~~#
\end{lstlisting}
\begin{lstlisting}
#~~~ User input possibly required ~~~#
\end{lstlisting}
\begin{lstlisting}
#~~~~~~~~~~~~~~~~~~~~~~~~~~~~~~~~~~~~#
\end{lstlisting}
\begin{lstlisting}
# Please copy the path of where the current data-set is located and insert it here
\end{lstlisting}
Monthly\_EA <- read\_excel("EA\_data\_monthly.xlsx")

\begin{lstlisting}
# Step 3. Prepare the Data #####################################################
\end{lstlisting}
\begin{lstlisting}
# Exclude all the Data, that carries NA in Dez 1998 (Start of 2 Yr Swap)
\end{lstlisting}
Monthly\_EA\_numeric <- Monthly\_EA[, -1]  \# Exclude the first column

Monthly\_EA\_ts <- ts(Monthly\_EA\_numeric, \#Convert into a Time-Series

start = c(1986, 12), \# Start in December 1986

frequency = 12)      \# Monthly data

time\_index <- time(Monthly\_EA\_ts)  \# Get the time index

years <- floor(time\_index)

months <- round((time\_index - years) * 12 + 1)

index\_start <- which(years == 1999 \& months == 1) \#+ 2 \# Change Starting Date of Data-Set

index\_start

columns\_with\_na <- is.na(Monthly\_EA\_ts[index\_start, ])

Monthly\_EA\_cleaned\_ts <- Monthly\_EA\_ts[, !columns\_with\_na]

Monthly\_EA\_cleaned\_ts <- Monthly\_EA\_cleaned\_ts[time(Monthly\_EA\_cleaned\_ts)

>= time(Monthly\_EA\_ts)[index\_start], ]

data <- ts(Monthly\_EA\_cleaned\_ts, start = c(1999, 1), frequency = 12)

data

data <- window(data, start = c(2016,03), end = c(2022,03))

\begin{lstlisting}
# Save Data into individual Series and Transform them to Log-Levels (except Interest-Rates)
\end{lstlisting}
\begin{lstlisting}
# Taking Logs allows us to interpret Model output as Elasticities / % Changes
\end{lstlisting}
infl\_us <- na.omit(data[,colnames(data) == "infl\_US"]) \#Inflation

Price <- na.omit(data[,colnames(data) == "infl"])  \# Inflation\_ECB

fed <- na.omit(data[,colnames(data) == "fed\_funds\_r"]) \#Fed Funds

cdty <- log(na.omit(data[,colnames(data) == "Com\_Index"])) \#Commodity Price Index: - Taking Logs

eur\_stoxx <- log(na.omit(data[,colnames(data) == "euro\_stoxx"])) \# Euro-Stoxx Index: - Taking Logs

r\_b\_2 <-  na.omit(data[,colnames(data) == "ICEIB2Y"]) \# 2Yr Swap

r\_b\_5 <-  na.omit(data[,colnames(data) == "ICEIB5Y"]) \# 5Yr Swap

r\_b\_10 <-  na.omit(data[,colnames(data) == "ICEIB10"]) \# 10Yr Swap

r\_b\_30 <-  na.omit(data[,colnames(data) == "ICEIB30"]) \# 30Yr Swap

eco\_sent <- na.omit(data[,colnames(data)=="eco\_sent"])

M3 <- na.omit(log(data[, colnames(data)== "EA\_M3"]))

\begin{lstlisting}
# Take the first difference of the series
\end{lstlisting}
M3 <- diff(M3)

\begin{lstlisting}
# Plot the detrended series
\end{lstlisting}
\begin{lstlisting}
#plot(M3_detrended, type = "l", main = "Detrended M3 (First Differences)")
\end{lstlisting}
rep\_index <- na.omit(data[, colnames(data)== "REP\_Index"]) \# Transformed using min-max

\begin{lstlisting}
# Min and max of the original variable
\end{lstlisting}
min\_val <- min(rep\_index)

max\_val <- max(rep\_index)

\begin{lstlisting}
# Define the new range
\end{lstlisting}
new\_min <- 0

new\_max <- 100

\begin{lstlisting}
# Apply the transformation
\end{lstlisting}
rep\_index <- (rep\_index - min\_val) / (max\_val - min\_val) * (new\_max - new\_min) + new\_min

\begin{lstlisting}
# Define a Subset of Variables that are transformed, making sure the variables length matches.
\end{lstlisting}
\begin{lstlisting}
# ! The Ordering of Variables in the Data-Frame matters for Identification later on !
\end{lstlisting}
\begin{lstlisting}
# We will use exact Identification based a Cholesky Decomposition.
\end{lstlisting}
\begin{lstlisting}
# A1: Data ordered before the ECB Refi Rate is observed by the ECB when setting the Rate
\end{lstlisting}
\begin{lstlisting}
# A2: Data ordered before the ECB Refi Rate reacts with a time-lag to Monetary Policy
\end{lstlisting}
\begin{lstlisting}
# A3: Data ordered after the ECB Refi Rate reacts immediately to monetary policy
\end{lstlisting}
\begin{lstlisting}
# Assumptions are based on https://www.nber.org/system/files/working_papers/w6400/w6400.pdf
\end{lstlisting}
\begin{lstlisting}
# A discussion on these assumptions and alternative identification schemes are given by
\end{lstlisting}
\begin{lstlisting}
# https://www.sciencedirect.com/science/article/abs/pii/S1574004816000045
\end{lstlisting}
dat <- cbind(infl\_us,rep\_index, Price,M3, eur\_stoxx,cdty, r\_b\_10, r\_b\_2)

colnames(dat) <- c("Us\_Inflation","Rep\_index","Inflation","M3", "EUR\_Stoxx","Commodity","SWAP\_Rate\_10Yr", "SWAP\_Rate\_2Yr")

dat <- ts(dat, start = c(2016,03), frequency = 12)

\begin{lstlisting}
# Correlation matrix and p-values
\end{lstlisting}
res2 <- rcorr(as.matrix(dat[,-c(7:8)]))

\begin{lstlisting}
# Insignificant correlations are crossed
\end{lstlisting}
\begin{lstlisting}
# In case R displays an error message, please ignore it
\end{lstlisting}
corrplot(res2\$r, type = "lower", order = "hclust",

p.mat = res2\$P, sig.level = 0.05, insig = "blank",

tl.cex = 0.8, tl.col = "black", \# Adjust text size and color

tl.labels = colnames(dat)) \# Use math expression labels

\begin{lstlisting}
####### Step 4: Building the Model  ############################################
\end{lstlisting}
\begin{lstlisting}
# Type of Prior, setting Hyperparameter Tuning up.
\end{lstlisting}
mn <- bv\_minnesota(lambda = bv\_lambda(mode = 0.2, sd = 0.4, min = 0.0001, max = 5),

alpha = bv\_alpha(mode = 2), var = 1e07)

soc <- bv\_soc(mode = 1, sd = 1, min = 1e-04, max = 50)

sur <- bv\_sur(mode = 1, sd = 1, min = 1e-04, max = 50)

priors <- bv\_priors(hyper = "auto", mn = mn, soc = soc, sur = sur)

\begin{lstlisting}
# Settings for the Methropolis-Hastings Algorithm
\end{lstlisting}
mh <- bv\_metropolis(scale\_hess = c(0.05, 0.0001, 0.0001),

adjust\_acc = TRUE, acc\_lower = 0.25, acc\_upper = 0.45)

\begin{lstlisting}
# Fitting the Model
\end{lstlisting}
run <- bvar(dat[-1,], \#Call in Dataset (no NA, inclduing only Series that should be modeled)

lags = 5, \# If you ran the optimization above, otherwise set 3 or 12.

n\_draw = 50000, n\_burn = 25000, n\_thin = 3, \# Standard Settings for the MCMC

priors = priors, mh = mh, verbose = TRUE)

\begin{lstlisting}
####### Step 5: Diagnostics   ##################################################
\end{lstlisting}
\begin{lstlisting}
# Summary of the Model
\end{lstlisting}
summary(run)

\begin{lstlisting}
# Assess convergence of the MCMC algorithm
\end{lstlisting}
\begin{lstlisting}
# The Trace Plots should concentrate around specific values and not "wander off"
\end{lstlisting}
plot(run)

run\_mcmc <- coda::as.mcmc(run) \#Convert the Output of Prior-Optimization into coda

resids <- ts(residuals(run), start = c(1998,12), frequency =12)

resids\_df <- as.data.frame(resids)

\begin{lstlisting}
# Plot the Resiudals
\end{lstlisting}
plot(resids)

\begin{lstlisting}
# MCMC Convergence Metrics
\end{lstlisting}
run\_mcmc <- as.mcmc(run)

autocorr.plot(run\_mcmc) \#Autocorrelation in Prior Hyperparameters

\begin{lstlisting}
# Strong Autocorrelation = Possibly weak convergence of MCMC chain
\end{lstlisting}
geweke.plot(run\_mcmc) \#Geweke 1999

\begin{lstlisting}
#Most z-scores should lie close to zero-line atleast inside confidence bounds
\end{lstlisting}
crosscorr.plot(run\_mcmc)

\begin{lstlisting}
# High Correlation = Possible Issue with Colinearity
\end{lstlisting}
\begin{lstlisting}
# Effecitve Sample Size (ESS), accounting for Autocorrelation
\end{lstlisting}
effectiveSize(run\_mcmc)

\begin{lstlisting}
#ESS > 100: Good, sufficient for most inference.
\end{lstlisting}
\begin{lstlisting}
#ESS < 100: Concerning, suggests poor mixing or high autocorrelation.
\end{lstlisting}
\begin{lstlisting}
#ESS < 30: Serious issues; inference is unreliable.
\end{lstlisting}
\begin{lstlisting}
# In-Sample Predictions (one period foreward):
\end{lstlisting}
\begin{lstlisting}
# Give In-Sample Fitted Values and plot them to actually realized values.
\end{lstlisting}
fitted\_ts <- ts(fitted(run, type = "mean"),

start = c(2016,8), \#Actual Start + Number of Lags

frequency = 12)

actual\_ts <- ts(dat[-c(1:6),], start = c(2016,8), frequency = 12)

\begin{lstlisting}
# Calculate RMSE:
\end{lstlisting}
as.data.frame(rmse(run))

\begin{lstlisting}
# Example: Assuming fitted_ts and actual_ts are matrices or data frames
\end{lstlisting}
\begin{lstlisting}
# Convert fitted_ts and actual_ts into tidy data frames
\end{lstlisting}
fitted\_df <- as.data.frame(fitted\_ts)

actual\_df <- as.data.frame(actual\_ts)

\begin{lstlisting}
# Add a Time column (assumes identical time indices for all series)
\end{lstlisting}
fitted\_df\$Time <- time(fitted\_ts)

actual\_df\$Time <- time(fitted\_ts)

\begin{lstlisting}
# Reshape both data frames to long format and add a Type column
\end{lstlisting}
fitted\_long <- fitted\_df \%>\%

pivot\_longer(cols = -Time, names\_to = "Series", values\_to = "Value") \%>\%

mutate(Type = "Fitted")

actual\_long <- actual\_df \%>\%

pivot\_longer(cols = -Time, names\_to = "Series", values\_to = "Value") \%>\%

mutate(Type = "Actual")

\begin{lstlisting}
# Combine the two long data frames
\end{lstlisting}
combined\_long <- bind\_rows(fitted\_long, actual\_long)

\begin{lstlisting}
# Create the ggplot
\end{lstlisting}
p <- ggplot(combined\_long, aes(x = Time, y = Value, color = Type, group = Type)) +

geom\_line() +

facet\_wrap(\textasciitilde{} Series, ncol = 2, scales = "free\_y") +  \# Adjust y-axis scale for each facet

labs(title = "Fitted vs Actual Time Series",

x = "Time", y = "Value", color = "Type") +

theme\_minimal()

\begin{lstlisting}
# Convert the ggplot to an interactive plotly plot
\end{lstlisting}
interactive\_plot <- ggplotly(p)

\begin{lstlisting}
# Print the interactive plot
\end{lstlisting}
interactive\_plot

\begin{lstlisting}
######## Step 6: Simulation / Impulse Response Analysis #######################
\end{lstlisting}
\begin{lstlisting}
# With standard Cholesky-Decomposition
\end{lstlisting}
opt\_irf <- bv\_irf(horizon = 12, identification = TRUE, fevd = TRUE)

print(opt\_irf)

irf(run) <- irf(run, opt\_irf, conf\_bands = c(0.05, 0.16))

plot(irf(run), area = TRUE)

\section{R Script Econometric AI model: Increasing yields}
\begin{lstlisting}
################################################################################
\end{lstlisting}
\begin{lstlisting}
################ Multiscenario IR forecasting utilizing AI #####################
\end{lstlisting}
\begin{lstlisting}
########## Supporting a structured view of interest rate forecasts #############
\end{lstlisting}
\begin{lstlisting}
############################## zeb business school #############################
\end{lstlisting}
\begin{lstlisting}
################################################################################
\end{lstlisting}
\begin{lstlisting}
############ Econometric AI model: increasing yields (01/2021-today) ############
\end{lstlisting}
\begin{lstlisting}
# In case the user needs to manually change the code to run smoothly, the following
\end{lstlisting}
\begin{lstlisting}
# message is displayed:
\end{lstlisting}
\begin{lstlisting}
#~~~~~~~~~~~~~~~~~~~~~~~~~~~~~~~~~~~~#
\end{lstlisting}
\begin{lstlisting}
#~~~ User input possibly required ~~~#
\end{lstlisting}
\begin{lstlisting}
#~~~~~~~~~~~~~~~~~~~~~~~~~~~~~~~~~~~~#
\end{lstlisting}
\begin{lstlisting}
# Step 1. Load/Install Dependencies ############################################
\end{lstlisting}
\begin{lstlisting}
#~~~~~~~~~~~~~~~~~~~~~~~~~~~~~~~~~~~~#
\end{lstlisting}
\begin{lstlisting}
#~~~ User input possibly required ~~~#
\end{lstlisting}
\begin{lstlisting}
#~~~~~~~~~~~~~~~~~~~~~~~~~~~~~~~~~~~~#
\end{lstlisting}
\begin{lstlisting}
# if one of the following packages is not yet installed, please "uncomment" and run the respective line of code
\end{lstlisting}
\begin{lstlisting}
#install.packages("dplyr")
\end{lstlisting}
\begin{lstlisting}
#install.packages("ggplot2")
\end{lstlisting}
\begin{lstlisting}
#install.packages("readxl")
\end{lstlisting}
\begin{lstlisting}
#install.packages("forecast")
\end{lstlisting}
\begin{lstlisting}
#install.packages("Metrics")
\end{lstlisting}
\begin{lstlisting}
#install.packages("tseries")
\end{lstlisting}
\begin{lstlisting}
#install.packages("BVAR")
\end{lstlisting}
\begin{lstlisting}
#install.packages("plotly")
\end{lstlisting}
\begin{lstlisting}
#install.packages("coda")
\end{lstlisting}
\begin{lstlisting}
#install.packages("tidyr")
\end{lstlisting}
\begin{lstlisting}
#install.packages("data.table")
\end{lstlisting}
\begin{lstlisting}
#install.packages("knitr")
\end{lstlisting}
\begin{lstlisting}
#install.packages("corrplot")
\end{lstlisting}
\begin{lstlisting}
#install.packages("Hmisc")
\end{lstlisting}
\begin{lstlisting}
library(dplyr)
\end{lstlisting}
\begin{lstlisting}
library(ggplot2)
\end{lstlisting}
\begin{lstlisting}
library(readxl)
\end{lstlisting}
\begin{lstlisting}
library(forecast)
\end{lstlisting}
\begin{lstlisting}
library(Metrics)
\end{lstlisting}
\begin{lstlisting}
library(tseries)
\end{lstlisting}
\begin{lstlisting}
library(BVAR)
\end{lstlisting}
\begin{lstlisting}
library(plotly)
\end{lstlisting}
\begin{lstlisting}
library(coda)
\end{lstlisting}
\begin{lstlisting}
library(tidyr)
\end{lstlisting}
\begin{lstlisting}
library(data.table)
\end{lstlisting}
\begin{lstlisting}
library(knitr)
\end{lstlisting}
\begin{lstlisting}
library(corrplot)
\end{lstlisting}
\begin{lstlisting}
library(Hmisc)
\end{lstlisting}
\begin{lstlisting}
# Step 2. Load Data-Set ########################################################
\end{lstlisting}
\begin{lstlisting}
#~~~~~~~~~~~~~~~~~~~~~~~~~~~~~~~~~~~~#
\end{lstlisting}
\begin{lstlisting}
#~~~ User input possibly required ~~~#
\end{lstlisting}
\begin{lstlisting}
#~~~~~~~~~~~~~~~~~~~~~~~~~~~~~~~~~~~~#
\end{lstlisting}
\begin{lstlisting}
# Please copy the path of where the current data-set is located and insert it here
\end{lstlisting}
Monthly\_EA <- read\_excel("EA\_data\_monthly.xlsx")

\begin{lstlisting}
# Step 3. Prepare the Data #####################################################
\end{lstlisting}
\begin{lstlisting}
# Exclude all the Data, that carries NA in Dez 1998 (Start of 2 Yr Swap)
\end{lstlisting}
Monthly\_EA\_numeric <- Monthly\_EA[, -1]  \# Exclude the first column

Monthly\_EA\_ts <- ts(Monthly\_EA\_numeric, \#Convert into a Time-Series

start = c(1986, 12), \# Start in December 1986

frequency = 12)      \# Monthly data

time\_index <- time(Monthly\_EA\_ts)  \# Get the time index

years <- floor(time\_index)

months <- round((time\_index - years) * 12 + 1)

index\_start <- which(years == 1999 \& months == 1) \#+ 2 \# Change Starting Date of Data-Set

index\_start

columns\_with\_na <- is.na(Monthly\_EA\_ts[index\_start, ])

Monthly\_EA\_cleaned\_ts <- Monthly\_EA\_ts[, !columns\_with\_na]

Monthly\_EA\_cleaned\_ts <- Monthly\_EA\_cleaned\_ts[time(Monthly\_EA\_cleaned\_ts)

>= time(Monthly\_EA\_ts)[index\_start], ]

data <- ts(Monthly\_EA\_cleaned\_ts, start = c(1999, 1), frequency = 12)

data

data <- window(data, start = c(2021,01))

\begin{lstlisting}
# Save Data into individual Series and Transform them to Log-Levels (except Interest-Rates)
\end{lstlisting}
\begin{lstlisting}
# Taking Logs allows us to interpret Model output as Elasticities / % Changes
\end{lstlisting}
infl\_us <- na.omit(data[,colnames(data) == "infl\_US"]) \#Inflation

\begin{lstlisting}
#trade_balance <- log(na.omit(data[,colnames(data) == "Trade Balance"]))
\end{lstlisting}
Y <- na.omit(data[,colnames(data) == "unempl\_r"]) \# Unemployment Rate

Price <- na.omit(data[,colnames(data) == "infl"])  \# Inflation\_ECB

fed <- na.omit(data[,colnames(data) == "fed\_funds\_r"]) \#Fed Funds

r\_s <-  na.omit(data[,colnames(data) == "EUR\_m\_refi"]) \# Refi Rate

\begin{lstlisting}
#cdty <- log(na.omit(data[,colnames(data) == "Com_Index"])) #Commodity Price Index: - Taking Logs
\end{lstlisting}
\begin{lstlisting}
#eur_stoxx <- log(na.omit(data[,colnames(data) == "euro_stoxx"])) # Euro-Stoxx Index: - Taking Logs
\end{lstlisting}
pmi <- na.omit(data[,colnames(data) == "m\_PMI\_US"]) \#PMI

zew\_exp2 <- na.omit(data[,colnames(data)=="ZEW\_exp2"])

r\_b\_2 <-  na.omit(data[,colnames(data) == "ICEIB2Y"]) \# 2Yr Swap

r\_b\_5 <-  na.omit(data[,colnames(data) == "ICEIB5Y"]) \# 5Yr Swap

r\_b\_10 <-  na.omit(data[,colnames(data) == "ICEIB10"]) \# 10Yr Swap

r\_b\_30 <-  na.omit(data[,colnames(data) == "ICEIB30"]) \# 30Yr Swap

\begin{lstlisting}
# Define a Subset of Variables that are transformed, making sure the variables length matches.
\end{lstlisting}
\begin{lstlisting}
# ! The Ordering of Variables in the Data-Frame matters for Identification later on !
\end{lstlisting}
\begin{lstlisting}
# We will use exact Identification based a Cholesky Decomposition.
\end{lstlisting}
\begin{lstlisting}
# A1: Data ordered before the ECB Refi Rate is observed by the ECB when setting the Rate
\end{lstlisting}
\begin{lstlisting}
# A2: Data ordered before the ECB Refi Rate reacts with a time-lag to Monetary Policy
\end{lstlisting}
\begin{lstlisting}
# A3: Data ordered after the ECB Refi Rate reacts immediately to monetary policy
\end{lstlisting}
\begin{lstlisting}
# Assumptions are based on https://www.nber.org/system/files/working_papers/w6400/w6400.pdf
\end{lstlisting}
\begin{lstlisting}
# A discussion on these assumptions and alternative identification schemes are given by
\end{lstlisting}
\begin{lstlisting}
# https://www.sciencedirect.com/science/article/abs/pii/S1574004816000045
\end{lstlisting}
dat <- cbind(pmi, infl\_us, Price, r\_s,zew\_exp2, r\_b\_10, r\_b\_2)

colnames(dat) <- c("PMI","Inflation\_US","Inflation\_EA","ECB\_Refi",

"ZEW-Expect.","SWAP\_Rate\_10Yr", "SWAP\_Rate\_2Yr")

dat <- ts(dat, start = c(2021,01), frequency = 12)

\begin{lstlisting}
### Draw a Correlogramm to check for colinearity
\end{lstlisting}
\begin{lstlisting}
# Correlation matrix and p-values
\end{lstlisting}
res2 <- rcorr(as.matrix(dat[,-c(10,9,8,7,6)]))

\begin{lstlisting}
# Insignificant correlations are crossed
\end{lstlisting}
\begin{lstlisting}
# In case R displays an error message, please ignore it
\end{lstlisting}
corrplot(res2\$r, type = "lower", order = "hclust",

p.mat = res2\$P, sig.level = 0.05, insig = "blank",

tl.cex = 0.8, tl.col = "black", \# Adjust text size and color

tl.labels = colnames(dat)) \# Use math expression labels

\begin{lstlisting}
####### Step 4: Building the Model  ############################################
\end{lstlisting}
\begin{lstlisting}
# Type of Prior, setting Hyperparameter Tuning up.
\end{lstlisting}
mn <- bv\_minnesota(lambda = bv\_lambda(mode = 0.2, sd = 0.4, min = 0.0001, max = 5),

alpha = bv\_alpha(mode = 2), var = 1e07)

soc <- bv\_soc(mode = 1, sd = 1, min = 1e-04, max = 50)

sur <- bv\_sur(mode = 1, sd = 1, min = 1e-04, max = 50)

priors <- bv\_priors(hyper = "auto", mn = mn, soc = soc, sur = sur)

\begin{lstlisting}
# Settings for the Methropolis-Hastings Algorithm
\end{lstlisting}
mh <- bv\_metropolis(scale\_hess = c(0.05, 0.0001, 0.0001),

adjust\_acc = TRUE, acc\_lower = 0.25, acc\_upper = 0.45)

\begin{lstlisting}
# Fitting the Model
\end{lstlisting}
run <- bvar(dat, \#Call in Dataset (no NA, inclduing only Series that should be modeled)

lags = 5, \# If you ran the optimization above, otherwise set 3 or 12.

n\_draw = 50000, n\_burn = 25000, n\_thin = 3, \# Standard Settings for the MCMC

priors = priors, mh = mh, verbose = TRUE)

\begin{lstlisting}
####### Step 5: Diagnostics   ##################################################
\end{lstlisting}
\begin{lstlisting}
# Summary of the Model
\end{lstlisting}
summary(run)

\begin{lstlisting}
# Assess convergence of the MCMC algorithm
\end{lstlisting}
\begin{lstlisting}
# The Trace Plots should concentrate around specific values and not "wander off"
\end{lstlisting}
plot(run)

run\_mcmc <- coda::as.mcmc(run) \#Convert the Output of Prior-Optimization into coda

resids <- ts(residuals(run), start = c(1998,12), frequency =12)

resids\_df <- as.data.frame(resids)

\begin{lstlisting}
# Plot the Resiudals
\end{lstlisting}
plot(resids)

\begin{lstlisting}
# MCMC Convergence Metrics
\end{lstlisting}
run\_mcmc <- as.mcmc(run)

autocorr.plot(run\_mcmc) \#Autocorrelation in Prior Hyperparameters

\begin{lstlisting}
# Strong Autocorrelation = Possibly weak convergence of MCMC chain
\end{lstlisting}
geweke.plot(run\_mcmc) \#Geweke 1999

\begin{lstlisting}
#Most z-scores should lie close to zero-line atleast inside confidence bounds
\end{lstlisting}
crosscorr.plot(run\_mcmc)

\begin{lstlisting}
# High Correlation = Possible Issue with Colinearity
\end{lstlisting}
\begin{lstlisting}
# Effecitve Sample Size (ESS), accounting for Autocorrelation
\end{lstlisting}
effectiveSize(run\_mcmc)

\begin{lstlisting}
#ESS > 100: Good, sufficient for most inference.
\end{lstlisting}
\begin{lstlisting}
#ESS < 100: Concerning, suggests poor mixing or high autocorrelation.
\end{lstlisting}
\begin{lstlisting}
#ESS < 30: Serious issues; inference is unreliable.
\end{lstlisting}
\begin{lstlisting}
# In-Sample Predictions (one period foreward):
\end{lstlisting}
\begin{lstlisting}
# Give In-Sample Fitted Values and plot them to actually realized values.
\end{lstlisting}
fitted\_ts <- ts(fitted(run, type = "mean"),

start = c(2021,6), \#Actual Start + Number of Lags

frequency = 12)

actual\_ts <- ts(dat[-c(1:5),], start = c(2021,6), frequency = 12)

\begin{lstlisting}
# Calculate RMSE:
\end{lstlisting}
as.data.frame(rmse(run))

\begin{lstlisting}
# Example: Assuming fitted_ts and actual_ts are matrices or data frames
\end{lstlisting}
\begin{lstlisting}
# Convert fitted_ts and actual_ts into tidy data frames
\end{lstlisting}
fitted\_df <- as.data.frame(fitted\_ts)

actual\_df <- as.data.frame(actual\_ts)

\begin{lstlisting}
# Add a Time column (assumes identical time indices for all series)
\end{lstlisting}
fitted\_df\$Time <- time(fitted\_ts)

actual\_df\$Time <- time(fitted\_ts)

\begin{lstlisting}
# Reshape both data frames to long format and add a Type column
\end{lstlisting}
fitted\_long <- fitted\_df \%>\%

pivot\_longer(cols = -Time, names\_to = "Series", values\_to = "Value") \%>\%

mutate(Type = "Fitted")

actual\_long <- actual\_df \%>\%

pivot\_longer(cols = -Time, names\_to = "Series", values\_to = "Value") \%>\%

mutate(Type = "Actual")

\begin{lstlisting}
# Combine the two long data frames
\end{lstlisting}
combined\_long <- bind\_rows(fitted\_long, actual\_long)

\begin{lstlisting}
# Create the ggplot
\end{lstlisting}
p <- ggplot(combined\_long, aes(x = Time, y = Value, color = Type, group = Type)) +

geom\_line() +

facet\_wrap(\textasciitilde{} Series, ncol = 2, scales = "free\_y") +  \# Adjust y-axis scale for each facet

labs(title = "Fitted vs Actual Time Series",

x = "Time", y = "Value", color = "Type") +

theme\_minimal()

\begin{lstlisting}
# Convert the ggplot to an interactive plotly plot
\end{lstlisting}
interactive\_plot <- ggplotly(p)

\begin{lstlisting}
# Print the interactive plot
\end{lstlisting}
interactive\_plot

\begin{lstlisting}
######## Step 6: Simulation / Impulse Response Analysis #######################
\end{lstlisting}
\begin{lstlisting}
# With standard Cholesky-Decomposition
\end{lstlisting}
opt\_irf <- bv\_irf(horizon = 12, identification = TRUE, fevd = TRUE)

print(opt\_irf)

irf(run) <- irf(run, opt\_irf, conf\_bands = c(0.05, 0.16))

plot(irf(run), area = TRUE)

\section{R Script Econometric AI model: Decreasing yields}
\begin{lstlisting}
################################################################################
\end{lstlisting}
\begin{lstlisting}
################ Multiscenario IR forecasting utilizing AI #####################
\end{lstlisting}
\begin{lstlisting}
########## Supporting a structured view of interest rate forecasts #############
\end{lstlisting}
\begin{lstlisting}
############################## zeb business school #############################
\end{lstlisting}
\begin{lstlisting}
################################################################################
\end{lstlisting}
\begin{lstlisting}
########### Econometric AI model: decreasing yields (07/2008-03/2016) ##########
\end{lstlisting}
\begin{lstlisting}
# In case the user needs to manually change the code to run smoothly, the following
\end{lstlisting}
\begin{lstlisting}
# message is displayed:
\end{lstlisting}
\begin{lstlisting}
#~~~~~~~~~~~~~~~~~~~~~~~~~~~~~~~~~~~~#
\end{lstlisting}
\begin{lstlisting}
#~~~ User input possibly required ~~~#
\end{lstlisting}
\begin{lstlisting}
#~~~~~~~~~~~~~~~~~~~~~~~~~~~~~~~~~~~~#
\end{lstlisting}
\begin{lstlisting}
# Step 1. Load/Install Dependencies ############################################
\end{lstlisting}
\begin{lstlisting}
#~~~~~~~~~~~~~~~~~~~~~~~~~~~~~~~~~~~~#
\end{lstlisting}
\begin{lstlisting}
#~~~ User input possibly required ~~~#
\end{lstlisting}
\begin{lstlisting}
#~~~~~~~~~~~~~~~~~~~~~~~~~~~~~~~~~~~~#
\end{lstlisting}
\begin{lstlisting}
# if one of the following packages is not yet installed, please "uncomment" and run the respective line of code
\end{lstlisting}
\begin{lstlisting}
#install.packages("dplyr")
\end{lstlisting}
\begin{lstlisting}
#install.packages("ggplot2")
\end{lstlisting}
\begin{lstlisting}
#install.packages("readxl")
\end{lstlisting}
\begin{lstlisting}
#install.packages("forecast")
\end{lstlisting}
\begin{lstlisting}
#install.packages("Metrics")
\end{lstlisting}
\begin{lstlisting}
#install.packages("tseries")
\end{lstlisting}
\begin{lstlisting}
#install.packages("BVAR")
\end{lstlisting}
\begin{lstlisting}
#install.packages("plotly")
\end{lstlisting}
\begin{lstlisting}
#install.packages("coda")
\end{lstlisting}
\begin{lstlisting}
#install.packages("tidyr")
\end{lstlisting}
\begin{lstlisting}
#install.packages("data.table")
\end{lstlisting}
\begin{lstlisting}
#install.packages("knitr")
\end{lstlisting}
\begin{lstlisting}
#install.packages("corrplot")
\end{lstlisting}
\begin{lstlisting}
#install.packages("Hmisc")
\end{lstlisting}
\begin{lstlisting}
library(dplyr)
\end{lstlisting}
\begin{lstlisting}
library(ggplot2)
\end{lstlisting}
\begin{lstlisting}
library(readxl)
\end{lstlisting}
\begin{lstlisting}
library(forecast)
\end{lstlisting}
\begin{lstlisting}
library(Metrics)
\end{lstlisting}
\begin{lstlisting}
library(tseries)
\end{lstlisting}
\begin{lstlisting}
library(BVAR)
\end{lstlisting}
\begin{lstlisting}
library(plotly)
\end{lstlisting}
\begin{lstlisting}
library(coda)
\end{lstlisting}
\begin{lstlisting}
library(tidyr)
\end{lstlisting}
\begin{lstlisting}
library(data.table)
\end{lstlisting}
\begin{lstlisting}
library(knitr)
\end{lstlisting}
\begin{lstlisting}
library(corrplot)
\end{lstlisting}
\begin{lstlisting}
library(Hmisc)
\end{lstlisting}
\begin{lstlisting}
# Step 2. Load Data-Set ########################################################
\end{lstlisting}
\begin{lstlisting}
#~~~~~~~~~~~~~~~~~~~~~~~~~~~~~~~~~~~~#
\end{lstlisting}
\begin{lstlisting}
#~~~ User input possibly required ~~~#
\end{lstlisting}
\begin{lstlisting}
#~~~~~~~~~~~~~~~~~~~~~~~~~~~~~~~~~~~~#
\end{lstlisting}
\begin{lstlisting}
# Please copy the path of where the current data-set is located and insert it here
\end{lstlisting}
Monthly\_EA <- read\_excel("EA\_data\_monthly.xlsx")

\begin{lstlisting}
# Step 3. Prepare the Data #####################################################
\end{lstlisting}
\begin{lstlisting}
# exclude all the data, that carry NA in Dec 1998 (Start of 2 Yr Swap)
\end{lstlisting}
\begin{lstlisting}
# Step 3. Prepare the Data #####################################################
\end{lstlisting}
\begin{lstlisting}
# We will exclude all the Data, that carries NA in Dez 1998 (Start of 2 Yr Swap)
\end{lstlisting}
Monthly\_EA\_numeric <- Monthly\_EA[, -1]  \# Exclude the first column

Monthly\_EA\_ts <- ts(Monthly\_EA\_numeric, \#Convert into a Time-Series

start = c(1986, 12), \# Start in December 1986

frequency = 12)      \# Monthly data

time\_index <- time(Monthly\_EA\_ts)  \# Get the time index

years <- floor(time\_index)

months <- round((time\_index - years) * 12 + 1)

index\_start <- which(years == 1999 \& months == 1) \#+ 2 \# Change Starting Date of Data-Set

index\_start

columns\_with\_na <- is.na(Monthly\_EA\_ts[index\_start, ])

Monthly\_EA\_cleaned\_ts <- Monthly\_EA\_ts[, !columns\_with\_na]

Monthly\_EA\_cleaned\_ts <- Monthly\_EA\_cleaned\_ts[time(Monthly\_EA\_cleaned\_ts)

>= time(Monthly\_EA\_ts)[index\_start], ]

data <- ts(Monthly\_EA\_cleaned\_ts, start = c(1999, 1), frequency = 12)

data

data <- window(data, start = c(2008,07), end = c(2016,03))

\begin{lstlisting}
# Save Data into individual Series and Transform them to Log-Levels (except Interest-Rates)
\end{lstlisting}
\begin{lstlisting}
# Taking Logs allows us to interpret Model output as Elasticities / % Changes
\end{lstlisting}
Price <- na.omit(data[,colnames(data) == "Core\_Inflation\_ECB"])  \# Inflation\_ECB

fed <- na.omit(data[,colnames(data) == "fed\_funds\_r"]) \#Fed Funds

r\_s <-  na.omit(data[,colnames(data) == "EUR\_m\_refi"]) \# Refi Rate

cdty <- log(na.omit(data[,colnames(data) == "Com\_Index"])) \#Commodity Price Index: - Taking Logs

pmi <- na.omit(data[,colnames(data) == "m\_PMI\_US"]) \#PMI

r\_b\_2 <-  na.omit(data[,colnames(data) == "ICEIB2Y"]) \# 2Yr Swap

r\_b\_5 <-  na.omit(data[,colnames(data) == "ICEIB5Y"]) \# 5Yr Swap

r\_b\_10 <-  na.omit(data[,colnames(data) == "ICEIB10"]) \# 10Yr Swap

r\_b\_30 <-  na.omit(data[,colnames(data) == "ICEIB30"]) \# 30Yr Swap

\begin{lstlisting}
# Take the first difference of the series
\end{lstlisting}
FX <- na.omit(log(data[, colnames(data)== "EUR\_USD"]))

\begin{lstlisting}
# Plot the detrended series
\end{lstlisting}
rep\_index <- na.omit(data[, colnames(data)== "REP\_Index"]) \# Transformed using min-max

\begin{lstlisting}
# Min and max of the original variable
\end{lstlisting}
min\_val <- min(rep\_index)

max\_val <- max(rep\_index)

\begin{lstlisting}
# Define the new range
\end{lstlisting}
new\_min <- 0

new\_max <- 100

\begin{lstlisting}
# Apply the transformation
\end{lstlisting}
rep\_index <- (rep\_index - min\_val) / (max\_val - min\_val) * (new\_max - new\_min) + new\_min

\begin{lstlisting}
# Define a Subset of Variables that are transformed, making sure the variables length matches.
\end{lstlisting}
\begin{lstlisting}
# ! The Ordering of Variables in the Data-Frame matters for Identification later on !
\end{lstlisting}
\begin{lstlisting}
# We will use exact Identification based a Cholesky Decomposition.
\end{lstlisting}
\begin{lstlisting}
# A1: Data ordered before the ECB Refi Rate is observed by the ECB when setting the Rate
\end{lstlisting}
\begin{lstlisting}
# A2: Data ordere before the ECB Refi Rate reacts with a time-lag to Monetary Policy
\end{lstlisting}
\begin{lstlisting}
# A3: Data order after the ECB Refi Rate reacts immediately to monetary policy
\end{lstlisting}
\begin{lstlisting}
# Assumptions are based on https://www.nber.org/system/files/working_papers/w6400/w6400.pdf
\end{lstlisting}
\begin{lstlisting}
# A discussion on these assumptions and alternative identification shemes are given by
\end{lstlisting}
\begin{lstlisting}
# https://www.sciencedirect.com/science/article/abs/pii/S1574004816000045
\end{lstlisting}
\begin{lstlisting}
# The Code may allow for identification via sign-restrictions, this is infeasible however
\end{lstlisting}
\begin{lstlisting}
# for models with large sets of variables.
\end{lstlisting}
dat <- cbind(pmi, rep\_index, Price,r\_s, FX,cdty, r\_b\_10, r\_b\_2)

colnames(dat) <- c("PMI", "Rep\_index","Core\_Inflation\_EUR","ECB\_Refi", "FX","Commodity","SWAP\_Rate\_10Yr", "SWAP\_Rate\_2Yr")

dat <- ts(dat, start = c(2008,07), frequency = 12)

\begin{lstlisting}
# Correlation matrix and p-values
\end{lstlisting}
res2 <- rcorr(as.matrix(dat[,-c(7:8)]))

\begin{lstlisting}
# Insignificant correlations are crossed
\end{lstlisting}
\begin{lstlisting}
# In case R displays an error message, please ignore it
\end{lstlisting}
corrplot(res2\$r, type = "lower", order = "hclust",

p.mat = res2\$P, sig.level = 0.05, insig = "blank",

tl.cex = 0.8, tl.col = "black", \# Adjust text size and color

tl.labels = colnames(dat)) \# Use math expression labels

\begin{lstlisting}
####### Step 4: Building the Model  ############################################
\end{lstlisting}
\begin{lstlisting}
# Type of Prior, setting Hyperparameter Tuning up.
\end{lstlisting}
mn <- bv\_minnesota(lambda = bv\_lambda(mode = 0.2, sd = 0.4, min = 0.0001, max = 5),

alpha = bv\_alpha(mode = 2), var = 1e07)

soc <- bv\_soc(mode = 1, sd = 1, min = 1e-04, max = 50)

sur <- bv\_sur(mode = 1, sd = 1, min = 1e-04, max = 50)

priors <- bv\_priors(hyper = "auto", mn = mn, soc = soc, sur = sur)

\begin{lstlisting}
# Settings for the Methropolis-Hastings Algorithm
\end{lstlisting}
mh <- bv\_metropolis(scale\_hess = c(0.05, 0.0001, 0.0001),

adjust\_acc = TRUE, acc\_lower = 0.25, acc\_upper = 0.45)

\begin{lstlisting}
# Fitting the Model
\end{lstlisting}
run <- bvar(dat[,], \#Call in Dataset (no NA, inclduing only Series that should be modeled)

lags = 5, \# If you ran the optimization above, otherwise set 3 or 12.

n\_draw = 50000, n\_burn = 25000, n\_thin = 3, \# Standard Settings for the MCMC

priors = priors, mh = mh, verbose = TRUE)

\begin{lstlisting}
####### Step 5: Diagnostics   ##################################################
\end{lstlisting}
\begin{lstlisting}
# Summary of the Model
\end{lstlisting}
summary(run)

\begin{lstlisting}
# Assess convergence of the MCMC algorithm
\end{lstlisting}
\begin{lstlisting}
# The Trace Plots should concentrate around specific values and not "wander off"
\end{lstlisting}
plot(run)

run\_mcmc <- coda::as.mcmc(run) \#Convert the Output of Prior-Optimization into coda

resids <- ts(residuals(run), start = c(1998,12), frequency =12)

resids\_df <- as.data.frame(resids)

\begin{lstlisting}
# Plot the Resiudals
\end{lstlisting}
plot(resids)

\begin{lstlisting}
# MCMC Convergence Metrics
\end{lstlisting}
run\_mcmc <- as.mcmc(run)

autocorr.plot(run\_mcmc) \#Autocorrelation in Prior Hyperparameters

\begin{lstlisting}
# Strong Autocorrelation = Possibly weak convergence of MCMC chain
\end{lstlisting}
geweke.plot(run\_mcmc) \#Geweke 1999

\begin{lstlisting}
#Most z-scores should lie close to zero-line atleast inside confidence bounds
\end{lstlisting}
crosscorr.plot(run\_mcmc)

\begin{lstlisting}
# High Correlation = Possible Issue with Colinearity
\end{lstlisting}
\begin{lstlisting}
# Effecitve Sample Size (ESS), accounting for Autocorrelation
\end{lstlisting}
effectiveSize(run\_mcmc)

\begin{lstlisting}
#ESS > 100: Good, sufficient for most inference.
\end{lstlisting}
\begin{lstlisting}
#ESS < 100: Concerning, suggests poor mixing or high autocorrelation.
\end{lstlisting}
\begin{lstlisting}
#ESS < 30: Serious issues; inference is unreliable.
\end{lstlisting}
\begin{lstlisting}
# In-Sample Predictions (one period foreward):
\end{lstlisting}
\begin{lstlisting}
# Give In-Sample Fitted Values and plot them to actually realized values.
\end{lstlisting}
fitted\_ts <- ts(fitted(run, type = "mean"),

start = c(2000,1), \#Actual Start + Number of Lags

frequency = 12)

actual\_ts <- ts(dat[-c(1:5,309,310),], start = c(2000,1), frequency = 12)

\begin{lstlisting}
# Calculate RMSE:
\end{lstlisting}
as.data.frame(rmse(run))

\begin{lstlisting}
# Example: Assuming fitted_ts and actual_ts are matrices or data frames
\end{lstlisting}
\begin{lstlisting}
# Convert fitted_ts and actual_ts into tidy data frames
\end{lstlisting}
fitted\_df <- as.data.frame(fitted\_ts)

actual\_df <- as.data.frame(actual\_ts)

\begin{lstlisting}
# Add a Time column (assumes identical time indices for all series)
\end{lstlisting}
fitted\_df\$Time <- time(fitted\_ts)

actual\_df\$Time <- time(fitted\_ts)

\begin{lstlisting}
# Reshape both data frames to long format and add a Type column
\end{lstlisting}
fitted\_long <- fitted\_df \%>\%

pivot\_longer(cols = -Time, names\_to = "Series", values\_to = "Value") \%>\%

mutate(Type = "Fitted")

actual\_long <- actual\_df \%>\%

pivot\_longer(cols = -Time, names\_to = "Series", values\_to = "Value") \%>\%

mutate(Type = "Actual")

\begin{lstlisting}
# Combine the two long data frames
\end{lstlisting}
combined\_long <- bind\_rows(fitted\_long, actual\_long)

\begin{lstlisting}
# Create the ggplot
\end{lstlisting}
p <- ggplot(combined\_long, aes(x = Time, y = Value, color = Type, group = Type)) +

geom\_line() +

facet\_wrap(\textasciitilde{} Series, ncol = 2, scales = "free\_y") +  \# Adjust y-axis scale for each facet

labs(title = "Fitted vs Actual Time Series",

x = "Time", y = "Value", color = "Type") +

theme\_minimal()

\begin{lstlisting}
# Convert the ggplot to an interactive plotly plot
\end{lstlisting}
interactive\_plot <- ggplotly(p)

\begin{lstlisting}
# Print the interactive plot
\end{lstlisting}
interactive\_plot

\begin{lstlisting}
######## Step 6: Simulation / Impulse Response Analysis #######################
\end{lstlisting}
\begin{lstlisting}
# With standard Cholesky-Decomposition
\end{lstlisting}
opt\_irf <- bv\_irf(horizon = 12, identification = TRUE, fevd = TRUE)

print(opt\_irf)

irf(run) <- irf(run, opt\_irf, conf\_bands = c(0.05, 0.16))

plot(irf(run), area = TRUE)

\section{R Script Econometric AI model: Final BVAR entire period}
\begin{lstlisting}
################################################################################
\end{lstlisting}
\begin{lstlisting}
################ Multiscenario IR forecasting utilizing AI #####################
\end{lstlisting}
\begin{lstlisting}
########## Supporting a structured view of interest rate forecasts #############
\end{lstlisting}
\begin{lstlisting}
############################## zeb business school #############################
\end{lstlisting}
\begin{lstlisting}
################################################################################
\end{lstlisting}
\begin{lstlisting}
############## Econometric AI model: entire period (01/1999-11/2024) #############
\end{lstlisting}
\begin{lstlisting}
# In case the user needs to manually change the code to run smoothly, the following
\end{lstlisting}
\begin{lstlisting}
# message is displayed:
\end{lstlisting}
\begin{lstlisting}
#~~~~~~~~~~~~~~~~~~~~~~~~~~~~~~~~~~~~#
\end{lstlisting}
\begin{lstlisting}
#~~~ User input possibly required ~~~#
\end{lstlisting}
\begin{lstlisting}
#~~~~~~~~~~~~~~~~~~~~~~~~~~~~~~~~~~~~#
\end{lstlisting}
\begin{lstlisting}
# Step 1. Load/Install Dependencies ############################################
\end{lstlisting}
\begin{lstlisting}
#~~~~~~~~~~~~~~~~~~~~~~~~~~~~~~~~~~~~#
\end{lstlisting}
\begin{lstlisting}
#~~~ User input possibly required ~~~#
\end{lstlisting}
\begin{lstlisting}
#~~~~~~~~~~~~~~~~~~~~~~~~~~~~~~~~~~~~#
\end{lstlisting}
\begin{lstlisting}
# if one of the following packages is not yet installed, please "uncomment" and run the respective line of code
\end{lstlisting}
\begin{lstlisting}
#install.packages("dplyr")
\end{lstlisting}
\begin{lstlisting}
#install.packages("ggplot2")
\end{lstlisting}
\begin{lstlisting}
#install.packages("readxl")
\end{lstlisting}
\begin{lstlisting}
#install.packages("forecast")
\end{lstlisting}
\begin{lstlisting}
#install.packages("Metrics")
\end{lstlisting}
\begin{lstlisting}
#install.packages("tseries")
\end{lstlisting}
\begin{lstlisting}
#install.packages("BVAR")
\end{lstlisting}
\begin{lstlisting}
#install.packages("plotly")
\end{lstlisting}
\begin{lstlisting}
#install.packages("coda")
\end{lstlisting}
\begin{lstlisting}
#install.packages("tidyr")
\end{lstlisting}
\begin{lstlisting}
#install.packages("data.table")
\end{lstlisting}
\begin{lstlisting}
#install.packages(lmtest)
\end{lstlisting}
\begin{lstlisting}
#install.packages(knitr)
\end{lstlisting}
\begin{lstlisting}
#install.packages(corrplot)
\end{lstlisting}
\begin{lstlisting}
#install.packages(Hmisc)
\end{lstlisting}
\begin{lstlisting}
library(dplyr)
\end{lstlisting}
\begin{lstlisting}
library(ggplot2)
\end{lstlisting}
\begin{lstlisting}
library(readxl)
\end{lstlisting}
\begin{lstlisting}
library(forecast)
\end{lstlisting}
\begin{lstlisting}
library(Metrics)
\end{lstlisting}
\begin{lstlisting}
library(tseries)
\end{lstlisting}
\begin{lstlisting}
library(BVAR)
\end{lstlisting}
\begin{lstlisting}
library(plotly)
\end{lstlisting}
\begin{lstlisting}
library(coda)
\end{lstlisting}
\begin{lstlisting}
library(tidyr)
\end{lstlisting}
\begin{lstlisting}
library(data.table)
\end{lstlisting}
\begin{lstlisting}
library(lmtest)
\end{lstlisting}
\begin{lstlisting}
library(knitr)
\end{lstlisting}
\begin{lstlisting}
library(corrplot)
\end{lstlisting}
\begin{lstlisting}
library(Hmisc)
\end{lstlisting}
\begin{lstlisting}
# Step 2. Load Data-Set ########################################################
\end{lstlisting}
\begin{lstlisting}
#~~~~~~~~~~~~~~~~~~~~~~~~~~~~~~~~~~~~#
\end{lstlisting}
\begin{lstlisting}
#~~~ User input possibly required ~~~#
\end{lstlisting}
\begin{lstlisting}
#~~~~~~~~~~~~~~~~~~~~~~~~~~~~~~~~~~~~#
\end{lstlisting}
\begin{lstlisting}
# Please copy the path of where the current data-set is located and insert it here
\end{lstlisting}
Monthly\_EA <- read\_excel("EA\_data\_monthly.xlsx")

\begin{lstlisting}
# Step 3. Prepare the Data #####################################################
\end{lstlisting}
\begin{lstlisting}
# Exclude all the Data, that carries NA in Jan 1999 (Start of Swap-Data)
\end{lstlisting}
Monthly\_EA\_numeric <- Monthly\_EA[, -1]  \# Exclude the first column (dates)

Monthly\_EA\_ts <- ts(Monthly\_EA\_numeric, \#Convert into a Time-Series

start = c(1986, 12), \# Start of Data-Set in December 1986

frequency = 12)      \# Monthly data

time\_index <- time(Monthly\_EA\_ts)  \# Get the time index

years <- floor(time\_index)

months <- round((time\_index - years) * 12 + 1)

index\_start <- which(years == 1999 \& months == 01) \#Swap Data Starts 1999

index\_start \# Nr. of Rows in Data-Set where Swaps start.

columns\_with\_na <- is.na(Monthly\_EA\_ts[index\_start, ])

Monthly\_EA\_cleaned\_ts <- Monthly\_EA\_ts[, !columns\_with\_na]

Monthly\_EA\_cleaned\_ts <- Monthly\_EA\_cleaned\_ts[time(Monthly\_EA\_cleaned\_ts)

>= time(Monthly\_EA\_ts)[index\_start], ]

data <- ts(Monthly\_EA\_cleaned\_ts, start = c(1999, 1), frequency = 12)

data

\begin{lstlisting}
# Save Data into individual Series and Transform them to Log-Levels (except Rates)
\end{lstlisting}
\begin{lstlisting}
# Taking Logs allows us to interpret Model output as Elasticities / % Changes
\end{lstlisting}
infl\_us <- na.omit(data[,colnames(data) == "infl\_US"]) \# US-Inflation Rate

Y <- na.omit(data[,colnames(data) == "ind\_prod"]) \# Growth-Rate Industrial Prod. (EA)

Price <- na.omit(data[,colnames(data) == "infl"])  \# Inflation (EA)

r\_s <-  na.omit(data[,colnames(data) == "EUR\_m\_refi"]) \# EZB Refinancing Rate

cdty <- log(na.omit(data[,colnames(data) == "Com\_Index"])) \# Commodity Price Index: - Taking Logs

eur\_stoxx <- log(na.omit(data[,colnames(data) == "euro\_stoxx"])) \# Euro-Stoxx50 Index: - Taking Logs

pmi <- na.omit(data[,colnames(data) == "m\_PMI\_US"]) \#PMI

r\_b\_2 <-  na.omit(data[,colnames(data) == "ICEIB2Y"]) \# 2Yr Swap

r\_b\_5 <-  na.omit(data[,colnames(data) == "ICEIB5Y"]) \# 5Yr Swap

r\_b\_10 <-  na.omit(data[,colnames(data) == "ICEIB10"]) \# 10Yr Swap

r\_b\_30 <-  na.omit(data[,colnames(data) == "ICEIB30"]) \# 30Yr Swap

\begin{lstlisting}
# Define a Subset of Variables that are transformed, making sure the variables length matches.
\end{lstlisting}
\begin{lstlisting}
# ! The Ordering of Variables in the Data-Frame matters for Identification later on !
\end{lstlisting}
\begin{lstlisting}
# We will use exact Identification based a Cholesky Decomposition.
\end{lstlisting}
\begin{lstlisting}
# A1: Data ordered before the ECB Refi Rate is observed by the ECB when setting the Rate
\end{lstlisting}
\begin{lstlisting}
# A2: Data ordered before the ECB Refi Rate reacts with a time-lag to Monetary Policy
\end{lstlisting}
\begin{lstlisting}
# A3: Data ordered after the ECB Refi Rate reacts immediately to monetary policy
\end{lstlisting}
\begin{lstlisting}
# Assumptions are based on https://www.nber.org/system/files/working_papers/w6400/w6400.pdf
\end{lstlisting}
\begin{lstlisting}
# A discussion on these assumptions and alternative identification schemes are given by
\end{lstlisting}
\begin{lstlisting}
# https://www.sciencedirect.com/science/article/abs/pii/S1574004816000045
\end{lstlisting}
\begin{lstlisting}
# The Code may allow for identification via sign-restrictions, this is infeasible however
\end{lstlisting}
\begin{lstlisting}
# for models with large sets of variables.
\end{lstlisting}
\begin{lstlisting}
#dat <- cbind(infl_us[1:308], Y[1:308], Price[1:308], r_s[1:308], pmi[1:308],cdty[1:308], eur_stoxx[1:308], r_b_30[1:308], r_b_10[1:308], r_b_5[1:308],r_b_2[1:308])
\end{lstlisting}
dat <- cbind(infl\_us , Y, Price , r\_s , pmi ,cdty, eur\_stoxx , r\_b\_30 , r\_b\_10 , r\_b\_5 ,r\_b\_2 )

colnames(dat) <- c("Inflation\_US", "Industrial\_Prod", "Inflation", "ECB\_Refi","Manufacturing\_PMI(US)", "Commodity\_Indx", "EUR\_Stoxx50", "SWAP\_Rate\_30Yr","SWAP\_Rate\_10Yr", "SWAP\_Rate\_5Yr", "SWAP\_Rate\_2Yr")

dat <- ts(dat, start = c(1999,01), frequency = 12)

\begin{lstlisting}
### Draw a Correlogramm to check for colinearity
\end{lstlisting}
\begin{lstlisting}
# Correlation matrix and p-values
\end{lstlisting}
res2 <- rcorr(as.matrix(dat[,-c(8:11)]))

\begin{lstlisting}
# Insignificant correlations are crossed
\end{lstlisting}
\begin{lstlisting}
# In case R displays an error message, please ignore it
\end{lstlisting}
corrplot(res2\$r, type = "lower", order = "hclust",

p.mat = res2\$P, sig.level = 0.05, insig = "blank",

tl.cex = 0.8, tl.col = "black", \# Adjust text size and color

tl.labels = colnames(dat)) \# Use math expression labels

\begin{lstlisting}
####### Step 4: Building the Model  ############################################
\end{lstlisting}
\begin{lstlisting}
# Setting the Minnesota Prior up.
\end{lstlisting}
mn <- bv\_minnesota(lambda = bv\_lambda(mode = 0.2, sd = 0.4, min = 0.0001, max = 5),

alpha = bv\_alpha(mode = 2), var = 1e07)

\begin{lstlisting}
# Setting up sum of coefficients and unit-root prior = Assumes that, given a
\end{lstlisting}
\begin{lstlisting}
# weak signal, that the model converges / shrinks towards a random walk.
\end{lstlisting}
\begin{lstlisting}
# Usually a sensible assumption for the regularization of financial time-series.
\end{lstlisting}
soc <- bv\_soc(mode = 1, sd = 1, min = 1e-04, max = 50) \#sum-of-coefficients prior.

sur <- bv\_sur(mode = 1, sd = 1, min = 1e-04, max = 50) \#unit-root prior.

\begin{lstlisting}
# Setting up automatic hierarchical prior - selection.
\end{lstlisting}
priors <- bv\_priors(hyper = "auto", mn = mn, soc = soc, sur = sur)

\begin{lstlisting}
# Settings for the Methropolis-Hastings Algorithm, standard settings as in
\end{lstlisting}
\begin{lstlisting}
# Kushnig & Vashold (2021)
\end{lstlisting}
mh <- bv\_metropolis(scale\_hess = c(0.05, 0.0001, 0.0001),

adjust\_acc = TRUE, acc\_lower = 0.25, acc\_upper = 0.45)

\begin{lstlisting}
# Fitting the Model
\end{lstlisting}
mod <- bvar(dat, \#Call in Dataset (no NA, including only Series that should be modeled)

lags = 5, \# 12 Lags, you may play with it if there are convergence problems.

n\_draw = 50000, n\_burn = 25000, n\_thin = 5, \# Higher degree of thinning to aid stability

priors = priors, mh = mh, verbose = TRUE)

\begin{lstlisting}
####### Step 5: Diagnostics   ##################################################
\end{lstlisting}
\begin{lstlisting}
# Summary of the Model
\end{lstlisting}
summary(mod)

\begin{lstlisting}
# Assess convergence of the MCMC algorithm
\end{lstlisting}
\begin{lstlisting}
# The Trace Plots should concentrate around specific values and not "wander off"
\end{lstlisting}
plot(mod)

mod\_mcmc <- coda::as.mcmc(mod, vars = "lambda") \#Convert the Output of Prior-Optimization into coda

resids <- ts(residuals(mod), start = c(1998,12), frequency =12)

resids\_df <- as.data.frame(resids)

\begin{lstlisting}
# Plot the Residuals
\end{lstlisting}
plot(resids[,1:6])

plot(resids[,7:11])

\begin{lstlisting}
# MCMC Convergence Metrics
\end{lstlisting}
mod\_mcmc <- as.mcmc(mod)

autocorr.plot(mod\_mcmc) \#Autocorrelation in Prior Hyperparameters

\begin{lstlisting}
# Strong Autocorrelation = Possibly weak convergence of MCMC chain
\end{lstlisting}
geweke.plot(mod\_mcmc) \#Geweke 1999

\begin{lstlisting}
#Most z-scores should lie close to zero-line at least inside confidence bounds
\end{lstlisting}
crosscorr.plot(mod\_mcmc)

\begin{lstlisting}
# High Correlation = Possible Issue with Colinearity
\end{lstlisting}
\begin{lstlisting}
# Effecitve Sample Size (ESS), accounting for Autocorrelation
\end{lstlisting}
effectiveSize(mod\_mcmc)

\begin{lstlisting}
#ESS > 100: Good, sufficient for most inference.
\end{lstlisting}
\begin{lstlisting}
#ESS < 100: Concerning, suggests poor mixing or high autocorrelation.
\end{lstlisting}
\begin{lstlisting}
#ESS < 30: Serious issues; inference is unreliable.
\end{lstlisting}
\begin{lstlisting}
# In-Sample Predictions (one period foreward):
\end{lstlisting}
\begin{lstlisting}
# Give In-Sample Fitted Values and plot them to actually realized values.
\end{lstlisting}
fitted\_ts <- ts(fitted(mod, type = "mean"),

start = c(2000,1), \#Actual Start + Number of Lags

frequency = 12)

actual\_ts <- ts(dat[-c(1:5,309,310),], start = c(2000,1), frequency = 12)

\begin{lstlisting}
# Calculate RMSE (forecast errors):
\end{lstlisting}
rmse(mod)

\begin{lstlisting}
# R-Squared and adjusted R-Squared
\end{lstlisting}
\begin{lstlisting}
## Explained Variance after accounting for amount of parameters in one equation
\end{lstlisting}
resid <- residuals(mod)

avrg <- mean(dat[,11])

\begin{lstlisting}
# 2Y
\end{lstlisting}
r2\_2Y <- 1-sum((resid[,11])\textasciicircum{}2)/sum((dat[,11]-avrg)\textasciicircum{}2)

r2\_2Y

ar2\_2Y <- 1-(1-r2\_2Y)*(length(Y)-1)/(length(Y)-5*11-1)

ar2\_2Y

\begin{lstlisting}
# 5Y
\end{lstlisting}
r2\_5Y <- 1-sum((resid[,10])\textasciicircum{}2)/sum((dat[,10]-avrg)\textasciicircum{}2)

r2\_5Y

ar2\_5Y <- 1-(1-r2\_5Y)*(length(Y)-1)/(length(Y)-5*10-1)

ar2\_5Y

\begin{lstlisting}
# 10Y
\end{lstlisting}
r2\_10Y <- 1-sum((resid[,9])\textasciicircum{}2)/sum((dat[,9]-avrg)\textasciicircum{}2)

r2\_10Y

ar2\_10Y <- 1-(1-r2\_10Y)*(length(Y)-1)/(length(Y)-5*9-1)

ar2\_10Y

\begin{lstlisting}
# 30Y
\end{lstlisting}
r2\_30Y <- 1-sum((resid[,8])\textasciicircum{}2)/sum((dat[,8]-avrg)\textasciicircum{}2)

r2\_30Y

ar2\_30Y <- 1-(1-r2\_30Y)*(length(Y)-1)/(length(Y)-5*8-1)

ar2\_30Y

\begin{lstlisting}
######## Backtest #############
\end{lstlisting}
\begin{lstlisting}
# Parameters
\end{lstlisting}
horizon <- 12

backtest\_obs <- 240 \# 20 Years inside Test-Set

\begin{lstlisting}
# Get the total number of observations
\end{lstlisting}
n <- nrow(dat)

\begin{lstlisting}
# Start backtesting from (n - backtest_obs + 1) to the end
\end{lstlisting}
start\_idx <- n - backtest\_obs + 1

\begin{lstlisting}
# Initialize a list to store forecasts
\end{lstlisting}
forecasts <- list()

forecast\_2Yr <- data.frame()

forecast\_5Yr <-  data.frame()

forecast\_10Yr <-  data.frame()

forecast\_30Yr <-  data.frame()

\begin{lstlisting}
# Perform rolling forecasts
\end{lstlisting}
for (i in start\_idx:(n - horizon)) \{

\begin{lstlisting}
# Define the training data up to the current point
\end{lstlisting}
train\_data <- dat[(i-5):i, ]

\begin{lstlisting}
# Generate forecast for the next 'horizon' periods
\end{lstlisting}
forecast <- predict(mod, horizon = horizon, newdata = train\_data, conf\_bands = c(0.05))

forecast\_2Yr <- rbind(forecast\$quants[,12,11], forecast\_2Yr)

forecast\_5Yr <- rbind(forecast\$quants[,12,10], forecast\_5Yr)

forecast\_10Yr <- rbind(forecast\$quants[,12,9], forecast\_10Yr)

forecast\_30Yr <- rbind(forecast\$quants[,12,8], forecast\_30Yr)

\begin{lstlisting}
# Store the forecast results
\end{lstlisting}
forecasts[[i - start\_idx + 1]] <- forecast

\}

df\_fcast\_2Yr <- as.data.frame(forecast\_2Yr)

df\_fcast\_5Yr <- as.data.frame(forecast\_5Yr)

df\_fcast\_10Yr <- as.data.frame(forecast\_10Yr)

df\_fcast\_30Yr <- as.data.frame(forecast\_30Yr)

actual\_2Yr <- dat[c(start\_idx+horizon):length(dat[,11]),11]

actual\_5Yr <- dat[c(start\_idx+horizon):length(dat[,11]),10]

actual\_10Yr <- dat[c(start\_idx+horizon):length(dat[,11]),9]

actual\_30Yr <- dat[c(start\_idx+horizon):length(dat[,11]),8]

rmse\_2Yr <- sqrt(1/length(actual\_2Yr)*sum((actual\_2Yr - rev(forecast\_2Yr[,2]))\textasciicircum{}2))

error\_bvar <- actual\_2Yr - rev(forecast\_2Yr[,2])

\begin{lstlisting}
# DM Test #############
\end{lstlisting}
\begin{lstlisting}
# Test if predictions are better than e.g. arima's predictions
\end{lstlisting}
\begin{lstlisting}
# Arima_error must be created inside the ARIMA Script (Backtest_VAR_ARIMA) and must be inside your Environment
\end{lstlisting}
\begin{lstlisting}
# dm.test(error_bvar, arima_error, alternative = "less")
\end{lstlisting}
rmse\_5Yr <- sqrt(1/length(actual\_5Yr)*sum((actual\_5Yr - rev(forecast\_5Yr[,2]))\textasciicircum{}2))

rmse\_10Yr <- sqrt(1/length(actual\_10Yr)*sum((actual\_10Yr - rev(forecast\_10Yr[,2]))\textasciicircum{}2))

rmse\_30Yr <- sqrt(1/length(actual\_30Yr)*sum((actual\_30Yr - rev(forecast\_30Yr[,2]))\textasciicircum{}2))

\begin{lstlisting}
# List of forecast Tenors and corresponding actual data
\end{lstlisting}
forecast\_Tenors <- list(

"2Yr" = list(forecast = forecast\_2Yr, actual = actual\_2Yr),

"5Yr" = list(forecast = forecast\_5Yr, actual = actual\_5Yr),

"10Yr" = list(forecast = forecast\_10Yr, actual = actual\_10Yr),

"30Yr" = list(forecast = forecast\_30Yr, actual = actual\_30Yr)

)

\begin{lstlisting}
# Initialize a combined data frame for plotting
\end{lstlisting}
combined\_data <- data.frame()

\begin{lstlisting}
# Loop through each Tenor to prepare data
\end{lstlisting}
for (Tenor\_name in names(forecast\_Tenors)) \{

\begin{lstlisting}
# Extract forecast and actual data for this Tenor
\end{lstlisting}
forecast\_data <- forecast\_Tenors[[Tenor\_name]]\$forecast

actual\_data <- forecast\_Tenors[[Tenor\_name]]\$actual

\begin{lstlisting}
# Reverse the forecast data (mean, lower, upper)
\end{lstlisting}
forecast\_mean <- rev(forecast\_data[, 2])

forecast\_lower <- rev(forecast\_data[, 1])

forecast\_upper <- rev(forecast\_data[, 3])

\begin{lstlisting}
# Calculate squared errors for this Tenor
\end{lstlisting}
squared\_errors <- (actual\_data - forecast\_mean)\textasciicircum{}2

\begin{lstlisting}
# Append to the combined data frame
\end{lstlisting}
combined\_data <- rbind(

combined\_data,

data.frame(

Time = time(dat)[(length(dat[,1]) - length(actual\_data) + 1):length(dat[,1])],

Actual = actual\_data,

Forecast = forecast\_mean,

Lower = forecast\_lower,

Upper = forecast\_upper,

Squared\_Error = squared\_errors,

Tenor = Tenor\_name

)

)

\}

\begin{lstlisting}
# Plot all forecasts in a single page using facets
\end{lstlisting}
ggplot(combined\_data, aes(x = Time)) +

geom\_ribbon(aes(ymin = Lower, ymax = Upper, fill = Tenor), alpha = 0.3) +

geom\_line(aes(y = Actual, color = "Actual"), size = 1) +

geom\_line(aes(y = Forecast, color = Tenor), size = 1) +

facet\_wrap(\textasciitilde{}Tenor, scales = "free\_y") +

labs(

y = "Value",

x = "Time",

color = "Point Values",

fill = "Confidence Intervalls"

) +

theme\_minimal() +

scale\_color\_manual(values = c("Actual" = "blue", "2Yr" = "red", "5Yr" = "green", "10Yr" = "purple", "30Yr" = "orange")) +

theme(legend.position = "bottom")

\begin{lstlisting}
######## Step 6: Simulation / Impulse Response Analysis #######################
\end{lstlisting}
\begin{lstlisting}
# With standard Cholesky-Decomposition
\end{lstlisting}
opt\_irf <- bv\_irf(horizon = 12, identification = TRUE, fevd = TRUE)

print(opt\_irf)

irf(mod) <- irf(mod, opt\_irf, conf\_bands = c(0.05, 0.16))

plot(irf(mod), area = TRUE)

\begin{lstlisting}
########## Step 7: Save IRFs ###################################################
\end{lstlisting}
str(mod\$irf\$quants)

\begin{lstlisting}
# Possibly change name if identy of variable / shock 1 changed
\end{lstlisting}
us\_infl\_shock <- data.frame(mod\$irf\$quants[, , , 1]) \#All quantiles, all Periods, all variables (responses), just 1 shock (ordering like in dat - ts)

industrial\_ouput\_shock <- data.frame(mod\$irf\$quants[, , , 2])

inflation\_shock <- data.frame(mod\$irf\$quants[, , , 3])

ecb\_shock <- data.frame(mod\$irf\$quants[, , , 4])

pmi\_shock <- data.frame(mod\$irf\$quants[, , , 5])

commodity\_shock <- data.frame(mod\$irf\$quants[, , , 6])

eurstoxx\_shock <- data.frame(mod\$irf\$quants[, , , 7])

\begin{lstlisting}
# Save IRFs
\end{lstlisting}
writexl::write\_xlsx(us\_infl\_shock, path = "irfs\_US\_infl.xlsx")

writexl::write\_xlsx(pmi\_shock, path = "irfs\_PMI\_shock.xlsx")

writexl::write\_xlsx(industrial\_ouput\_shock, path = "irfs\_ind\_prod.xlsx")

writexl::write\_xlsx(inflation\_shock, path = "irfs\_infl.xlsx")

writexl::write\_xlsx(ecb\_shock, path = "irfs\_ecb.xlsx")

writexl::write\_xlsx(commodity\_shock, path = "irfs\_commodity.xlsx")

writexl::write\_xlsx(eurstoxx\_shock, path = "irfs\_eurstoxx.xlsx")
\end{document}